\documentclass[10pt,a4paper]{article}
\usepackage[utf8]{inputenc}
\usepackage{a4wide}
\usepackage{graphicx}
\usepackage{multirow}
\usepackage{cite}
\usepackage{hyperref}
\usepackage{color}
\usepackage{cancel}
\usepackage{adjustbox}

\usepackage[table]{xcolor}

\usepackage[title, titletoc]{appendix}

\usepackage{amsmath}
\usepackage{verbatim} 

\definecolor{boxcolor}{RGB}{235,245,255}
\newcommand{\mybox}[1]{\begin{center}\fcolorbox{black}{boxcolor}{\parbox[c]{16.5cm}{#1}}\end{center}}
\newcommand{\inbox}[2]{\begin{tabular}{rl}{\tt In[#1]:= } #2\end{tabular}}
\newcommand{\out}[2]{\begin{tabular}{rl}{\tt Out[#1]:= } #2\end{tabular}}
\newcommand{\outbox}[2]{\begin{tabular}{rl}{Prints $\Rightarrow$\hspace{0.85mm}} #2\end{tabular}}
\newcommand{\boxsplit}{\rule{16.5cm}{0.5pt}}
\newcommand{\gibspace}[2]{\begin{tabular}{rl}{\tt \hphantom{aaaaaaa} } #2\end{tabular}}

\begin{document}

\begin{titlepage}
\begin{flushright}
\end{flushright}
\begin{flushright}
\today
\end{flushright}

\begin{flushright}
ZU-TH 84/25
\end{flushright}
\begin{flushright}
\end{flushright}
\vspace{4truecm}

\begin{center}

{\Large\bf Sunset integrals with up to three mass scales in chiral perturbation theory: a comparative study of the Mellin-Barnes representation technique} 
\bigskip

{\bf  Balasubramanian Ananthanarayan$^a$, Sumit Banik$^{b,c}$, V\'eronique Bernard$^d$,\\ Samuel Friot$^{d,e}$, Shayan Ghosh and Ulf-G. Mei{\ss}ner$^{f,g,h}$}\\[0.5cm]
{$^a$ Centre for High Energy Physics, Indian Institute of Science, \\
Bangalore-560012, Karnataka, India}\\[0.25cm]
{$^b$ Physik-Institut, Universit\"at Z\"urich, Winterthurerstrasse 190, 8057 Z\"urich, Switzerland}\\[0.25cm]
{$^c$ PSI Center for Neutron and Muon Sciences, 5232 Villigen PSI, Switzerland}\\[0.25cm]
{$^d$ Universit\'e Paris-Saclay, CNRS/IN2P3, IJCLab, 91405 Orsay Cedex, France} \\[0.25cm]
{$^e$ Univ Lyon, Univ Claude Bernard Lyon 1, CNRS/IN2P3 IP2I UMR5822,\\
 F-69622 Villeurbanne Cedex, France} \\[0.25cm]
{$^f$ Helmholtz-Institut f\"ur Strahlen- und Kernphysik and Bethe Center for Theoretical Physics, \\ Universit\"at Bonn, D-53115 Bonn, Germany}\\[0.25cm]
{$^g$Institute for Advanced Simulation (IAS-4), Forschungszentrum J\"ulich, D-52425 J\"ulich, Germany} \\[0.25cm]
{$^h$ Peng Huanwu Collaborative Center for Research and Education, International Institute for Interdisciplinary and Frontiers, Beihang University, Beijing 100191, China}
\end{center}
\vspace{1.0cm}
\begin{abstract}

Sunset integrals are among the simplest of two-loop integrals that appear in perturbative quantum field theories and possess up to four distinct mass scales. By means of integration by parts identities, they can be written in terms of four distinct master integrals. In this article, we discuss the independent configurations of on-shell and off-shell sunset master integrals with one, two and three mass scales that arise in chiral perturbation theory. We derive Mellin-Barnes integral representations of these integrals and analytically solve them using various methods to obtain exact results in the form of single and double convergent series of the hypergeometric type, for the values of the mass parameters that allow us to do so. We then discuss how to analytically continue the results to other regions of the parameters and conclude by discussing a few applications in chiral perturbation theory. 
\end{abstract}

\pagebreak

\tableofcontents
\vfill
\end{titlepage}

\section{Introduction}

\begin{figure}[hbtp]
\centering
\includegraphics[scale=0.55]{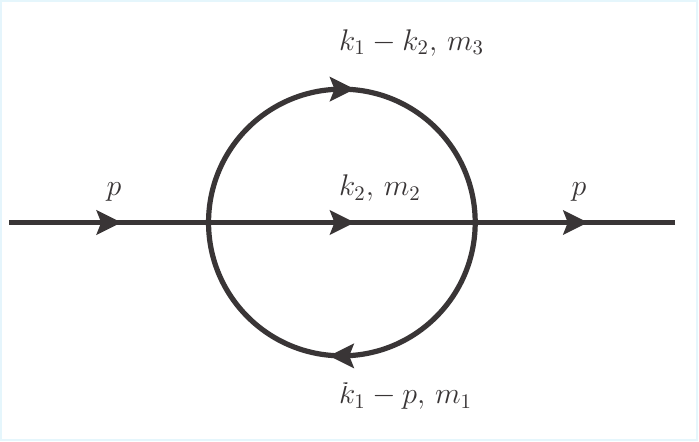}
\caption{The sunset diagram}
\label{Fig:Sunset}
\end{figure}

The self-energy diagram shown in Fig.~\ref{Fig:Sunset} is one of the simplest of the two-loop integrals that appear in the perturbation expansion of quantities in various quantum field theories. Known in the literature as the \textit{sunset} (or sunrise) integral,  it is defined as:
\begin{align}
	H_{\{a_1, a_2, a_3\}}(m_1, m_2, m_3; p^2) \equiv \int  \frac{d^dk_1 d^dk_2}{\left[ (k_1-p)^2 - m_1^2 +i0\right]^{\alpha} \left[ k_2^2 - m_2^2 +i0\right]^{\beta} \left[ (k_1-k_2)^2 - m_3^2 +i0\right]^{\gamma}}
	\label{Eq:Sunset}
\end{align}
where the external momentum $p^2$ can take any value. The most general sunset integral, therefore, can have up to four independent mass scales and arbitrary powers of the propagators $(a_1,a_2,a_3)$ can be reduced using integration by parts into a linear combination of a maximum of four master integrals \cite{Tarasov:1997kx} with $(a_1,a_2,a_3) = (1,1,1),(2,1,1),(1,2,1),(1,1,2)$. The number of master integrals is directly related to the number of mass scales, namely $n$ mass scales correspond to $n$ master integrals. Hereafter, we will discuss the master integrals and will refer to them as sunset integrals.

Much work has been done on the evaluation of the sunset integrals both numerically and analytically~\cite{Amoros:1999dp, Kaiser:2007kf, Berends:1997vk, Davydychev:1992mt, Aguilar:2008qj, Gasser:1998qt, Czyz:2002re, Martin:2003qz, Berends:1993ee, Adams:2015gva,Bogner:2019lfa}. A review of some of the methods used in the evaluation of sunset (and other Feynman) integrals is given in~\cite{Ananthanarayan2025:EPJSTReview}. Analytic results for the one mass scale sunset integral and the two mass scale sunsets that fall into configurations known as the threshold and pseudo-threshold have been given in \cite{Berends:1997vk, Kaiser:2007kf}. For two mass scale configurations that do not fall in the threshold and pseudo-threshold categories, analytic results in the form of single and double infinite series have been given in \cite{Ananthanarayan:2016pos}. The analytical results above are those in the limit $\epsilon \rightarrow 0$, where $\epsilon$ is the small parameter that is usually used in dimensional regularisation. The most general four-mass scale sunset integral has been calculated analytically using Mellin-Barnes (MB) representations in \cite{Berends:1993ee} and \cite{Adams:2015gva}. In \cite{Berends:1993ee}, the full $\epsilon$ dependence is given, and Lauricella functions have been used, while in \cite{Adams:2015gva}, the evaluation is made in terms of newly defined generalised elliptic functions. 

In this pedagogical article, we detail the calculations performed to obtain and extend some of the results used in \cite{Ananthanarayan:2016pos,Ananthanarayan:2017yhz,Ananthanarayan:2017qmx,Ananthanarayan:2018irl}. That is, we give full $\epsilon$-dependent analytical results for the master integrals of the sunset diagram with up to three independent mass scales using the MB representation technique. For pedagogical purposes, we employ different approaches to derive MB representations of sunsets, as well as to analytically evaluate the corresponding integrals. In particular, we show in a detailed way how to use $\texttt{AMBRE}$ \cite{Gluza:2007rt} and other techniques such as the one inspired by the method of brackets \cite{Prausa:2017frh} or the recent work \cite{Belitsky:2022gba} to derive MB representations where, in general, non-straight contours appear in the corresponding MB integrals. We will also show how one can obtain MB representations where the contours of the MB integrals are, from the very beginning and at all steps of the derivation, kept as straight lines parallel to the imaginary axis in the complex planes of the integration variables. This, in turn, allows one to use the method described in \cite{Friot:2011ic} to evaluate the MB integrals, while MB representations with non-straight contours obtained from the previously mentioned methods will be computed using the more recent and general techniques of conic hulls and triangulations \cite{Ananthanarayan:2020fhl,Banik:2022bmk,Banik:2023rrz,Banik:2024jqj,Banik:2024ann,MBCH} implemented in \texttt{MBConicHulls}~\cite{Ananthanarayan:2020fhl}.

We believe that mixing and comparing all these different techniques, applied to the relatively simple yet non-trivial example of sunsets, can provide, in addition to the cross-checks they offer each other, enlightening pedagogical applications of some of the various MB calculational methods available on the market.

After providing a detailed derivation of the MB representations of the master integrals and the analytic results for the latter, we address a second problem. Direct evaluation of the MB representations of the sunsets does not permit one to obtain series solutions of these integrals that converge for all possible values of the mass parameters. Therefore, using the method presented in~\cite{Ananthanarayan:2020xut} and already applied in the case of the four-mass scale sunset in~\cite{Ananthanarayan:2019icl}, and the method of \cite{Friot:2022dme}, we show how to analytically continue the results so as to obtain expressions which do converge for these values. and therefore how to obtain complete results (\textit{i.e.} for nearly all possible values of the mass parameters) in terms of single and double infinite series of the hypergeometric type. We then provide an overview of how these results are applied in both mesonic and baryonic chiral perturbation theory ($\chi$-PT), and list some important results.  


This paper is organised as follows. In Section~\ref{Sec:MB}, we give an overview of the MB method
and the different ways to resolve its singularities in the dimensional-regularisation parameter $\epsilon$. Next, in Section~\ref{AMBRE}, we show how to derive the MB representation of sunset integrals using \texttt{AMBRE}. Then, in Section~\ref{SunsetHyperRep1}, we discuss how to evaluate the sunset MB integral and obtain analytic series solutions using \texttt{MBConicHulls}. In Section~\ref{Discussions}, we explain why it is important to compare the result obtained in the previous section with an alternative approach, which we present in Sections~\ref{Sec:TwoMassScale} and~\ref{Sec:ThreeMassScale} for two-mass- and three-mass-scale sunset integrals. In the three-mass-scale case, we derive multiple MB representations for each master integral to demonstrate the use of our new approach to resolving singularities and to place it in the context of currently existing approaches. In Section~\ref{Sec:AC}, we extend the results of the previous section by analytic continuation to obtain series solutions for the sunsets in regions of the mass ratios not previously covered. We end the paper with a summary in Section~\ref{Sec:Applications} of how the results of this work may be applied in $\chi$PT contexts. In Appendix~\ref{MoB}, we provide an explicit calculation of how to derive the sunset MB integral using an alternative technique. In Appendix~\ref{SunsetHyperRep2}, we provide explicit series solutions of various MB integrals considered in the main text. Finally, in Appendices~\ref{Sec:H111kResults},~\ref{Sec:H211Result}, and~\ref{Sec:H211Results}, we list the full results for the three-mass-scale sunsets.

\section{The Mellin-Barnes method \label{Sec:MB}}

We begin with a brief overview of the MB approach to solving Feynman integrals. For a more comprehensive review see \cite{Smirnov:2009up, Smirnov:2012gma, Tsikh:1998, Friot:2011ic}. The Mellin transform is defined as follows:
\begin{align}
	[M(f)](s) = \int\limits_0^\infty f(t)t^{s-1}dt, \hspace{0.2in} s \in \mathcal{C}
\end{align}
Its inverse is given by:
\begin{align}
	[M^{-1} (g)](x) = \frac{1}{2\pi i} \int\limits_{c-i\infty}^{c+i\infty}  x^{-s}g(s)ds
\end{align}

The following formula derived from the inverse Mellin transform is used in high energy physics to write massive propagators as combinations of massless propagators:
\begin{align} \frac{1}{(1+X)^\nu}=\frac{1}{2\pi i}\int^{c+i \infty}_{c-i\infty} dz \, X^z \frac{\Gamma(-z) \Gamma(\nu+z)}{\Gamma(\nu)}	\hspace{0.5cm} \text{if } |\text{arg } X|<\pi \text{ and } c \equiv \text{Re}(z) \in ]-\text{Re}(\nu),0[
\label{Eq:MBformula}
\end{align}
The (possibly multi-fold) complex plane integral obtained after (possibly repeated) application of this formula, and evaluation of all the internal momenta integrals, is known as the MB representation of a Feynman integral. The integrand of the representation consists of products and ratios of gamma and polygamma functions whose arguments contain the MB integration parameters.

The evaluation of MB integrals may then be performed either numerically, or analytically by taking a series of residues. To analytically solve multi-fold MB integrals, we use the method of \cite{Friot:2011ic}. The advantage of this method over the conventional sequential closing of contours and adding of residues is that it often allows one to obtain more `cones' than the former, in addition to being more graphical and intuitive. The summing of residues in the process of analytically solving multi-fold MB integrals produce results that are in the form of multi-fold infinite series, dependent on multiple mass ratio parameters. The numerical values of these parameters dictate which residues are to be summed, and therefore the exact form of the series. The range of values of the parameters over which a sum converges defines a `cone'. The results of each cone are analytic continuations of each other.

To apply the method of \cite{Friot:2011ic}, straight line contours (parallel to imaginary axes) are needed for MB representations. The traditional approach to deriving MB representations requires that the contour separate the poles of the $\Gamma(...+z_i)$ from those of the $\Gamma(...-z_i)$ appearing in the numerator thereof, which may lead to non-straight contours. Instead, straight line contours are obtained at the stage of resolving the $\epsilon$ singularity. This refers to transforming or obtaining a representation so that one may take a Laurent expansion of the integrand around $\epsilon=0$. This is necessary to obtain the finite part of an integral in dimensional regularisation.

In what is called Strategy A in \cite{Smirnov:2009up, Smirnov:2012gma}, resolving the singularities in $\epsilon$ begins with a representation with non-straight contours, and then taking residues at each singularity that crosses the contours in the $\epsilon \rightarrow 0$ limit. In Strategy B of \cite{Smirnov:2009up, Smirnov:2012gma}, one begins with an MB representation and straight line contours, by assuming that the real parts of the arguments of all gamma functions in the numerator are positive, and solving for values of $\epsilon$ and the real part of the contour that satisfy this requirement. If $\epsilon=0$ is not part of the solution set, one analytically continues the MB representation in $\epsilon$ by taking residues of it at singularities that cross the straight line contour in the $\epsilon \rightarrow 0$ limit.

We adopt a strategy of mathematically requiring both straight line contours and the possibility of setting $\epsilon$ to be infinitesimally small throughout the derivation process. This ensures that the MB representation we obtain is ab initio suitable for solving using the method of \cite{Friot:2011ic}, and that we can solve for the finite part of the divergent integral by taking a Laurent series in $\epsilon$ at the (MB representation) integrand level. Though not necessary, it is sometimes easier to work with a representation where one can directly set $\epsilon=0$ in the integrals instead of taking a Laurent series. To obtain such a representation, depending on the case, it may be necessary to shift contours at the end of the above derivation process.

The result of resolving the singularities in all approaches is an MB representation of a Feynman integral consisting of several terms, each of which is a complex plane integral (not necessarily over the same number of variables) or an isolated non integral. We demonstrate our approach in the next section by deriving an MB representation for the single mass scale sunset.

\section{Mellin-Barnes representations of sunsets using $\texttt{AMBRE}$\label{AMBRE}}

The $\texttt{AMBRE}$ package \cite{Gluza:2007rt} allows to automatically derive MB representations of Feynman integrals using \textit{Mathematica}.

We are going to show how to use it for the cases considered in this article, namely $H_{\{1,1,1\}}(m_1,m_2,m_3;m_1^2)$, $H_{\{2,1,1\}}(m_1,m_2,m_3;m_1^2)$, $H_{\{1,1,1\}}(m_1,m_2,m_2;m_1^2)$, $H_{\{2,1,1\}}(m_1,m_2,m_2;m_1^2)$, $H_{\{1,1,1\}}(m_1,m_2,m_2;m_3^2)$ and $H_{\{1,1,1\}}(m_1,m_1,m_1;m_1^2)$. Other ways to derive MB representations can be used, as described in Sec.~\ref{Discussions} and App.~\ref{MoB}.

With the exception of $H_{\{1,1,1\}}(m_1,m_2,m_3;m_1^2)$, all sunsets above can be derived as particular or limiting cases of $H_{\{a_1,a_2,a_3\}}(m_1,m_2,m_3;m_1^2)$, therefore we will first focus on the latter\footnote{In fact, all sunsets can be obtained from the four scale case $H_{\{a_1,a_2,a_3\}}(m_1,m_2,m_3;p^2)$, see the discussion at the end of the present section as well as in the next one.}. 

In order to obtain the MB representation of $H_{\{a_1,a_2,a_3\}}(m_1,m_2,m_3;m_1^2)$ where the mass-shell constraint $p^2=m_1^2$ is added to Eq.~\ref{Eq:Sunset}, one simply enters the following code in a \textit{Mathematica} notebook where the package $\texttt{AMBRE}$ is first loaded 

\mybox{
\inbox{1}{& {\tt Get["AMBREv2.1.1.m"];}}\\
\boxsplit\\\\
\outbox{2}{& {\tt AMBRE by K.Kajda   ver: 2.1.1 
}\\
& {{\tt last modified Aug 2017}}}
}
and then executing
\mybox{
\inbox{2}{& ${\tt Propagators  = \{PR[k_1 - p, \, m_1, a_1]*PR[k_2,\,  m_2, a_1]*
    PR[k_1 - k_2, \,m_3, a_3] \}; }$}\\
\gibspace{}{& ${\tt invariants = \{p^2 \to m^2_1\};
}$}\\
\gibspace{}{& ${\tt PreFactor = \{1\};
}$}\\
\gibspace{}{& ${\tt Rep = MBrepr[PreFactor, 
  Propagators, \{k_2, k_1\}, Text \to False, 
  BarnesLemma \to True]
}$}\\
\boxsplit\\\\
\out{1}{& ${\tt \dfrac{(-1)^{a_1+a_2+a_3} \,m^4_1 \, m^4_2 \, (m^2_3)^{z1} \, (m^2_1)^{-a_1 - eps + z2} \, (m^2_2)^{-a_2 - a_3 - eps - z1 - z2}}{\Gamma[a_1] \Gamma[a_2] \Gamma[a_3]}
 \,\Gamma[- z1 ] \, \Gamma[- z2 ] }$\\
& ${\tt \dfrac{\Gamma[-eps - z1 -a_3 + 2] \Gamma[-2\,eps + 2 \, z2 -a_1 + 4] \, \Gamma[eps - z2 + a_1 - 2] \, \Gamma[z1 + z2 + a_3]}{\Gamma[-eps + z2 + 2] \Gamma[-2 \, eps + z2 - a_1 + 4]}  }$\\
& ${\tt \Gamma[eps +z1 + z2 + a_2 + a_3 -2] } $}\\
}

Let us note that in the above instructions, the integrations over momenta have been performed in the $k_2, k_1$ order and that the first Barnes lemma
\begin{multline}
	\int_{-i \infty}^{+i \infty}  \frac{dz}{2 \pi i} \Gamma(\alpha_1+z)\Gamma(\alpha_2+z) \Gamma(\beta_1-z) \Gamma(\beta_2-z)=\frac{ \Gamma(\alpha_1+\beta_1) \Gamma(\alpha_1+\beta_2)\Gamma(\alpha_2+\beta_1)\Gamma(\alpha_2+\beta_2)}{\Gamma(\alpha_1+\alpha_2+\beta_1+\beta_2)} 
\label{BL1}
\end{multline}
has been successfully applied to lower the number of MB folds during the steps of derivation of this MB representation, which reads\footnote{In $\texttt{AMBRE}$, each momentum integration of a Feynman integral comes with an $i\pi^{2-\epsilon}$ factor in the denominator. However, the latter do not appear in our definition of the sunsets given in Eq.~\ref{Eq:Sunset}, which is the reason why a squared factor of this type has been added overall to the output shown above (see Eq.~\ref{MBrepHa1a2a3}).}
\begin{multline}
	H_{\{a_1,a_2,a_3\}}(m_1,m_2,m_3;m_1^2) = (i\pi^{2-\epsilon})^2\frac{(-1)^{a_1+a_2+a_3}}{\Gamma(a_1)\Gamma(a_2)\Gamma(a_3)}(m_1^2)^{2-a_1-\epsilon} (m_2^2)^{2-a_2-a_3-\epsilon}  \\
	\times \int_{-i \infty}^{+i \infty}  \frac{dz_1}{2 \pi i} \int_{-i \infty}^{+i \infty}  \frac{dz_2}{2 \pi i}\left( \frac{m_3^2}{m_2^2} \right)^{z_1} \left( \frac{m_1^2}{m_2^2} \right)^{z_2}\Gamma(-z_1)\Gamma(-z_2) \Gamma(2-a_3-\epsilon-z_1) \Gamma(-2+a_1+\epsilon-z_2)\\ 
 \times\frac{ \Gamma(a_3+z_1+z_2) \Gamma(-2+a_2+a_3+\epsilon+z_1+z_2) \Gamma(4-a_1-2\epsilon+2z_2)}{\Gamma(4-a_1-2\epsilon+z_2) \Gamma(2-\epsilon+z_2)} 
\label{MBrepHa1a2a3}
\end{multline}
where the contours of integration, depending on the chosen values of the $a_i$, are in general non-straight and such that they do not split the set of poles of each of the gamma functions of the numerator of the integrand in different subsets.

We note here that a reverse ordering of the loop momenta in $\texttt{AMBRE}$, followed by an application of the second Barnes lemma
\begin{multline}
	\int_{-i \infty}^{+i \infty}  \frac{dz}{2 \pi i} \frac{\Gamma(\alpha_1+z)\Gamma(\alpha_2+z) \Gamma(\beta_1-z) \Gamma(\beta_2-z)\Gamma(\beta_3-z)}{\Gamma(\alpha_1+\alpha_2+\beta_1+\beta_2+\beta_3+z)}\\=\frac{ \Gamma(\alpha_1+\beta_1) \Gamma(\alpha_1+\beta_2)\Gamma(\alpha_1+\beta_3)\Gamma(\alpha_2+\beta_1) \Gamma(\alpha_2+\beta_2)\Gamma(\alpha_2+\beta_3)}{\Gamma(\alpha_1+\alpha_2+\beta_1+\beta_2)\Gamma(\alpha_1+\alpha_2+\beta_1+\beta_3)\Gamma(\alpha_1+\alpha_2+\beta_2+\beta_3)} 
\label{BL2}
\end{multline}
instead of the first one,
yields a slightly different two-fold MB representation (which in fact is related to the previous one by the change of variable $z_2= 2-a_2-a_3-\epsilon-z_1-z_3$):  
\begin{multline}
	H_{\{a_1,a_2,a_3\}}(m_1,m_2,m_3;m_1^2) = (i\pi^{2-\epsilon})^2\frac{(-1)^{a_1+a_2+a_3}}{\Gamma(a_1)\Gamma(a_2)\Gamma(a_3)}(m_1^2)^{4-a_1-a_2-a_3-2\epsilon}   \\
	\times \int_{-i \infty}^{+i \infty}  \frac{dz_1}{2 \pi i} \int_{-i \infty}^{+i \infty}  \frac{dz_3}{2 \pi i}\left( \frac{m_3^2}{m_1^2} \right)^{z_1} \left( \frac{m_2^2}{m_1^2} \right)^{z_3}\Gamma(-z_1)\Gamma(-z_3)  \Gamma(8-a_1-2a_2-2a_3-4\epsilon-2z_1-2z_3)\\ 
 \times\frac{ \Gamma(2-a_3-\epsilon-z_1)\Gamma(2-a_2-\epsilon-z_3) \Gamma(-2+a_2+a_3+\epsilon+z_1+z_3) \Gamma(-4+a_1+a_2+a_3+2\epsilon+z_1+z_3)}{\Gamma(6-a_1-a_2-a_3-3\epsilon-z_1-z_3) \Gamma(4-a_2-a_3-2\epsilon-z_1-z_3)} 
\label{MBrepHa1a2a3_v2}
\end{multline}
We will use this second MB representation at a later stage in this article, but from now on, let us focus on the first one.

From Eq.~\ref{MBrepHa1a2a3}, one can readily find the MB representations of $H_{\{1,1,1\}}(m_1,m_2,m_3;m_1^2)$ and $H_{\{2,1,1\}}(m_1,m_2,m_3;m_1^2)$ which respectively read
\begin{multline}
	H_{\{1,1,1\}}(m_1,m_2,m_3;m_1^2) = -(i\pi^{2-\epsilon})^2(m_1^2)^{1-\epsilon} (m_2^2)^{-\epsilon}  \\
	\times \int_{-i \infty}^{+i \infty}  \frac{dz_1}{2 \pi i} \int_{-i \infty}^{+i \infty}  \frac{dz_2}{2 \pi i}\left( \frac{m_3^2}{m_2^2} \right)^{z_1} \left( \frac{m_1^2}{m_2^2} \right)^{z_2}\Gamma(-z_1)\Gamma(-z_2) \Gamma(1-\epsilon-z_1) \Gamma(-1+\epsilon-z_2)\\ 
 \times\frac{ \Gamma(1+z_1+z_2) \Gamma(\epsilon+z_1+z_2) \Gamma(3-2\epsilon+2z_2)}{\Gamma(3-2\epsilon+z_2) \Gamma(2-\epsilon+z_2)} 
\label{MBrepH111}
\end{multline}
and
\begin{multline}
	H_{\{2,1,1\}}(m_1,m_2,m_3;m_1^2) = (i\pi^{2-\epsilon})^2(m_1^2)^{-\epsilon} (m_2^2)^{-\epsilon}  \\
	\times \int_{-i \infty}^{+i \infty}  \frac{dz_1}{2 \pi i} \int_{-i \infty}^{+i \infty}  \frac{dz_2}{2 \pi i}\left( \frac{m_3^2}{m_2^2} \right)^{z_1} \left( \frac{m_1^2}{m_2^2} \right)^{z_2}\Gamma(-z_1)\Gamma(-z_2) \Gamma(1-\epsilon-z_1) \Gamma(\epsilon-z_2)\\ 
 \times\frac{ \Gamma(1+z_1+z_2) \Gamma(\epsilon+z_1+z_2) \Gamma(2-2\epsilon+2z_2)}{\Gamma(2-2\epsilon+z_2) \Gamma(2-\epsilon+z_2)} 
\label{MBrepH211}
\end{multline}
Putting $m_2=m_3\doteq M$ and $m_1\doteq m$ in the latter two equations, one can apply the first Barnes lemma to obtain
\begin{multline}
	H_{\{1,1,1\}}(m,M,M;m^2) = -(i\pi^{2-\epsilon})^2(2m^2)^{1-2\epsilon} \left(\frac{M^2}{m^2}\right)^{-\epsilon}  \\
	\times \int_{-i \infty}^{+i \infty}  \frac{dz}{2 \pi i} \left( \frac{m^2}{M^2} \right)^{z}\frac{ \Gamma(-z)\Gamma(\frac{3}{2}-\epsilon+z) \Gamma(2-\epsilon+z) \Gamma(1+z) \Gamma(\epsilon+z) \Gamma(-1+\epsilon-z)}{\Gamma(3-2\epsilon+z) \Gamma(\frac{3}{2}+z)} 
\label{MBrepH111mM}
\end{multline}
and
\begin{multline}
	H_{\{2,1,1\}}(m,M,M;m^2) = (i\pi^{2-\epsilon})^2(4m^2M^2)^{-\epsilon}\\
	\times \int_{-i \infty}^{+i \infty}  \frac{dz}{2 \pi i} \left( \frac{m^2}{M^2} \right)^{z}\frac{ \Gamma(-z)\Gamma(\frac{3}{2}-\epsilon+z)\Gamma(1-\epsilon+z)  \Gamma(1+z) \Gamma(\epsilon+z) \Gamma(\epsilon-z)}{\Gamma(2-2\epsilon+z) \Gamma(\frac{3}{2}+z)} 
\label{MBrepH211mM}
\end{multline}
The case of $H_{\{1,1,1\}}(m,m,m;m^2)$, whose MB representation has no scale, can be trivially derived from Eq.~\ref{MBrepH111mM} by removing the dependency in the mass ratio.

Having considered all sunsets of interest except $H_{\{1,1,1\}}(m_1,m_2,m_2;m_3^2)$, it remains now to derive the MB representation of the latter. Using $\texttt{AMBRE}$ once again we get
\begin{multline}
	H_{\{1,1,1\}}(m_1,m_2,m_2;m_3^2) = -(i\pi^{2-\epsilon})^2(-m_3^2)^{1-2\epsilon}  \int_{-i \infty}^{+i \infty}  \frac{dz_1}{2 \pi i} \int_{-i \infty}^{+i \infty}  \frac{dz_3}{2 \pi i}\left(-\frac{m_2^2}{m_3^2} \right)^{z_1} \left(- \frac{m_1^2}{m_3^2} \right)^{z_3}\\
	\times \frac{\Gamma(-z_1)\Gamma(-z_3)\Gamma(2-2\epsilon-z_1)\Gamma(1-\epsilon-z_1)^2 \Gamma(1-\epsilon-z_3) \Gamma(-1+2\epsilon+z_1+z_3)}{\Gamma(2-2\epsilon-2z_1)\Gamma(3-3\epsilon-z_1-z_3)}
\label{MBrepH111m1m2m2m3}
\end{multline}
which, using the duplication formula
\begin{align}
	\Gamma(2z)=\Gamma(z)\Gamma\left(z+\frac{1}{2}\right)\frac{2^{2z-1}}{\sqrt{\pi}}
\label{duplication}
\end{align}
and the successive changes of variables $z_1=-\epsilon-z_1$ and $z_2=1-\epsilon+z_1-z_3$, we rewrite for later convenience as
\begin{multline}
	H_{\{1,1,1\}}(m_1,m_2,m_2;m_3^2) = -(i\pi^{2-\epsilon})^2\frac{\sqrt{\pi}}{2}(m_2^2)^{-\epsilon}(m_1^2)^{1-\epsilon}  \int_{-i \infty}^{+i \infty}  \frac{dz_1}{2 \pi i} \int_{-i \infty}^{+i \infty}  \frac{dz_2}{2 \pi i}\\
	\times \left(\frac{m_1^2}{4m_2^2} \right)^{z_1} \left(- \frac{m_3^2}{m_1^2} \right)^{z_2}\frac{\Gamma(-z_2)\Gamma(2-\epsilon+z_1)\Gamma(1+z_1)\Gamma(\epsilon+z_1) \Gamma(-z_1+z_2) \Gamma(-1+\epsilon-z_1+z_2)}{\Gamma(\frac{3}{2}+z_1)\Gamma(2-\epsilon+z_2)}
\label{MBrepH111m1m2m2m3v2}
\end{multline}
We conclude this section by the following remark: all the MB representations presented above are in fact special cases of the 3-fold MB representation of the most general four scale (off-shell) sunset $H_{\{a_1,a_2,a_3\}}(m_1,m_2,m_3;p^2)$, one possible expression of the latter being given by $\texttt{AMBRE}$ as 
\begin{multline}
	H_{\{a_1,a_2,a_3\}}(m_1,m_2,m_3;p^2) = (i\pi^{2-\epsilon})^2\frac{(-1)^{a_1+a_2+a_3}}{\Gamma(a_1)\Gamma(a_2)\Gamma(a_3)}(m_2^2p^2)^2(m_2^2)^{-a_2-a_3-\epsilon} (-p^2)^{-a_1-\epsilon}  \\
	\times \int_{-i \infty}^{+i \infty}  \frac{dz_1}{2 \pi i} \int_{-i \infty}^{+i \infty}  \frac{dz_2}{2 \pi i} \int_{-i \infty}^{+i \infty}  \frac{dz_4}{2 \pi i} \left( \frac{m_3^2}{m_2^2} \right)^{z_1} \left( \frac{-p^2}{m_2^2} \right)^{z_2} \left( \frac{m_1^2}{-p^2} \right)^{z_4}\Gamma(-z_1)\Gamma(-z_4) \Gamma(2-a_3-\epsilon-z_1) \\ 
 \times\frac{\Gamma(-2+a_1+\epsilon-z_2+z_4) \Gamma(a_3+z_1+z_2) \Gamma(-2+a_2+a_3+\epsilon+z_1+z_2) \Gamma(2-a_1-\epsilon-z_4)}{\Gamma(4-a_1-2\epsilon+z_2-z_4) } 
\label{MBrepHa1a2a3FourScales}
\end{multline}
Indeed, using the change of variable $z_3=2-a_1-\epsilon+z_2-z_4$, the mass-shell condition $p^2=m_1^2$ and another lemma of the Barnes type whose expression reads \cite{Jantzen:2012cb}
\begin{multline}
	\int_{-i \infty}^{+i \infty}  \frac{dz}{2 \pi i} e^{\pm i\pi z}\frac{\Gamma(\alpha-z)\Gamma(\beta_1+z) \Gamma(\beta_2+z)}{\Gamma(\gamma+z)}=e^{\pm i\pi\alpha}\frac{ \Gamma(\alpha+\beta_1) \Gamma(\alpha+\beta_2)\Gamma(\gamma-\alpha-\beta_1-\beta_2)}{\Gamma(\gamma-\beta_1)\Gamma(\gamma-\beta_2)} 
\label{BL3}
\end{multline}
one obtains Eq.~\ref{MBrepHa1a2a3}.

Similarly, putting $m_3=m_2$ and $p^2=m_3^2$ in Eq.~\ref{MBrepHa1a2a3FourScales}, using the first Barnes lemma shown in Eq.~\ref{BL1}, the change of variable $z_1= -\epsilon-z_2$ and the relabelling $z_4\rightarrow z_2$, one can easily derive Eq.~\ref{MBrepH111m1m2m2m3v2} when $a_1=a_2=a_3=1$.

This cascading behavior can also be checked at the level of the hypergeometric representations presented in the next section: all sunsets are in fact linked together and can be obtained from the four mass sunset, using some properties of the hypergeometric functions such as the Gauss summation theorem and other less-known reduction formulas. This will be explained in details in the next section.

We finally note that after the change of variable $z_2\rightarrow-z_1-z_2-\epsilon$ in Eq.~\ref{MBrepHa1a2a3FourScales}, followed by the re-labelling $z_1\rightarrow z_3$ and $z_4\rightarrow z_1$, we get, when $a_1=a_2=a_3=1$, the more symmetrical expression
\begin{multline}
	H_{\{1,1,1\}}(m_1,m_2,m_3;p^2) =-(i\pi^{2-\epsilon})^2(-p^2)^{1-2\epsilon}\\
 \times\int_{-i \infty}^{+i \infty}  \frac{dz_1}{2 \pi i} \int_{-i \infty}^{+i \infty}  \frac{dz_2}{2 \pi i} \int_{-i \infty}^{+i \infty}  \frac{dz_3}{2 \pi i} \left( \frac{m_1^2}{-p^2} \right)^{z_1} \left( \frac{m_2^2}{-p^2} \right)^{z_2} \left( \frac{m_3^2}{-p^2} \right)^{z_4}\Gamma(-z_1)\Gamma(-z_2) \Gamma(-z_3) \\ 
 \times\frac{\Gamma(1-\epsilon-z_1) \Gamma(1-\epsilon-z_2)\Gamma(1-\epsilon-z_3) \Gamma(2\epsilon-1+z_1+z_2+z_3)}{\Gamma(3-2\epsilon-z_1-z_2-z_3) } 
\label{MBrepH111FourScales}
\end{multline}
which agrees with the MB representation of $H_{\{1,1,1\}}(m_1,m_2,m_3;p^2)$ given in \cite{Berends:1993ee} once the overall factor $(i\pi^2(2\pi\mu)^{-2\epsilon})^{-2}$ coming from the convention of these authors is taken into account.
\section{Hypergeometric representations using $\texttt{MBConicHulls}$\label{SunsetHyperRep1}}

Having derived the MB representations of the sunsets in the previous section it is straightforward to obtain their exact hypergeometric representations, keeping full $\epsilon$ dependency, using the $\texttt{MBConicHulls}$ \textit{Mathematica} package \cite{Ananthanarayan:2020fhl,Banik:2022bmk,Banik:2023rrz,Banik:2024jqj,Banik:2024ann,MBCH}. The corresponding results are presented in the different subsections of the present section where it is also explained how some analytic cross-checks have been performed.

Let us do one calculation explicitly, taking the simple example of $H_{\{1,1,1\}}(m,M,M;m^2)$, whose MB representation is given in Eq.~\ref{MBrepH111mM}. After loading the $\texttt{MBConicHulls}$ package in a \textit{Mathematica} notebook,

\mybox{
\inbox{1}{& {\tt Get["MBConicHulls.wl"];}}\\
\boxsplit\\\\
\outbox{2}{& {\tt Last Updated: 8th December, 2023
}\\
& {{\tt Version 1.2.2 by S.Banik, S. Friot}}}
}

one enters the following instructions

\mybox{
\inbox{2}{& ${\tt SunsetMBRepOut  = MBRep\Big[-\bigl(\mathrm{i}\,\pi^{\,2-\epsilon}\bigr)^{2}
(2m^{2})^{\,1-2\epsilon} \left(\frac{M^{2}}{m^{2}}\right)^{-\epsilon},
\ \{z\},\ \left\{\frac{m^{2}}{M^{2}}\right\}, \ \Big\{\{-z,\tfrac{3}{2},-\epsilon+z,  }$}\\
\gibspace{}{& ${\tt  \qquad \qquad  \qquad \qquad  \qquad \qquad \,2-\epsilon+z,\,1+z,\,\epsilon+z,\,-1+\epsilon-z\},
\ \{\,3-2\epsilon+z,\,\tfrac{3}{2}+z\,\}\Big\}\Big];
}$}\\
\boxsplit\\\\
\outbox{1}{& ${\tt Non-Straight \,\, Contours. }$\\
& ${\tt Take \,\, Taken \, \, 0.17952  \,\, seconds }$\\
}\\
}

which define the MB representation of $H_{\{1,1,1\}}(m,M,M;m^2)$. One then proceeds to the derivation of the series representations of this MB integral by first finding the relevant sets of poles of the MB integrand:

\mybox{
\inbox{2}{& ${\tt ResolveSunsetMBRepOut  = ResolveMB\Big[ SunsetMBRepOut \Big];  }$}\\
\boxsplit\\\\
\outbox{1}{& ${\tt Degenerate \,\, case \,\, with \,\, 8 \,\, conic \,\, hulls }$\\
& ${\tt Found \,\, 2 \,\, series \,\, solutions. }$\\
& ${\tt Cardinality \,\, 2 :: \,\, Solution \,\, found \,\, 1. }$\\
& ${\tt Cardinality \,\, 4 :: \,\, Solution \,\, found \,\, 1. }$\\[8pt]
& ${\tt \textbf{Series Solution  1} :: \,\, Cardinality \,\, 2. \,\, Intersecting \,\, Conic \,\, Hulls \,\, \{C_1,\, C_6\}. }$\\
& ${\tt Set \,\, of \,\, poles :: \,\, \{\{\, n_1 \,\}, \{\, -1 + \epsilon + n_1 \,\}\}
     \,\, with \,\, master \,\, series \,\, characteristic \,\, list }$\\
& ${\tt  and \,\, variables \,\, \{\, n_1, \{\, m^{2}/M^{2} \,\} \}. }$\\[8pt]
& ${\tt \textbf{Series Solution 2} :: \,\, Cardinality \,\, 4. \,\, Intersecting \,\, Conic \,\, Hulls \,\, \{C_2,\, C_3,\, C_4,\, C_5\}. }$\\
& ${\tt Set \,\, of \,\, poles :: \,\, \{\{\, -\tfrac{3}{2} + \epsilon - n_1 \,\}, \{\, -2 + \epsilon - n_1 \,\}, \{\, -1 - n_1 \,\}, \{\, -\epsilon - n_1 \,\}\}
     \,\, with \,\, master }$\\
& ${\tt  series \,\, characteristic \,\, list \,\, and \,\, variables \,\, \{\, n_1, \{\, M^{2}/m^{2} \,\} \}. }$\\[8pt]
& ${\tt Time \,\, Taken \,\, 0.227345 \,\, seconds }$\\
}\\
}

The package tells us that two different series representations of $H_{\{1,1,1\}}(m,M,M;m^2)$ can be obtained from its MB representation. The first one has two terms (cardinality 2) whereas the other one has four terms. The package also gives the associated sets of poles of the MB integrand, as well as some informations about the master series (their characteristic list and variables). We refer the reader to \cite{Ananthanarayan:2020fhl} for more details about the latter.

To obtain for instance the first series representation explicitly, one can proceed as follows

\mybox{
\inbox{2}{
& ${\tt SeriesNumber = 1;}$\\
& ${\tt EvaluateSeriesOut = EvaluateSeries\Big[ ResolveSunsetMBRepOut, \{ \}, SeriesNumber \Big] ; }$
}\\
\boxsplit\\\\
\outbox{1}{
& ${\tt The \,\, series \,\, solution \,\, is \,\, a \,\, sum \,\, of \,\, the \,\, following \,\, 2 \,\, series. }$\\[8pt]

& ${\tt \textbf{Series Number  1} :: \, \frac{(-1)^{n_{1}}\,
2^{\,1-2\epsilon}\,
(m^{2})^{\,1-2\epsilon}\,
\left(\frac{m^{2}}{M^{2}}\right)^{n_{1}}\,
\left(\frac{M^{2}}{m^{2}}\right)^{-\epsilon} \pi^{\,4-2\epsilon}\,
\Gamma\!\big[-1+\epsilon+n_{1}\big]}{\Gamma\!\big[\tfrac{3}{2}+n_{1}\big]\,
 \Gamma\!\big[3-2\epsilon+n_{1}\big]}}$\\[6pt]
& $\displaystyle
\qquad \qquad \qquad \qquad \qquad{\,
\Gamma\!\big[\tfrac{3}{2}-\epsilon+n_{1}\big]\,
\Gamma\!\big[2-\epsilon+n_{1}\big]\,
\Gamma\!\big[\epsilon+n_{1}\big]}
\quad {\tt valid \,\, for } \, n_{1}\ge 0
$\\[10pt]

& ${\tt \textbf{Series  Number 2 }:: \, \frac{(-1)^{n_{1}}\,
2^{\,1-2\epsilon}\,
(m^{2})^{-2\epsilon}\,
\left(\frac{m^{2}}{M^{2}}\right)^{\epsilon+n_{1}}\,
M^{2}\,
\left(\frac{M^{2}}{m^{2}}\right)^{-\epsilon}\,\pi^{\,4-2\epsilon}\,
\Gamma\!\big[1-\epsilon-n_{1}\big]\,
\Gamma\!\big[\tfrac{1}{2}+n_{1}\big]\,}{\Gamma\!\big[2-\epsilon+n_{1}\big]\,
 \Gamma\!\big[\tfrac{1}{2}+\epsilon+n_{1}\big]}}$\\[6pt]
& $\displaystyle
{ \qquad \qquad \qquad \qquad \qquad
\Gamma\!\big[\epsilon+n_{1}\big]\,
\Gamma\!\big[-1+2\epsilon+n_{1}\big]}
\quad {\tt valid \,\, for } \, n_{1}\ge 0
$\\[8pt]

& ${\tt Time \,\, Taken \,\, 0.440079 \,\, seconds }$\\
}
}

The above result  gives us the expressions of each of the two terms that constitute the series representation.

Now, applying the generalized reflection formula
\begin{equation}
	\Gamma(z-n)=\frac{\Gamma(z)\Gamma(1-z)(-1)^n}{\Gamma(n+1-z)}
\label{reflection}
\end{equation}
valid for $n\in \mathbf{Z}$, these two terms can be written in a more convenient form and one is led to the final expression
\begin{multline}
	H_{\{1,1,1\}}(m,M,M;m^2) = -(i\pi^{2-\epsilon})^2 \left(m^2\right)^{1-\epsilon } \left(M^2\right)^{-\epsilon } \Gamma (\epsilon -1) \Gamma (\epsilon ) \\
	\times \Bigg(
	\, {}_3F_2 \left[ \begin{array}{c}
	1,\frac{3}{2}-\epsilon ,\epsilon \\
	\frac{3}{2},3-2 \epsilon \\
	\end{array}	\bigg| \frac{m^2}{M^2} \right] -\frac{\Gamma (2 \epsilon -1)}{\Gamma (2 \epsilon)} \left(\frac{m^2}{M^2}\right)^{\epsilon -1} \,
	{}_3F_2 \left[ \begin{array}{c}
	\frac{1}{2},1,2 \epsilon -1 \\
	2-\epsilon, \epsilon +\frac{1}{2} \\
	\end{array}	\bigg| \frac{m^2}{M^2} \right] \Bigg)
 \label{SeriesH111TwoMass1}
\end{multline}
which involves the well-known $_3F_2$ hypergeometric series.

A similar instruction in $\texttt{MBConicHulls}$ yields the other series representation consisting of four terms:
\begin{align}
	H_{\{1,1,1\}}&(m,M,M;m^2) =\nonumber\\
-(i\pi^{2-\epsilon})^2&\left(\pi \left(2m^2\right)^{1-2 \epsilon } \left(\frac{M^2}{m^2}\right)^{\frac{3}{2}-2 \epsilon }\frac{\Gamma \left(\epsilon -\frac{1}{2}\right) \Gamma \left(2
   \epsilon -\frac{3}{2}\right)}{\Gamma (\epsilon )}  \, {}_2F_1 \left[ \begin{array}{c}
	1-\epsilon ,\epsilon
   -\frac{1}{2}  \\
	\frac{5}{2}-2 \epsilon \\
	\end{array}	\bigg| \frac{M^2}{m^2} \right]\right.\nonumber\\
&+2\sqrt{\pi } 
   \left(2m^2\right)^{1-2 \epsilon } \left(\frac{M^2}{m^2}\right)^{2-2 \epsilon }\frac{\Gamma (\epsilon ) \Gamma (2 \epsilon -2)}{\Gamma \left(\epsilon
   -\frac{1}{2}\right)}  \,
   {}_3F_2 \left[ \begin{array}{c}
	1,\frac{3}{2}-\epsilon ,\epsilon   \\
	\frac{3}{2},3-2 \epsilon  \\
	\end{array}	\bigg| \frac{M^2}{m^2} \right]\nonumber\\
   &+M^2 \left(m^2\right)^{-2 \epsilon }  \left(\frac{M^2}{m^2}\right)^{-\epsilon }\frac{\Gamma \left(\frac{1}{2}-\epsilon \right) \Gamma (\epsilon
   -1) \Gamma (\epsilon )}{\Gamma \left(\frac{3}{2}-\epsilon \right)} \, {}_3F_2 \left[ \begin{array}{c}
	\frac{1}{2},1,2 \epsilon -1  \\
	2-\epsilon
   ,\epsilon +\frac{1}{2} \\
	\end{array}	\bigg| \frac{M^2}{m^2} \right]\nonumber\\
   &\left.+\left(2m^2\right)^{1-2 \epsilon } \frac{\Gamma \left(\frac{3}{2}-2 \epsilon \right) \Gamma (2-2 \epsilon ) \Gamma (1-\epsilon )
   \Gamma (\epsilon ) \Gamma (2 \epsilon -1)}{\Gamma (3-3 \epsilon ) \Gamma \left(\frac{3}{2}-\epsilon \right)} \, {}_2F_1 \left[ \begin{array}{c}
	\epsilon -\frac{1}{2},3 \epsilon -2  \\
	2 \epsilon
   -\frac{1}{2} \\
	\end{array}	\bigg| \frac{M^2}{m^2} \right]\right)
 \label{SeriesH111TwoMass2}
\end{align}
The hypergeometric representation in Eq.~\ref{SeriesH111TwoMass1}, seen as a combination of series, converges for $\vert \frac{m^2}{M^2}\vert\leq1$ while the one in Eq.~\ref{SeriesH111TwoMass2} converges for $\vert \frac{m^2}{M^2}\vert\geq1$: they are analytic continuations of one another, as is well-known from the linear transformation \cite{Bateman}

\begin{align}
{}_3F_2 &\left[ \begin{array}{c}
	a_1,a_2,a_3  \\
	b_1, b_2 \\
	\end{array}	\bigg| x \right]
 = \frac{\Gamma(b_1)\Gamma(b_2)}{\Gamma(a_1)\Gamma(a_2)\Gamma(a_3)}\nonumber\\
\times &\left(
\frac{\Gamma(a_1)\Gamma(a_2 - a_1)\Gamma(a_3 - a_1)}{\Gamma(b_1 - a_1)\Gamma(b_2 - a_1)}
(-x)^{-a_1} {}_3F_2 \left[ \begin{array}{c}
	a_1,a_1-b_1+1,a_1-b_2+1  \\
	a_1-a_2+1, a_1-a_3+1 \\
	\end{array}	\bigg| \frac{1}{x} \right]
\right.
\nonumber\\
&+ \frac{\Gamma(a_2)\Gamma(a_1 - a_2)\Gamma(a_3 - a_2)}{\Gamma(b_1 - a_2)\Gamma(b_2 - a_2)}
(-x)^{-a_2}{}_3F_2 \left[ \begin{array}{c}
	a_2,a_2-b_1+1,a_2-b_2+1  \\
	a_2-a_1+1, a_2-a_3+1 \\
	\end{array}	\bigg| \frac{1}{x} \right]
\nonumber\\
&+ \left. \frac{\Gamma(a_3)\Gamma(a_1 - a_3)\Gamma(a_2 - a_3)}{\Gamma(b_1 - a_3)\Gamma(b_2 - a_3)}
(-x)^{-a_3} {}_3F_2 \left[ \begin{array}{c}
	a_3,a_3-b_1+1,a_3-b_2+1  \\
	a_3-a_1+1, a_3-a_2+1 \\
	\end{array}	\bigg| \frac{1}{x} \right]
\right)
 \label{LT3F2}
\end{align}
which, by the way, can be easily derived from the MB representation
\begin{multline}
	{}_3F_2 \left[ \begin{array}{c}
	a_1,a_2,a_3  \\
	b_1, b_2 \\
	\end{array}	\bigg| x \right] = \frac{\Gamma(b_1)\Gamma(b_2)}{\Gamma(a_1)\Gamma(a_2)\Gamma(a_3)}\int_{-i \infty}^{+i \infty}  \frac{dz}{2 \pi i}\ x^{z}\  \Gamma(-z)\frac{\Gamma(a_1+z)\Gamma(a_2+z)\Gamma(a_3+z)}{\Gamma(b_1+z)\Gamma(b_2+z)}
 \label{MBrep3F2}
\end{multline}
Let us here recall that for $x=1$, ${}_3F_2 \left[ \begin{array}{c}
	a_1,a_2,a_3  \\
	b_1, b_2 \\
	\end{array}	\bigg| x \right]$ converges if $\Re(b_1+b_2-a_1-a_2-a_3)>0$ which is the case for the two functions appearing in Eq.~\ref{SeriesH111TwoMass1}. Therefore, one can trivially get the expression of the one-mass sunset $H_{\{1,1,1\}} \left(m, m, m; m^2 \right)$ by replacing $M$ by $m$ in Eq.~\ref{SeriesH111TwoMass1}. An equivalent formula can be derived from Eq.~\ref{SeriesH111TwoMass2}.
    
In a completely similar way as what has been presented in details in this section, one can derive the hypergeometric representations of the other sunsets. The most compact ones are presented in the following subsections while other results can be found in App.~\ref{SunsetHyperRep2}. 

\subsection{$H_{\{2,1,1\}} \left(m, M, M; m^2 \right)$ \label{Sec:H211ResultsmM}}
From Eq.~\ref{MBrepH211mM} we get two hypergeometric representations of $H_{\{2,1,1\}} \left(m, M, M; m^2 \right)$ which are analytic continuations of one another, exactly as in the case of $H_{\{1,1,1\}} \left(m, M, M; m^2 \right)$ shown above.

The most compact representation reads
\begin{multline}
	H_{\{2,1,1\}}(m,M,M;m^2) = (i\pi ^{2- \epsilon })^2 \left(m^2\right)^{-\epsilon} \left(M^2\right)^{-\epsilon } \Gamma (\epsilon )^2 \\
	\times \Bigg( {}_3F_2 \left[ \begin{array}{c}
	1,\frac{3}{2}-\epsilon ,\epsilon \\
	\frac{3}{2},2-2 \epsilon \\
	\end{array}	\bigg| \frac{m^2}{M^2} \right] + \frac{1}{4} \frac{\Gamma (\epsilon -1) \Gamma \left(\epsilon +\frac{1}{2}\right)}{\Gamma (\epsilon ) \Gamma \left(\epsilon +\frac{3}{2}\right)}  \left(\frac{m^2}{M^2}\right)^{\epsilon }
	{}_3F_2 \left[ \begin{array}{c}
		1,\frac{3}{2},2 \epsilon \\
		2-\epsilon ,\epsilon +\frac{3}{2} \\
	\end{array}	\bigg| \frac{m^2}{M^2} \right] \Bigg)
 \label{SeriesH211TwoMass1}
\end{multline}

\subsection{$H_{\{1,1,1\}} \left(m_1, m_2, m_3; m_1^2 \right)$ \label{Sec:H111Results}}

From Eq.~\ref{MBrepHa1a2a3_v2}, we get the following MB representation:
\begin{multline}
	H_{\{1,1,1\}} \left(m_1, m_2, m_3; m_1^2 \right) \equiv 
	-(i\pi^{2-\epsilon})^2(m_1^2)^{1-2\epsilon}
	 \int_{-i\infty}^{+i\infty} \frac{dz_1}{2i\pi}\int_{-i\infty}^{+i\infty} \frac{dz_2}{2i\pi} \left( \frac{m_3^2-i0}{m_1^2} \right)^{z_1} \left( \frac{m_2^2-i0}{m_1^2} \right)^{z_2}  \\
	 \times  \frac{\Gamma(-z_1) \Gamma(-z_2)\Gamma(-1+2\epsilon+z_1+z_2) \Gamma(\epsilon+z_1+z_2) \Gamma(1-\epsilon-z_1) \Gamma(1-\epsilon-z_2) \Gamma(3-4\epsilon-2z_1-2z_2)}{\Gamma(3-3\epsilon-z_1-z_2) \Gamma(2-2\epsilon-z_1-z_2)} 
  \label{MBrepH111v2}
\end{multline}
We choose this MB representation as a starting point of the calculations instead of the one given in Eq.~\ref{MBrepH111} because it has the same integrand as that of $I_4$, defined in Eq.~\ref{I4}. It will give us the opportunity to better show how the matching between the two MB computational methods discussed in this paper\footnote{These two methods are the by hand method with straight contours presented in Section 6 and 7, and the automated method using conic hulls/triangulations which can also be used for non-straight contour and from which we give all the results of Section~\ref{SunsetHyperRep1} and of App.~\ref{SunsetHyperRep2}} occurs. Note however that, at the very end of the calculations, both MB representations in Eqs.~\ref{MBrepH111} and ~\ref{MBrepH111v2} give the exact same result.

Keeping the full $\epsilon$ dependency, four hypergeometric representations of $H_{\{1,1,1\}} \left(m_1, m_2, m_3; m_1^2 \right)$ can be obtained from its MB representation, which are analytic continuations of one another. The most compact one is
\begin{multline}
	H_{\{1,1,1\}}(m_1,m_2,m_3;m_1^2) = \\(i\pi^{2-\epsilon})^2 \left(\left(m_1^2\right)^{1-2\epsilon }\left(\frac{m_2^2}{m_1^2}\right)^{-\epsilon}  \Gamma (\epsilon-1)^2
	F{}^{2:0;1}_{0:1;2}
  \left[
   \setlength{\arraycolsep}{0pt}
   \begin{array}{c@{{}:{}}c@{;{}}c}
 1,\epsilon & - & \frac{3}{2}-\epsilon\\[1ex]
  - & \epsilon & 2-\epsilon,3-2\epsilon
   \end{array}
   \;\middle|\;
\frac{m_3^2}{m_2^2},\frac{4m_1^2}{m_2^2}
 \right]\right.\\
 +\left(m_2^2\right)^{1-2\epsilon }   \Gamma (1- \epsilon)\Gamma (\epsilon-1)\Gamma (2\epsilon-1)\,
	F{}^{2:0;1}_{0:1;2}
  \left[
   \setlength{\arraycolsep}{0pt}
   \begin{array}{c@{{}:{}}c@{;{}}c}
 \epsilon,2\epsilon-1 & - & \frac{1}{2}\\[1ex]
  - & \epsilon & \epsilon,2-\epsilon
   \end{array}
   \;\middle|\;
\frac{m_3^2}{m_2^2},\frac{4m_1^2}{m_2^2}
 \right]\\
 -\left(m_1^2\right)^{1-2\epsilon }\frac{m_3^2}{m_2^2}\left(\frac{m_3^2}{m_1^2}\right)^{-\epsilon} \Gamma (\epsilon-1)^2\,
	F{}^{2:0;1}_{0:1;2}
  \left[
   \setlength{\arraycolsep}{0pt}
   \begin{array}{c@{{}:{}}c@{;{}}c}
 1,2-\epsilon & - & \frac{3}{2}-\epsilon\\[1ex]
  - & 2-\epsilon & 2-\epsilon,3-2\epsilon
   \end{array}
   \;\middle|\;
\frac{m_3^2}{m_2^2},\frac{4m_1^2}{m_2^2}
 \right]
 \\
 \left. +\left(m_1^2\right)^{1-2\epsilon }  \left(\frac{m_2^2}{m_1^2}\right)^{-\epsilon}\left(\frac{m_3^2}{m_1^2}\right)^{1-\epsilon} \Gamma (\epsilon-1)^2\,
	F{}^{2:0;1}_{0:1;2}
  \left[
   \setlength{\arraycolsep}{0pt}
   \begin{array}{c@{{}:{}}c@{;{}}c}
 1,\epsilon & - & \frac{1}{2}\\[1ex]
  - & 2-\epsilon & \epsilon,2-\epsilon
   \end{array}
   \;\middle|\;
\frac{m_3^2}{m_2^2},\frac{4m_1^2}{m_2^2}
 \right]
 \right)
 \label{SeriesH111}
\end{multline}
where the $F{}^{2:0;1}_{0:1;2}$ Kamp\'e de F\'eriet series converge for $\sqrt{\left\vert\frac{m_3^2}{m_2^2}\right\vert}+\sqrt{\left\vert\frac{4m_1^2}{m_2^2}\right\vert}<1$ \cite{Srivastava}. The three other series representations of $H_{\{1,1,1\}}(m_1,m_2,m_3;m_1^2)$ converge in other ranges of masses (see App.~\ref{SunsetHyperRep2}).

Let us note that putting $m_3=m_2$ in Eq.~\ref{SeriesH111} and using Gauss summation theorem
\begin{equation}
{}_2F_1 \left[ \begin{array}{c}
	a,b  \\
	c \\
	\end{array}	\bigg| 1 \right]=\frac{\Gamma(c)\Gamma(c-a-b)}{\Gamma(c-a)\Gamma(c-b)}
 \label{Gauss_summation}
 \end{equation}
one can check Eq.~\ref{SeriesH111TwoMass1}. Indeed, in this limit, the first and third terms of Eq.~\ref{SeriesH111} yield the first term of Eq.~\ref{SeriesH111TwoMass1} while the last term of Eq.~\ref{SeriesH111} reduces to the second term of Eq.~\ref{SeriesH111TwoMass1} (the second term of Eq.~\ref{SeriesH111} cancels). 
\subsection{$H_{\{2,1,1\}} \left(m_1, m_2, m_3; m_3^2 \right)$ \label{H211Results}}
From Eq.~\ref{MBrepHa1a2a3_v2}, we get the following MB representation:
\begin{multline}
	H_{\{2,1,1\}} \left(m_1, m_2, m_3; m_3^2 \right) \equiv 
	\frac{(i\pi^{2-\epsilon})^2}{(m_1^2)^{2\epsilon}}
	 \int_{-i\infty}^{+i\infty} \frac{dz_1}{2i\pi}\int_{-i\infty}^{+i\infty} \frac{dz_2}{2i\pi} \left( \frac{m_3^2-i0}{m_1^2} \right)^{z_1} \left( \frac{m_2^2-i0}{m_1^2} \right)^{z_2}  \\
	 \times  \frac{\Gamma(-z_1) \Gamma(-z_2)\Gamma(2\epsilon+z_1+z_2) \Gamma(\epsilon+z_1+z_2) \Gamma(1-\epsilon-z_1) \Gamma(1-\epsilon-z_2) \Gamma(2-4\epsilon-2z_1-2z_2)}{\Gamma(2-2\epsilon-z_1-z_2) \Gamma(2-3\epsilon-z_1-z_2)} 
  \label{MBrepH211v2}
\end{multline}
As before, we choose this MB representation as a starting point of the calculations instead of the one given in Eq.~\ref{MBrepH211} only because it has the same integrand as the one of $J_4$, defined in Eq.~\ref{Eq:J4}, but both MB representations in Eqs.~\ref{MBrepH211} and ~\ref{MBrepH211v2} give the exact same result.

Keeping the full $\epsilon$ dependency, four hypergeometric representations of $H_{\{2,1,1\}} \left(m_1, m_2, m_3; m_1^2 \right)$ can be obtained from its MB representation, which are analytic continuations of one another. The most compact one is
\begin{multline}
	H_{\{2,1,1\}}(m_1,m_2,m_3;m_1^2) = \\-(i\pi^{2-\epsilon})^2 \left(\left(m_3^2\right)^{-2\epsilon }  \frac{\Gamma (1-\epsilon)^2 \Gamma (\epsilon )\Gamma(2\epsilon) }{\Gamma(2-\epsilon)}
	F{}^{2:0;2}_{0:1;3}
  \left[
   \setlength{\arraycolsep}{0pt}
   \begin{array}{c@{{}:{}}c@{;{}}c}
 1+\epsilon,2\epsilon & - & 1,\frac{3}{2}\\[1ex]
  - & \epsilon & 1+\epsilon,2,2-\epsilon
   \end{array}
   \;\middle|\;
\frac{m_2^2}{m_3^2},\frac{4m_1^2}{m_3^2}
 \right]\right.\\
 -\left(m_1^2\right)^{-2\epsilon }  \left(\frac{m_3^2}{m_1^2}\right)^{-\epsilon} \frac{\Gamma (1- \epsilon)\Gamma (\epsilon)^2}{\Gamma (2 -\epsilon)}\,
	F{}^{2:0;1}_{0:1;2}
  \left[
   \setlength{\arraycolsep}{0pt}
   \begin{array}{c@{{}:{}}c@{;{}}c}
 1,\epsilon & - & \frac{3}{2}-\epsilon\\[1ex]
  - & \epsilon & 2-2\epsilon,2-\epsilon
   \end{array}
   \;\middle|\;
\frac{m_2^2}{m_3^2},\frac{4m_1^2}{m_3^2}
 \right]\\
 -\frac{(m_2^2)^{1-\epsilon}}{(m_3^2)^{1+\epsilon}} \Gamma (\epsilon-1)^2\,
	F{}^{2:0;2}_{0:1;3}
  \left[
   \setlength{\arraycolsep}{0pt}
   \begin{array}{c@{{}:{}}c@{;{}}c}
 2,1+\epsilon & - & 1,\frac{3}{2}\\[1ex]
  - & 2-\epsilon & 1+\epsilon,2,2-\epsilon
   \end{array}
   \;\middle|\;
\frac{m_2^2}{m_3^2},\frac{4m_1^2}{m_3^2}
 \right]
 \\
 \left. -\left(m_1^2\right)^{-2\epsilon }  \left(\frac{m_2^2}{m_1^2}\right)^{1-\epsilon}\left(\frac{m_3^2}{m_1^2}\right)^{-1} \Gamma (\epsilon)\Gamma (\epsilon-1)\,
	F{}^{2:0;1}_{0:1;2}
  \left[
   \setlength{\arraycolsep}{0pt}
   \begin{array}{c@{{}:{}}c@{;{}}c}
 1,2-\epsilon & - & \frac{3}{2}-\epsilon\\[1ex]
  - & 2-\epsilon & 2-2\epsilon,2-\epsilon
   \end{array}
   \;\middle|\;
\frac{m_2^2}{m_3^2},\frac{4m_1^2}{m_3^2}
 \right]
 \right)
 \label{SeriesH211}
\end{multline}
where the $F{}^{2:0;2}_{0:1;3}$ and $F{}^{2:0;1}_{0:1;2}$ Kamp\'e de F\'eriet series converge in the same range of mass values, namely $\sqrt{\left\vert\frac{m_2^2}{m_3^2}\right\vert}+\sqrt{\left\vert\frac{4m_1^2}{m_3^2}\right\vert}<1$ \cite{Srivastava}. The three other series representations of $H_{\{2,1,1\}}(m_1,m_2,m_3;m_1^2)$ converge in other ranges of masses (see App.~\ref{SunsetHyperRep2}).

As in the case of $H_{\{1,1,1\}}(m_1,m_2,m_3;m_1^2)$, putting $m_3=m_2$ in Eq.~\ref{SeriesH211} and using Gauss summation theorem (see Eq.~\ref{Gauss_summation}), one can easily check Eq.~\ref{SeriesH211TwoMass1}.

\subsection{$H_{\{1,1,1\}} \left(m_1, m_2, m_2; m_3^2 \right)$ \label{Sec:H111m2m2Results}}
From the MB representation given in Eq.~\ref{MBrepH111m1m2m2m3v2} and keeping the full $\epsilon$ dependency, we can extract three hypergeometric representations of $H_{\{1,1,1\}} \left(m_1, m_2, m_2; m_3^2 \right)$. 

The most compact one is
\begin{multline}
	H_{\{1,1,1\}}(m_1,m_2,m_2;m_3^2) = \\-(i\pi^{2-\epsilon})^2 \left(m_1^2\right)^{1-\epsilon } \left(m_2^2\right)^{-\epsilon } \Gamma (\epsilon -1) \Gamma (\epsilon ) 
	 \Bigg(
	F{}^{3:0;0}_{1:1;1}
  \left[
   \setlength{\arraycolsep}{0pt}
   \begin{array}{c@{{}:{}}c@{;{}}c}
 1,2-\epsilon,\epsilon & - & -\\[1ex]
  \frac{3}{2} & 2-\epsilon & 2-\epsilon
   \end{array}
   \;\middle|\;
\frac{m_3^2}{4m_2^2},\frac{m_1^2}{4m_2^2}
 \right]\\\left. -\frac{\Gamma (2 \epsilon -1)}{\Gamma (2 \epsilon)} \left(\frac{m_1^2}{m_2^2}\right)^{\epsilon -1} \,
	F{}^{3:0;0}_{1:1;1}
  \left[
   \setlength{\arraycolsep}{0pt}
   \begin{array}{c@{{}:{}}c@{;{}}c}
 1,\epsilon,2\epsilon-1 & - & -\\[1ex]
  \frac{1}{2}+\epsilon & 2-\epsilon & \epsilon
   \end{array}
   \;\middle|\;
\frac{m_3^2}{4m_2^2},\frac{m_1^2}{4m_2^2}
 \right]\right)
 \label{SeriesH111m2m21}
\end{multline}
where the Kamp\'e de F\'eriet hypergeometric series $F{}^{3:0;0}_{1:1;1}$ converge for $\sqrt{\left\vert\frac{m_3^2}{4m_2^2}\right\vert}+\sqrt{\left\vert\frac{m_1^2}{4m_2^2}\right\vert}<1$.
The two other series representations of $H_{\{1,1,1\}}(m_1,m_2,m_2;m_3^2)$ which converge for other ranges of masses can be found in App.~\ref{SunsetHyperRep2}.

Let us note here that putting $m_3=m_1$ in Eq.~\ref{SeriesH111m2m21} and using the reduction formula \cite{Srivastava}
\begin{equation}
	F{}^{p:0;0}_{q:1;1}
  \left[
   \setlength{\arraycolsep}{0pt}
   \begin{array}{c@{{}:{}}c@{;{}}c}
 a_1,...,a_p & - & -\\[1ex]
 b_1,...,b_q & \nu & \sigma
   \end{array}
   \;\middle|\;
x,x
 \right]={}_{p+2}F_{q+3} \left[ \begin{array}{c}
	a_1,...,a_p,\Delta(2;\nu+\sigma-1) \\
	b_1,...,b_q,\nu,\sigma,\nu+\sigma-1 \\
	\end{array}	\bigg| 4x \right]
 \label{reduction}
\end{equation}
where $\Delta(l;\lambda)$ abbreviates the array of $l$ parameters $\left\{\frac{\lambda}{l},\frac{\lambda+1}{l},...,\frac{\lambda+l-1}{l}\right\}$,
it is straightforward to derive Eq.~\ref{SeriesH111TwoMass1} which, therefore, gives us a second analytic cross-check of this formula.

\subsection{$H_{\{1,1,1\}} \left(m_1, m_2, m_3; p^2 \right)$ \label{Sec:H111FourScalesResults}}
In order to further check the hypergeometric representations of the three mass scale sunsets given above, it is interesting to consider the four mass scale case, as one can obtain the former by taking appropriate limits of the latter. Let us see this briefly for the results presented in Eqs.~\ref{SeriesH111} and ~\ref{SeriesH111m2m21}. The MB representation given in Eq.~\ref{MBrepH111FourScales} provides four hypergeometric representations of $H_{\{1,1,1\}} \left(m_1, m_2, m_3; p^2 \right)$, the most compact one reading
\begin{multline}
	H_{\{1,1,1\}}(m_1,m_2,m_3;p^2) = (i\pi^{2-\epsilon})^2 \Bigg\{\\
 \left(m_3^2\right)^{1-2\epsilon }\Gamma (1-\epsilon ) \Gamma (\epsilon -1) \Gamma (2\epsilon-1 ) 
	F_C
  \left[
   \setlength{\arraycolsep}{0pt}
   \begin{array}{c@{{}{}}c@{;{}}c}
 \epsilon,2\epsilon-1 \\[1ex]
 \epsilon, \epsilon, 2-\epsilon
   \end{array}
   \;\middle|\;
\frac{m_1^2}{m_3^2},\frac{m_2^2}{m_3^2},\frac{p^2}{m_3^2}
 \right]\\
 +\left(m_2^2\right)^{1-\epsilon}\left(m_3^2\right)^{-\epsilon }\Gamma (\epsilon-1)^2 
	F_C
  \left[
   \setlength{\arraycolsep}{0pt}
   \begin{array}{c@{{}{}}c@{;{}}c}
 1,\epsilon \\[1ex]
 \epsilon, 2-\epsilon, 2-\epsilon
   \end{array}
   \;\middle|\;
\frac{m_1^2}{m_3^2},\frac{m_2^2}{m_3^2},\frac{p^2}{m_3^2}
 \right]\\
 +\left(m_1^2\right)^{1-\epsilon}\left(m_3^2\right)^{-\epsilon }\Gamma (\epsilon-1)^2 
	F_C
  \left[
   \setlength{\arraycolsep}{0pt}
   \begin{array}{c@{{}{}}c@{;{}}c}
 1,\epsilon \\[1ex]
 2-\epsilon, \epsilon, 2-\epsilon
   \end{array}
   \;\middle|\;
\frac{m_1^2}{m_3^2},\frac{m_2^2}{m_3^2},\frac{p^2}{m_3^2}
 \right]\\
 -\left.\frac{(m_1^2m_2^2)^{1-\epsilon}}{m_3^2} \Gamma (\epsilon-1)^2
	F_C
  \left[
   \setlength{\arraycolsep}{0pt}
   \begin{array}{c@{{}{}}c@{;{}}c}
 1,2-\epsilon \\[1ex]
 2-\epsilon, 2-\epsilon, 2-\epsilon
   \end{array}
   \;\middle|\;
\frac{m_1^2}{m_3^2},\frac{m_2^2}{m_3^2},\frac{p^2}{m_3^2}
 \right]\right\}
 \label{SeriesH111FourScales}
\end{multline}
where the triple hypergeometric Lauricella $F_C$  series converge for $\sqrt{\left\vert\frac{m_1^2}{m_3^2}\right\vert}+\sqrt{\left\vert\frac{m_2^2}{m_3^2}\right\vert}+\sqrt{\left\vert\frac{p^2}{m_3^2}\right\vert}<1$.
This expression, once multiplied by the conventional overall factor $\frac{1}{(i\pi^2(2\pi\mu)^{-2\epsilon})^2}$ of \cite{Berends:1993ee}, agrees with the result given this reference for the four mass scale sunset. Now, putting $p^2=m_1^2$ in Eq.~\ref{SeriesH111FourScales} and using Eq.~\ref{reduction}, it is easy to find Eq~\ref{SeriesH111}. 
Similarly, putting $m_3=m_2$ in Eq.~\ref{SeriesH111FourScales} (followed by the replacement $p^2=m_3^2$), one can obtain Eq.~\ref{SeriesH111m2m21} after the use of Gauss summation theorem shown in Eq.~\ref{Gauss_summation}. 

Eq.~\ref{SeriesH211} can be checked in a similar way starting from Eq.~\ref{MBrepHa1a2a3FourScales} with $a_1=2$ and $a_2=a_3=1$.

\section{Discussion \label{Discussions}}

 At the time of writing of \cite{Ananthanarayan:2016pos,Ananthanarayan:2017yhz,Ananthanarayan:2017qmx,Ananthanarayan:2018irl}, we did not yet develop the analytical computational method of multifold MB integral based on conic hulls/triangulations, which was published later in \cite{Ananthanarayan:2020fhl,Banik:2022bmk,Banik:2023rrz}. Therefore, in order to obtain the analytic expressions of the sunsets from their MB representations, we instead used the techniques of \cite{Tsikh:1998,Friot:2011ic} which can be applied to two-fold MB integrals with straight contours only.

We thus had to obtain MB representations of the sunsets with straight contours and at most two folds. Keeping straight contours at all steps of the transformation of a Feynman integral into its MB representation while having at the same time arbitrary small values of $\epsilon$ is more complicated than the traditional approach described in Section~\ref{AMBRE}, where in general non-straight contours are tacitly used. Indeed, the former (non-conventional) approach requires to take into account the mathematical constraints coming from intermediate steps of the calculations\footnote{This in particular requires to fix the values of the powers of the propagators.}, which can be ignored in the non-straight contours approach.  However, it is pedagogically interesting to show how both methods yield the same analytical results at the end, which is what we will show in details in this section. The straight contours approach also provides a further mathematical justification (and some additional analytic cross-checks) of the results given by the non-straight contours approach, for those of the beginners in the field who would be questioned about the validity of the MB representations of Feynman integrals and results derived from them in the literature, or as obtained for instance using the \textit{Mathematica} packages presented in Sec.~\ref{AMBRE} and Sec.~\ref{SunsetHyperRep1}. Let us note also that, in some cases, the straight contours approach can directly provide the resolution of $\epsilon$-singularities.
As a last remark, the straight contours approach often yields MB representations made of split MB integrals whose series representations may render the equivalence with the non-straight contours results a bit difficult to see, due to the necessity, in order to prove this equivalence, to find cancellations of some of the straight contours terms, as well as rewritings of some terms of the non-straight contours approach, as we will see later with the example of $H_{\{1,1,1\}}(m_1,m_2,m_3;m_1^2)$. The conic hulls/triangulation method \cite{Ananthanarayan:2020fhl,Banik:2022bmk,Banik:2023rrz}, which allows to compute MB integrals with non-straight or straight contours and with an arbitrary number of folds, and which directly gives the most compact results, is thus much more efficient than the method presented here only for pedagogical purpose.

Let us now show how the straight contours approach works in the cases of the two mass scale sunsets $H_{\{1,1,1\}}(m,M,M;m^2)$ and $H_{\{2,1,1\}}(m,M,M;m^2)$ and, in a subsequent section, in the three mass scale case.


\section{Two mass scale sunsets \label{Sec:TwoMassScale}}

There are seven independent mass configurations of the sunset master integrals with two mass scales. Two of these fall into the pseudothreshold configurations, in which $p^2=m_1^2+m_2^2-m_3^2$. These are the only two mass scale configurations that appear in the expressions of pseudoscalar meson masses and decay constants, as well as in those of the nucleon-axial coupling $g_A$, at two loops. Results for the pseudothresholds in the $\epsilon \rightarrow 0$ limit are given in \cite{Berends:1997vk}. We present results for the full $\epsilon$ dependent two mass scale pseudothresholds in this work, while results for the five non-pseudothreshold configurations in $\epsilon \rightarrow 0$ limit are given in \cite{Ananthanarayan:2016pos}.

The sunset two mass scale pseudothreshold configuration is $(m_1^2,m_2^2,m_2^2;p^2=m_1^2)$. Of the four master integrals, the ones with propagator powers $(\alpha,\beta,\gamma)=(1,2,1)$ and $(1,1,2)$ are equivalent, and the $(1,1,2)$ master integral may be expressed in terms of the $(1,1,1)$ and $(2,1,1)$ master integral by means of the relation:
\begin{align}
	H_{\{1,1,2\}} & (m,M,M;m^2) = - \frac{1}{(16 \pi^2)^2} \left(\frac{5}{2} + \frac{\pi^2}{12} + \frac{3}{8} \frac{m^2}{M^2} \right) \nonumber \\
	& + \frac{1}{M^2} H_{\{1,1,1\}} (m,M,M;m^2) - \frac{m^2}{M^2} H_{\{2,1,1\}} (m,M,M;m^2) 
 \label{eq:h211mMMm}
\end{align}

There are therefore two independent pseudothreshold with two masses. For concreteness, we calculate $H_{\{1,1,1\}}(m,M,M;m^2)$ and $H_{\{2,1,1\}}(m,M,M;m^2)$.

\subsection{$H_{\{1,1,1\}}(m,M,M;m^2)$ in the straight contours approach}


From Eq.~\ref{Eq:Sunset}, the sunset master integral with two masses reads
\begin{align}
	H_{\{1,1,1\}}(m,M,M;m^2) = \int \frac{d^dk_1 d^dk_2}{\left[ (k_1-p)^2 - m^2 + i0\right] \left[ k_2^2 - M^2 + i0 \right] \left[ (k_1-k_2)^2 - M^2 + i0 \right]} \Bigg|_{p^2 = m^2} &&
\end{align}

Taking a loop by loop approach, we first consider the $k_2$ momentum loop, and collect the propagators with $k_2$ dependence to define:
\begin{align}
	I_{k_2} \equiv& \int \frac{d^d k_2}{[ k_2^2 - M^2 + i0 ] [ (k_1-k_2)^2 - M^2 + i0]}
\end{align}

We then apply the well-known Feynman Parametrization formula:
\begin{align}
    \frac{1}{A^\alpha B^\beta} 
    = \frac{\Gamma(\alpha+\beta)}{\Gamma(\alpha) \Gamma(\beta)} \int_0^1 dx \frac{x^{\alpha-1 (1-x)^{\beta-1}}}{[xA + (1-x)B]^{\alpha+\beta}}
\label{Eq:FP}
\end{align}
to $I_{k_2}$ to obtain:
\begin{align}
	I_{k_2} = \int d^d k_2 \int_0^1 dx \frac{1}{[ x \{k_2^2 - M^2 + i0\} + (1-x) \{ (k_1-k_2)^2 - M^2 + i0 \} ]^2}
\end{align}
This can then be re-expressed as:
\begin{align}
    I_{k_2} = \int_0^1 dx \int \frac{d^d\tilde{k}}{ [ \tilde{k}^2 - \Delta ]^2}
\end{align}
where $\tilde{k} \equiv k_2 - k_1 (1-x)$ and $\Delta \equiv - k_1^2 x (1-x) + M^2 - i0$. We perform the integration over $\tilde{k}$ using the formula~\cite{Peskin:1995ev}: 
\begin{align}
    \int \frac{d^d k}{[k-\Delta]^\alpha} 
    = \frac{(-1)^{\alpha} i \pi^{d/2}}{\Delta^{\alpha-d/2}} \frac{\Gamma(\alpha-d/2)}{\Gamma(\alpha)} 
\label{Eq:SinglePropagator}
\end{align}
to get:
\begin{align}
    I_{k_2} &= i \pi^{2-\epsilon} \Gamma(\epsilon) \int_0^1 dx \frac{1}{\left( M^2 - i0 -k_1^2 x (1-x) \right)^\epsilon} \nonumber \\
            &= i \pi^{2-\epsilon} \frac{\Gamma(\epsilon)}{(M^2 - i0)^\epsilon} \int_0^1 dx \frac{1}{\left( 1 + \frac{-k_1^2 x (1-x)}{M^2 - i0} \right)^\epsilon}
\end{align}
which holds if $\text{Re}(\epsilon)>0$.

We now apply Eq.~\ref{Eq:MBformula} to the integrand above to obtain:
\begin{align}
	I_{k_2} = \frac{i \pi^{2-\epsilon}}{(M^2 - i0)^\epsilon} \frac{1}{2 \pi i} \int_0^1 dx \int_{c-i\infty}^{c+i\infty} dz \, \left( \frac{-k_1^2 x(1-x)}{M^2-i0} \right)^z \Gamma(-z) \Gamma(\epsilon+z)
\label{Eq:OneMassMB}
\end{align}
where $c \equiv \text{Re}(z) \in ] -\text{Re}(\epsilon),0 [ $. The first condition of Eq.~\ref{Eq:MBformula} is satisfied thanks to the $-i0$ term.

The integral over $x$ is a beta function that can be expressed as:
\begin{align}
	\int_0^1 dx \, x^z (1-x)^z = \frac{\Gamma^2(1+z)}{\Gamma(2+2z)}
\end{align}
provided that $\text{Re}(1+z)>0$ is satisfied. This constraint and the previous one, $\text{Re}(z) \in ] -\text{Re}(\epsilon),0 [ $, are compatible with each other and with our requirement of allowing infinitesimally small values of $\epsilon$, i.e. it is possible to find some value of $\text{Re}(z)$ such that both conditions are satisfied, including for arbitrarily small positive values of $\epsilon$. Therefore, we may continue with the calculation to obtain:
\begin{align}
	I_{k_2} = \frac{i \pi^{2-\epsilon}}{(M^2-i0)^\epsilon} \frac{ \sqrt{\pi}}{2} \frac{1}{2 \pi i} \int_{c-i\infty}^{c+i\infty} dz \left( \frac{-k_1^2}{4(M^2-i0)} \right)^z \frac{\Gamma(-z) \Gamma(\epsilon+z) \Gamma(1+z)}{\Gamma \left( \frac{3}{2}+z \right)}.
\label{Eq:Ik2solved}
\end{align}
where we used the Gamma function duplication formula:
\begin{align}
    \Gamma(2z) = \Gamma(z) \Gamma(z+\tfrac{1}{2}) \frac{2^{2z-1}}{\sqrt{\pi}}
\end{align}
to re-write the $\Gamma(2+2z)$ factor that appears in the denominator of the integral.

The complete integral can now be written as:
\begin{align}
    H_{\{1,1,1\}}& (m,M,M;m^2) \nonumber \\
    &= \frac{i \pi^{2-\epsilon}}{(M^2-i0)^\epsilon} \frac{\sqrt{\pi}}{2} \int \frac{d^dk_1}{\left[ (k_1-p)^2 - m^2 \right]} \frac{1}{2 \pi i} \int_{c-i\infty}^{c+i\infty} dz \left( \frac{-k_1^2}{4(M^2-i0)} \right)^z \frac{\Gamma(-z) \Gamma(\epsilon+z) \Gamma(1+z)}{\Gamma \left( \frac{3}{2}+z \right)} \nonumber \\
    & = \frac{i \pi^{2-\epsilon}}{(M^2-i0)^\epsilon} \frac{\sqrt{\pi}}{2}
    \frac{1}{2 \pi i}\int_{c-i\infty}^{c+i\infty} dz \left( \frac{1}{4(M^2-i0)} \right)^z \frac{\Gamma(-z) \Gamma(\epsilon+z) \Gamma(1+z)}{\Gamma \left( \frac{3}{2}+z \right)} I_{k_1}
\label{Eq:H111mMMm_int1}
\end{align}
where $I_{k_1}$ represents the $k_1$ loop, in which we have combined all the $k_1$ dependent factors from the above to define:
\begin{align}
	I_{k_1} \equiv \int \frac{d^d k_1}{\left[ (k_1-p)^2 - m^2 + i0 \right] \left[ - k_1^2 \right]^{-z}}
\end{align}
We once again apply the Feynman parametrization formula of Eq.~\ref{Eq:FP} to rewrite $I_{k_1}$ as:
\begin{align}
    I_{k_1} = \frac{1}{(-1)^{-z}}\frac{\Gamma(1-z)}{\Gamma(-z)} \int_0^1 dx x^{2-2\epsilon+2z} (1-x)^{-z-1}  \int \frac{d^d\tilde{k}}{[\tilde{k}^2 - \Delta]^{1-z}}
\end{align}
where $\tilde{k} \equiv k_1-px$ and $\Delta \equiv m^2 x^2-i0$. In this step, we make use of the fact that the external momentum is given by $p^2=m^2$. Evaluating the $k_1$ integral using Eq.~\ref{Eq:SinglePropagator}, we obtain:
\begin{align}
	I_{k_1} = \frac{-i \pi^{2-\epsilon}}{(m^2-i0)^{\epsilon-1-z}} \frac{\Gamma(\epsilon-1-z)}{\Gamma(-z)} \int_0^1 dx \, x^{2-2\epsilon+2z} (1-x)^{-1-z}
\end{align}
This last equality holds provided $\text{Re}(\epsilon-1-z)>0$. This constraint is compatible with the previous ones, but not for infinitesimally small values of $\epsilon$. This problem would be solved if instead of $\text{Re}(z) \in ] -\text{Re}(\epsilon),0 [ $, we had $\text{Re}(z) \in ] -1-\text{Re}(\epsilon),-\text{Re}(\epsilon) [ $ as a constraint. This is achieved by returning to Eq.~\ref{Eq:OneMassMB} and shifting the contour by taking the residue at $z=-\epsilon$. We then have for $I_{k_2}$:
\begin{align}
	I_{k_2} &= \frac{i \pi^{2-\epsilon}}{(M^2 - i0)^\epsilon} \Gamma(\epsilon)   \int_0^1 dx \left( \frac{-k_1^2 x(1-x)}{M^2-i0} \right)^{-\epsilon} \nonumber \\
	& + \frac{i \pi^{2-\epsilon}}{(M^2 - i0)^\epsilon} \frac{1}{2 \pi i} \int_0^1 dx \int_{c'-i\infty}^{c'+i\infty} dz \, \left( \frac{-k_1^2 x(1-x)}{M^2-i0} \right)^z \Gamma(-z) \Gamma(\epsilon+z)
\label{Eq:Ik2p}
\end{align}
subject to $c'=\text{Re}(z) \in ] -1-\text{Re}(\epsilon),-\text{Re}(\epsilon)[$ and $\text{Re}(z+1)>0$. Let us denote the first term by $I_{k_2}'$ and the second by $I_{k_2}''$, i.e.
\begin{align}
    H_{\{1,1,1\}} & (m,M,M;m^2) = \int \frac{d^dk_1}{\left[ (k_1-p)^2 - m^2 + i0 \right]} \left( I_{k_2}' + I_{k_2}''  \right)
\label{Eq:H111mMMm_int2}
\end{align}
Written out in full, this is equivalent to:
\begin{align}
	H_{\{1,1,1\}} & (m,M,M;m^2) = \frac{i \pi^{2-\epsilon} \Gamma(\epsilon)}{(M^2 - i0)^\epsilon} \int \frac{d^dk_1}{\left[ (k_1-p)^2 - m^2 \right]}  \int_0^1 dx \left( \frac{-k_1^2 x(1-x)}{M^2-i0} \right)^{-\epsilon} \nonumber \\
    & + \frac{i \pi^{2-\epsilon}}{(M^2 - i0)^\epsilon}\frac{1}{2 \pi i}  \int \frac{d^dk_1}{\left[ (k_1-p)^2 - m^2 \right]} \int_0^1 dx \int_{c'-i\infty}^{c'+i\infty} dz \, \left( \frac{-k_1^2 x(1-x)}{M^2-i0} \right)^z \Gamma(-z) \Gamma(\epsilon+z)
\label{Eq:H111mMMm_int3}
\end{align}

The second term above is equivalent to Eq.~\ref{Eq:OneMassMB} with $c \rightarrow c'$. Therefore, we can continue the calculation for $I_{k_1}$, with the new constraint arising from $c'$ instead of $c$, to get:
\begin{align}
	I_{k_1} = \frac{-i\pi^{2-\epsilon}}{(m^2-i0)^{\epsilon-1-z}} \frac{\Gamma(\epsilon-1-z) \Gamma(3-2\epsilon+2z)}{\Gamma(3-2\epsilon+z)} 
	\label{Eq:Ik1solved}
\end{align}
The conditions that permit the equality above to be true are $\text{Re}(z)<0$ and $\text{Re}(3-2\epsilon+2z)>0$, which are compatible with the prior constraints, and also permit $\text{Re}(\epsilon)$ to be infinitesimally small. Using the above result for $I_{k_1}$ in the expression of Eq.~\ref{Eq:H111mMMm_int1} with $c \rightarrow c'$, we get for the second term of Eq.~\ref{Eq:H111mMMm_int3}:
\begin{align}
     & \frac{\pi^{4-2\epsilon}}{(m^2-i0)^{\epsilon-1} (M^2-i0)^{\epsilon}} \frac{\sqrt{\pi}}{2}  \nonumber \\
    & \times \frac{1}{2 \pi i} \int_{c'-i \infty}^{c'+i \infty} dz \left( \frac{m^2 - i0}{4(M^2 - i0)} \right)^z \frac{\Gamma(-z) \Gamma(1+z) \Gamma(\epsilon+z) \Gamma(\epsilon-1-z) \Gamma(3-2\epsilon+2z)}{\Gamma(3-2\epsilon+z) \Gamma(\frac{3}{2}+z)}
\end{align}

To evaluate the first term of Eq.~\ref{Eq:H111mMMm_int3}, we first evaluate $I_{k_2}'$ to get:
\begin{align}
	I_{k_2}' \equiv i \pi^{2-\epsilon} \frac{\Gamma(\epsilon)}{(M^2 - i0)^\epsilon} \int_0^1 dx \left( \frac{-k_1^2 x(1-x)}{M^2-i0} \right)^{-\epsilon} = \frac{i \pi^{2-\epsilon} \Gamma(\epsilon)}{(-k_1^2)^\epsilon} \frac{\Gamma^2(1-\epsilon)}{\Gamma(2-2\epsilon)} 
\end{align}
subject to $0<\text{Re}(\epsilon)<1$. From the first term of Eq.~\ref{Eq:H111mMMm_int2}, we combine the $k_1$ dependent terms to get:
\begin{align}
	I_{k_1}' \equiv \int \frac{d^d k_1}{\left[ (k_1-p)^2 - m^2 + i0 \right] \left[ k_1^2 \right]^\epsilon} = \frac{(-1)^{1+\epsilon} i \pi^{2-\epsilon}} {(m^2-i0)^{2\epsilon-1}} \frac{\Gamma(2\epsilon-1) \Gamma(3-4\epsilon)}{\Gamma(3-3\epsilon)}
	\label{Eq:Ik1psolved}
\end{align}
using $p^2 = m^2$, and with $\frac{1}{2}<\text{Re}(\epsilon)<\frac{3}{4}$. Although this does not satisfy our requirement that $\text{Re}(\epsilon)$ be allowed to be infinitesimally small, as the constraint above does not involve the contour $z$, we may analytically continue in $\epsilon$ the expression $I_{k_1}'$ to infinitesimally small values without changing its form. Combining these results, we get the expression for the first term of Eq.~\ref{Eq:H111mMMm_int3}:
\begin{align}
    \frac{\pi^{4-2\epsilon}}{(m^2-i0)^{2\epsilon-1}} \frac{\Gamma(\epsilon) \Gamma^2(1-\epsilon) \Gamma(2\epsilon-1) \Gamma(3-4\epsilon)}{\Gamma(2-2\epsilon) \Gamma(3-3\epsilon)}
\end{align}

Combining these, the final MB representation for $H_{\{1,1,1\}}(m,M,M;m^2)$ is given by:
\begin{align}
	& H_{\{1,1,1\}} (m,M,M;m^2) = \frac{\pi^{4-2\epsilon}}{(m^2-i0)^{2\epsilon-1}}
	\frac{\Gamma(\epsilon) \Gamma^2(1-\epsilon) \Gamma(2\epsilon-1) \Gamma(3-4\epsilon)}{\Gamma(2-2\epsilon) \Gamma(3-3\epsilon)} \nonumber \\
	& + \frac{\pi^{4-2\epsilon} 2^{1-2\epsilon}}{(m^2-i0)^{\epsilon-1} (M^2-i0)^{\epsilon}} \frac{1}{2 \pi i}  \nonumber \\
    & \times \int_{c'-i \infty}^{c'+i \infty} dz \left( \frac{m^2 - i0}{M^2 - i0} \right)^z \frac{\Gamma(-z) \Gamma(1+z) \Gamma(\epsilon+z) \Gamma(\epsilon-1-z) \Gamma(\frac{3}{2}-\epsilon+z) \Gamma(2-\epsilon+z)}{\Gamma(3-2\epsilon+z) \Gamma(\frac{3}{2}+z)}
\label{Eq:H111SCMB}
\end{align}
with the second term subject to the constraints $c' = \text{Re}(z) \in ] -1-\text{Re}(\epsilon),-\text{Re}(\epsilon) [ $, $\text{Re}(1+z)>0$, $\text{Re}(\epsilon-1-z)>0$, $\text{Re}(z)<0$ and $\text{Re}(3-2\epsilon+2z)>0$ which can be summarized as $-1<c'<-1+\text{Re}(\epsilon)$ (for small positive values of $\text{Re}(\epsilon)$).


Note that the representation above satisfies our requirements of straight-line contours and can be used to compute the exact expression of $H_{\{1,1,1\}}(m,M,M;m^2)$ as $\epsilon$ can be taken at any infinitesimally small value. However, we cannot set $\epsilon=0$ due to the pinch singularity of the constraint, which can be seen in Fig.~\ref{Fig:HpppMBS} which shows the singularity structure of the MB integrand. To obtain a representation in which we may safely set $\epsilon=0$, we shift the contour of the second term of Eq.~\ref{Eq:H111SCMB} to the right by subtracting the residue at $z=-1+\epsilon$ to get 
\begin{align}
    & H_{\{1,1,1\}}(m,M,M;m^2) = \frac{\pi^{4-2\epsilon} 2^{1-2\epsilon}}{(m^2-i0)^{\epsilon-1} (M^2-i0)^{\epsilon}}   \nonumber \\
    & \times \frac{1}{2 \pi i} \int_{c''-i \infty}^{c''+i \infty} dz \left( \frac{m^2 - i0}{M^2 - i0} \right)^z \frac{\Gamma(-z) \Gamma(1+z) \Gamma(\epsilon+z) \Gamma(\epsilon-1-z) \Gamma(\frac{3}{2}-\epsilon+z) \Gamma(2-\epsilon+z)}{\Gamma(3-2\epsilon+z) \Gamma(\frac{3}{2}+z)} \nonumber \\    
    & + \frac{\pi^{4-2\epsilon}}{(m^2-i0)^{2\epsilon-1}} \frac{\Gamma(\epsilon) \Gamma^2(1-\epsilon) \Gamma(2\epsilon-1) \Gamma(3-4\epsilon)}{\Gamma(2-2\epsilon) \Gamma(3-3\epsilon)}  + \frac{\pi^{\frac{9}{2}-2 \epsilon}} {(2M^2-i0)^{2\epsilon-1}} \frac{\Gamma (1-\epsilon ) \Gamma (\epsilon ) \Gamma (2 \epsilon -1)}{\Gamma (2-\epsilon ) \Gamma \left(\epsilon +\frac{1}{2}\right)} \nonumber \\
\end{align}
where  $c'' = \text{Re}(z) \in ]-1+\text{Re}(\epsilon),-\text{Re}(\epsilon)[$ (still for small and positive values of $\text{Re}(\epsilon)$). In this representation, the singularity of $\epsilon$ around $0$ has been extracted to the two non-integral terms, and which therefore still require to be expanded as a Laurent series, but which allows us to put $\epsilon=0$ directly in the integral term, if one is interested in the finite part of the sunset only.

\subsection{Solving the MB representation \label{SMBSCH111mM}}

\begin{figure}[t]
\centering
\includegraphics[scale=0.7]{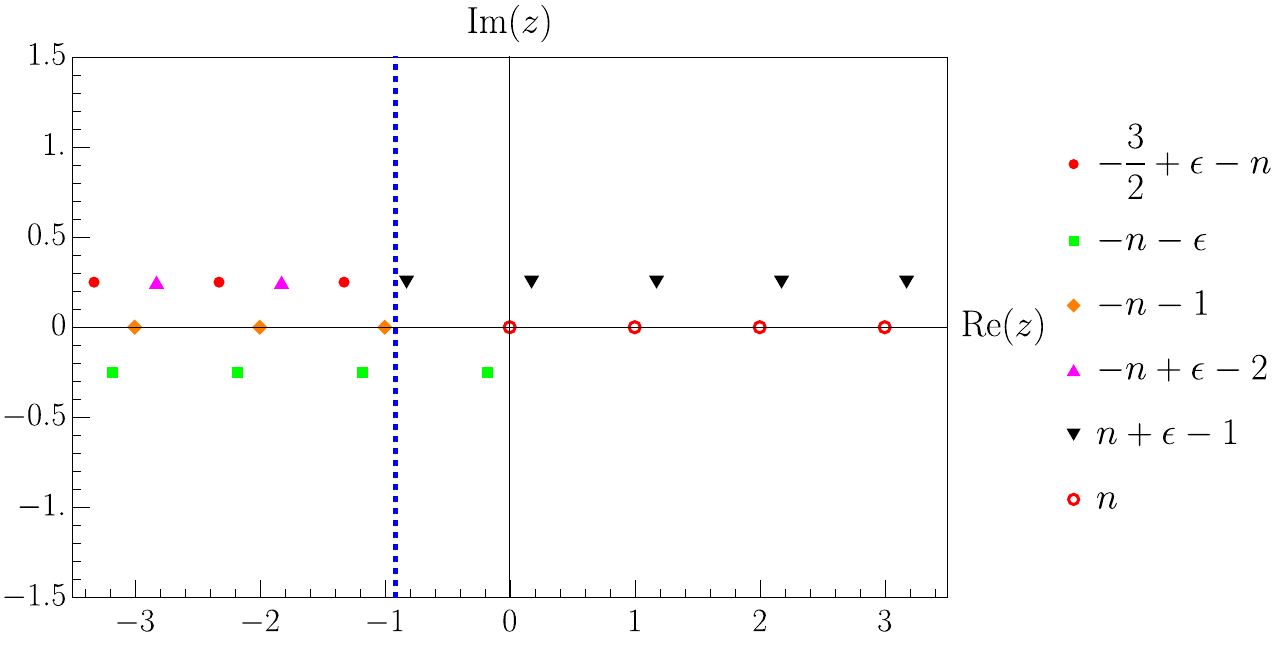}
\caption{Singularity structure, in the complex $z$-plane, of the MB integrand of the second term of the MB representation of $H_{\{1,1,1\}}(m,M,M;m^2)$ as given in Eq.~\ref{Eq:H111SCMB}, for (small) positive $\text{Re}(\epsilon)$. The dashed vertical line segment represents the MB straight contour.}
\label{Fig:HpppMBS}
\end{figure}

Using Cauchy's theorem, we evaluate the integral term of Eq.~\ref{Eq:H111SCMB} by closing the contour shown in Fig.~\ref{Fig:HpppMBS} either to the right side or to the left (see \cite{Paris} for details). As shown in the picture, the singularities on the right side of the contour are at $z=-\epsilon$, $n$ and $-1+\epsilon+n$, and on the left are at $z=-1-n$, $-\epsilon-1-n$, $-\frac{3}{2}+\epsilon-n$ and $-2+\epsilon-n$, where $n=0,1,2,3,...$. 

\begin{figure}[h]
\centering
\includegraphics[scale=0.7]{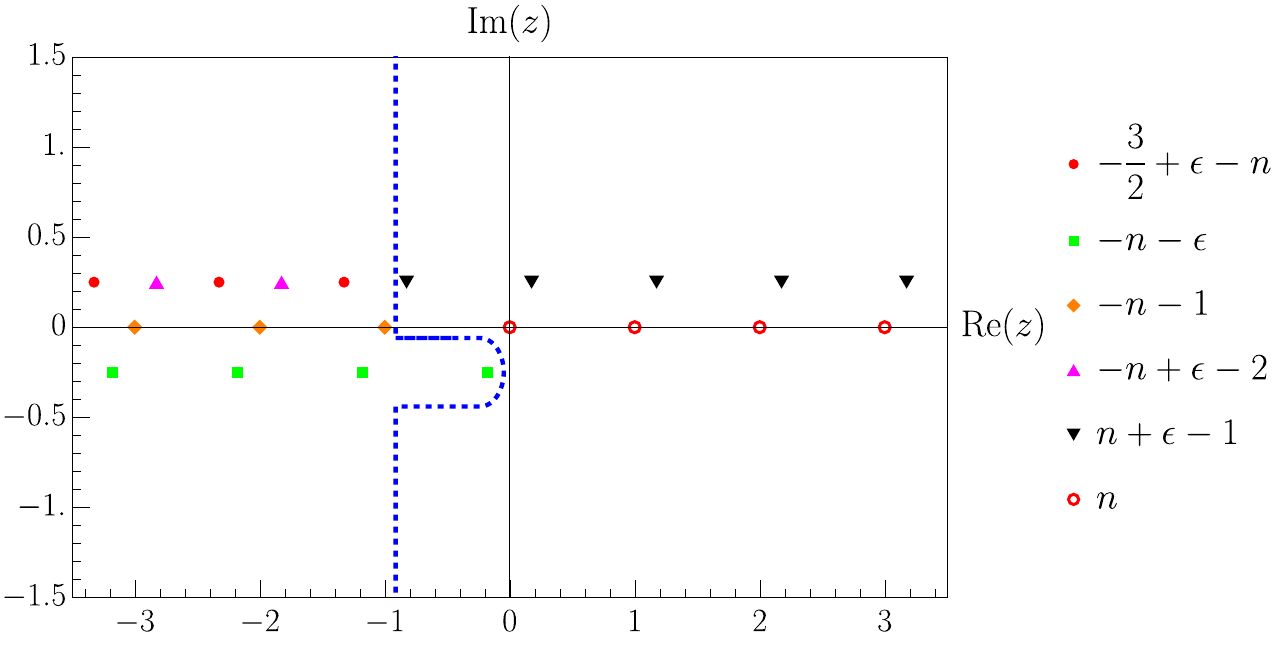}
\caption{Singularity structure, in the complex $z$-plane, of the MB integrand of the MB representation of $H_{\{1,1,1\}}(m,M,M;m^2)$ as given in Eq.~\ref{MBrepH111}, for positive real $\epsilon$. The blue dashed curve represents the MB non-straight contour}
\label{Fig:HpppMBnon-S}
\end{figure}

Note that the singularity at $z=-\epsilon$, which falls on the right of the contour, actually belongs to the set of singular points $z=-\epsilon-n$, the remaining singularities of which fall on the left. This is what differentiates this MB representation from the traditional one, given in Eq.~\ref{MBrepH111}, where the contour of integration is non-straight. Indeed, in the latter, the corresponding picture is shown in Fig.~\ref{Fig:HpppMBnon-S} where it is clear that the singularity at $z=-\epsilon$ is located to the left of the contour and thus should not be taken into account when closing the contour to the right. 
What happens in the straight contour approach is that, if one closes the contour to the right, the residue contribution at $z=-\epsilon$ will in fact be canceled by the contribution of the first term of Eq.~\ref{Eq:H111SCMB}, so that at the end one finds the same result as in the non-straight contour case, namely Eq.~\ref{SeriesH111TwoMass1}. In a similar way, if one closes the contour to the left, the first term of Eq.~\ref{Eq:H111SCMB} gives the same contribution as if one had included the pole at $z=-\epsilon$, so that this ``missing" residue is in fact taken into account (separately from the other singularities of the type $z=-\epsilon-m$) and, at the end of the calculations, once one has included this term into the corresponding series, one gets the exact same result as in the non-straight contour case, \textit{i.e} Eq.~\ref{SeriesH111TwoMass2}. This is why the straight contour approach is computationally heavier that the non-straight one, since, in addition to require a complicated derivation of the MB representation, it asks for the computation of more contributions. Moreover, the way to show the exact equivalence between the results derived from the straight contours approach and those obtained from the non-straight one  can sometimes be a bit tricky, as will be shown in the less simple example $H_{\{1,1,1\}}(m_1,m_2,m_3;m_1^2)$ in a subsequent section.

To get the final result for $H_{\{1,1,1\}}(m,M,M;m^2)$, we sum the residues on either side of the contour. We express the sums as generalized hypergeometric functions, since aside from familiarity and compactness, this also has the advantage that it may be easily analytically continued for all values of the mass ratio. Summing the residues on the RHS of the contour, and writing the sums as $_pF_q$ functions, we get:
\begin{multline}
	H_{\{1,1,1\}}(m,M,M;m^2) = \pi ^{4-2 \epsilon } \left(m^2\right)^{1-\epsilon } \left(M^2\right)^{-\epsilon } \Gamma (\epsilon -1) \Gamma (\epsilon ) \\
	\times \Bigg(
	\, {}_3F_2 \left[ \begin{array}{c}
	1,\frac{3}{2}-\epsilon ,\epsilon \\
	\frac{3}{2},3-2 \epsilon \\
	\end{array}	\bigg| \frac{m^2}{M^2} \right] -\frac{\Gamma (2 \epsilon -1)}{\Gamma (2 \epsilon)} \left(\frac{m^2}{M^2}\right)^{\epsilon -1} \,
	{}_3F_2 \left[ \begin{array}{c}
	\frac{1}{2},1,2 \epsilon -1 \\
	2-\epsilon, \epsilon +\frac{1}{2} \\
	\end{array}	\bigg| \frac{m^2}{M^2} \right] \Bigg)
 \label{H111mMMmHyperRep}
\end{multline}

Before we switch to other detailed examples, we would like to mention the following two points. The first point is that if one applies the straight contours approach to $H_{\{2,1,1\}}(m,M,M;m^2)$, one will get the following (single term) MB representation
\begin{multline}
	H_{\{2,1,1\}}(m,M,M;m^2) = 
	\frac{(i\pi^{2-\epsilon})^2}{[4(m^2-i0)(M^2-i0)]^\epsilon} \\
	\times \frac{1}{2 \pi i} \int_{c-i \infty}^{c+i \infty} dz \left( \frac{m^2 - i0}{M^2 - i0} \right)^z \frac{\Gamma(-z) \Gamma(1+z) \Gamma(\epsilon+z) \Gamma(\epsilon-z) \Gamma(\frac{3}{2}-\epsilon+z) \Gamma(1-\epsilon+z)}{\Gamma(2-2\epsilon+z) \Gamma(\frac{3}{2}+z)}
\label{Eq:H2pkkPTMB}
\end{multline}
subject to $c = \text{Re}(z) \in ]-\text{Re}(\epsilon),0[$, $\text{Re}(\epsilon-z)>0$, $\text{Re}(z+1)>0$, $\text{Re}(z)<0$ and $\text{Re}(2-2\epsilon+2z)>0$ (summarized as $-\text{Re}(\epsilon)<c<0$ for small positive $\text{Re}(\epsilon)$) and which in fact does not differ from the MB representation shown in Eq.~\ref{MBrepH211mM}. This is due to the fact that, contrary to what we were forced to do in the previous section for $H_{\{1,1,1\}}(m^2,M^2,M^2;m^2)$, there is no need to shift the contour in the intermediate steps of the derivation of the MB representation Eq.~\ref{Eq:H2pkkPTMB}: the mathematical constraints of the intermediate steps do not require it\footnote{The momentum integral in $I_{k_1}$ gives rise to the constraint $\text{Re}(\epsilon-z)>0$ instead of $\text{Re}(\epsilon-1-z)>0$ as with $H_{\{1,1,1\}}(m^2,M^2,M^2;m^2)$, due to the square power of the $(k_1-p)^2 - m^2 + i0$ propagator. This constraint is compatible with the previous ones $\text{Re}(z) \in ]-\text{Re}(\epsilon),0[$ and $\text{Re}(1+z)>0$ even for infinitesimally small values of $\epsilon$. Therefore, there is no need to restart the calculation with shifted contours, and the MB representation is a one term integral.}. Said differently, it is possible to find a fixed value of $\text{Re}(z)$ such that the real parts of the arguments of the gamma functions of the numerator of the MB integrand of Eq.~\ref{MBrepH211mM} can all be positive at the same time, for arbitrary small values of $\epsilon$. This means that it is possible to find a straight contour which does not split the set of poles of each gamma function in different subsets. As this is a necessary condition for the MB integral to represent the sunset, one concludes that the MB representations derived from AMBRE do not always require non-straight contours.

\begin{figure}[h]
\centering
\includegraphics[scale=0.6]{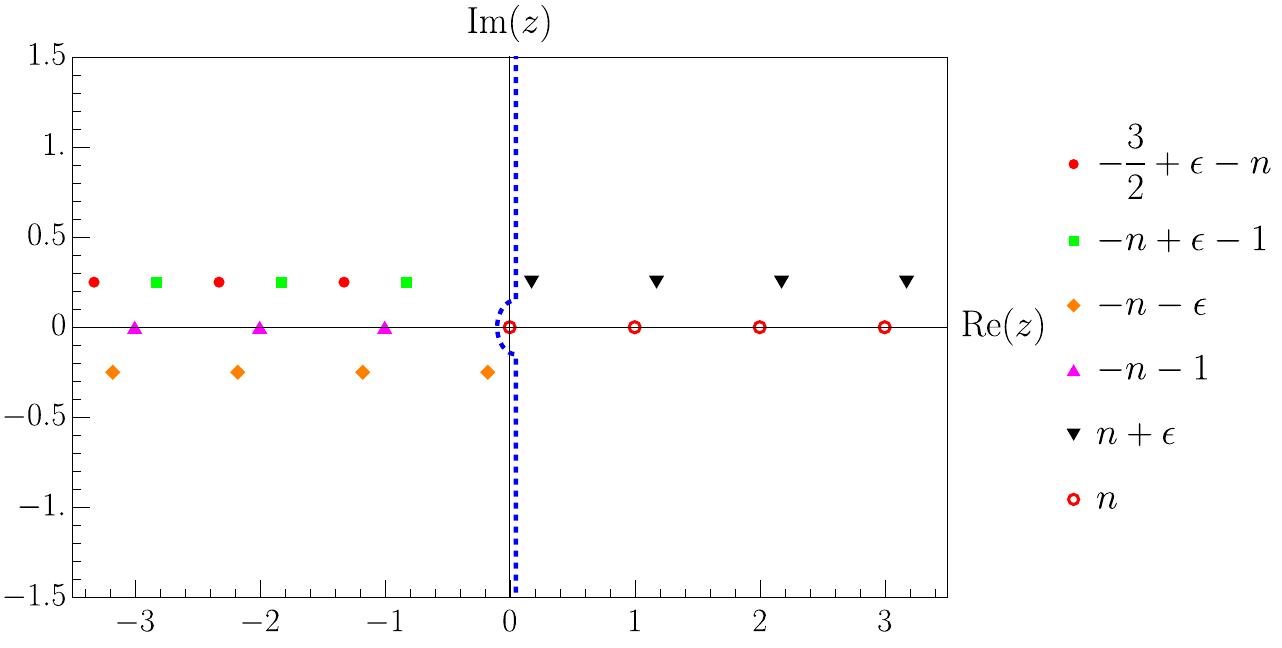}
\includegraphics[scale=0.6]{H111mMMm_nsc.pdf}
\caption{Singularity structure, in the complex $z$-plane, of the MB integrand of the MB representation of $H_{\{2,1,1\}}(m,M,M;m^2)$ as given by Eq.~\ref{MBrepH211mM} (upper figure) and by Eq.~\ref{MBrepH211mM} where Eq.~\ref{Gamma_relation} is used (lower figure), for small positive $\text{Re}(\epsilon)$. The blue dashed curves represents the MB non-straight contours. \label{Fig:H211mMMm}}
\end{figure}

The second point is that, looking at Eq.~\ref{Eq:H2pkkPTMB}, it can be tempting to use the relation \begin{equation}
\Gamma(\epsilon-z)\Gamma(1-\epsilon+z)=-\Gamma(\epsilon-1-z)\Gamma(2-\epsilon+z)
\label{Gamma_relation}
\end{equation}
in order to derive an MB representation which has exactly the same numerator integrand as the one of the second term of Eq.~\ref{Eq:H111SCMB}, allowing us to reuse part of the calculations that have been done for $H_{\{1,1,1\}}(m,M,M;m^2)$. This is perfectly justified in the straight contour case because the contour is fixed and the singular structure remains unchanged after applying the relation above. However, one cannot use this relation blindly in the non-straight contour case. Indeed, in the latter situation, this modification of the gamma functions in the integrand would also automatically lead to a modification of the contour, as shown in Fig.~\ref{Fig:H211mMMm}. Therefore, applying this relation blindly would give an incorrect analytic result at the end.

For $H_{\{2,1,1\}}(m,M,M;m^2)$, closing the contour to the right, gives us the following expression:
\begin{multline}
	H_{\{2,1,1\}}(m,M,M;m^2) = - \pi^{4-2 \epsilon} \left(m^2\right)^{-\epsilon} \left(M^2\right)^{-\epsilon} \Gamma (\epsilon)^2 \\
	\times \Bigg( {}_3F_2 \left[ \begin{array}{c}
	1,\frac{3}{2}-\epsilon ,\epsilon \\
	\frac{3}{2},2-2 \epsilon \\
	\end{array}	\bigg| \frac{m^2}{M^2} \right] + \frac{1}{4} \frac{\Gamma (\epsilon -1) \Gamma \left(\epsilon +\frac{1}{2}\right)}{\Gamma (\epsilon ) \Gamma \left(\epsilon +\frac{3}{2}\right)}  \left(\frac{m^2}{M^2}\right)^{\epsilon }
	{}_3F_2 \left[ \begin{array}{c}
		1,\frac{3}{2},2 \epsilon \\
		2-\epsilon ,\epsilon +\frac{3}{2} \\
	\end{array}	\bigg| \frac{m^2}{M^2} \right] \Bigg)
\end{multline}
which may be easily analytically continued for all values of the mass ratio.

In Sec.~\ref{Sec:Applications}, we expand these results after applying chiral subtraction, and then demonstrate their uses in chiral perturbation theory applications.

\section{Three mass scale sunsets \label{Sec:ThreeMassScale}}

In the three mass scale case, there are two distinct mass configurations that the sunset integrals can have, namely $(m_1, m_2, m_3; m_1^2)$ and $(m_1, m_2, m_2; m_3^2)$, and for each of these there are four possible master integrals. However by differentiating w.r.t. $m_2^2$ or $m_3^2$ the $(m_1, m_2, m_3; m_1^2)$ master integral with the $(1,1,1)$ propagator configuration, one can obtain the $(1,2,1)$ and $(1,1,2)$ propagator configurations, respectively, of this mass configuration. Similarly, by differentiating w.r.t. $m_1^2$ and $m_2^2$ the $(1,1,1)$ master integral with the $(m_1, m_2, m_2; m_3^2)$ mass configuration, one obtains all the other master integrals of this mass configuration. In total, therefore, there are only three independent sunset integrals that need to be evaluated in the three mass scale case: $H_{\{1,1,1\}} \left(m_1, m_2, m_3; m_1^2 \right)$, $H_{\{2,1,1\}} \left(m_1, m_2, m_3; m_1^2 \right)$, and $H_{\{1,1,1\}} \left(m_1, m_2, m_2; m_3^2  \right)$. In this section we find MB representations for each of these three integrals, and then solve them completely.

\subsection{$H_{\{1,1,1\}}(m_1,m_2,m_3;m_1^2)$ in the straight contours approach}

The derivation of the MB representation of
\begin{align}
	H_{\{1,1,1\}} \left(m_1, m_2, m_3; m_1^2 \right) \equiv
	\int \frac{d^d k_1 d^d k_2}{\left[ (k_1-p)^2 - m_1^2 + i 0 \right] \left[ k_2^2 - m_2^2 +i 0 \right] \left[ (k_1-k_2)^2 - m_3^2 + i 0 \right]} \Bigg|_{p^2 = m_1^2}
\label{Eq:H111kaon}
\end{align}
begins by evaluating the $k_2$ loop, which we call $I_{k_2}$:
\begin{align}
	I_{k_2} &\equiv \int \frac{d^d k_2}{\left[ k_2^2 - m_2^2 +i 0 \right] \left[ (k_1-k_2)^2 - m_3^2 + i 0 \right]} 
	= \int \frac{d^d k_2}{k_2^2 \left( 1 + \frac{-m_2^2+i0}{k_2^2} \right) (k_1-k_2)^2 \left( 1 + \frac{-m_3^2+i0}{(k_1-k_2)^2} \right)} 
\end{align}
We then apply the ``battle-horse" MB representation to both the propagators:
\begin{align}
	\left( 1 + \frac{-m_3^2+i0}{(k_1-k_2)^2} \right)^{-1} = \frac{1}{2\pi i} \int_{c-i\infty}^{c+i\infty}  dz_1 \left( \frac{-m_3^2+i0}{(k_1-k_2)^2} \right)^{z_1} \Gamma(-z_1) \Gamma(1+z_1)
	\label{Eq:MBsKaon}
\end{align}
\begin{align}
	\left( 1 + \frac{-m_2^2+i0}{k_2^2} \right)^{-1} = \frac{1}{2\pi i} \int_{d-i\infty}^{d+i\infty}  dz_2 \left( \frac{-m_2^2+i0}{k_2^2} \right)^{z_2} \Gamma(-z_2) \Gamma(1+z_2)
	\label{Eq:MBtKaon}
\end{align}
where the straight line contours parallel to the imaginary axis are given by $c=\text{Re}(z_1) \in  ]-1,0[$ and $d=\text{Re}(z_2) \in  ]-1,0[$. We therefore end up with:
\begin{multline}
	I_{k_2} = \frac{1}{(2\pi i)^2} 
	\int_{c-i\infty}^{c+i\infty} dz_1 \, \Gamma(-z_1) \Gamma(1+z_1)   \left( -m_3^2+i0 \right)^{z_1}	
	\int_{d-i\infty}^{d+i\infty}  dz_2 \,  \Gamma(-z_2) \Gamma(1+z_2) \left(-m_2^2+i0 \right)^{z_2} \\
	\times \int d^d k_2 \frac{1}{[k_2^2]^{1+z_2} [ (k_1-k_2)^2]^{1+z_1}} 
\end{multline}

We apply Feynman parameterization on the now massless propagators of $I_{k_2}$ to obtain:
\begin{align}
	\int \frac{d^d k_2}{[k_2^2]^{1+z_2} [ (k_1-k_2)^2]^{1+z_1} } = \frac{i \pi^{2-\epsilon} (-1)^{z_1+z_2}}{(-k_1^2)^{\epsilon+z_1+z_2}} \frac{\Gamma(\epsilon+z_1+z_2)}{\Gamma(1+z_1) \Gamma(1+z_2)} \int_0^1 dx \, x^{-\epsilon-z_1} (1-x)^{-\epsilon-z_2}
\end{align}
which holds if $\text{Re}\left(2+z_1+z_2-\frac{d}{2}\right)>0$. This constraint results from the convergence of the $k_2$ integral. If $\text{Re}(\epsilon)>0$ this constraint is compatible with the ones arising from application of the MB formula on the propagators: $\text{Re}(z_1) \in  ]-1,0[$ and $\text{Re}(z_2) \in  ]-1,0[$. However, the constraints are not simultaneously satisfiable for $\epsilon=0$. To allow for this, we need to replace either the constraint $\text{Re}(z_1) \in  ]-1,0[$ by $\text{Re}(z_1) \in  ]0,1[$, or $\text{Re}(z_2) \in  ]-1,0[$ by $\text{Re}(z_2) \in  ]0,1[$. This can be achieved by shifting the contour in Eq.~\ref{Eq:MBsKaon} or Eq.~\ref{Eq:MBtKaon} to the right. Let us choose the latter:
\begin{align}
	\left( 1 + \frac{-m_2^2+i0}{k_2^2} \right)^{-1} = 1 +  \frac{1}{2\pi i} \int_{d'-i\infty}^{d'+i\infty}  dz_2 \left( \frac{-m_2^2+i0}{k_2^2} \right)^{z_2} \Gamma(-z_2) \Gamma(1+z_2)
	\label{Eq:MBtKaon2}
\end{align}
where $d' = \text{Re}(z_2) \in  ]0,1[$. This will introduce an additional residue term, but the rest of the calculation follows through as previously, except with a shifted contour, $d'$ instead of $d$. Therefore, we continue with the calculation and compute the integral over the Feynman parameter $x$:
\begin{align}
	\int_0^1 dx \, x^{-\epsilon-z_1} (1-x)^{-\epsilon-z_2} = \frac{\Gamma(1-\epsilon-z_1) \Gamma(1-\epsilon-z_2)}{\Gamma(2-2\epsilon-z_1-z_2)}
\end{align}
where the equality holds provided that the constraints coming from the integral representation of the Beta function are satisfied, namely $\text{Re}(1-\epsilon-z_1) > 0$ and $\text{Re}(1-\epsilon-z_2) > 0$. We end up with $I_{k_2}$ now given by:
\begin{align}
	I_{k_2} &= \frac{1}{2\pi i} \int_{c-i\infty}^{c+i\infty} dz_1 \, \Gamma(-z_1) \Gamma(1+z_1) \left( -m_3^2+i0 \right)^{z_1} 
	\int \frac{d^d k_2}{ k_2^2 [ (k_1-k_2)^2]^{1+z_1} } \nonumber \\
	& + \frac{1}{(2\pi i)^2} \frac{i \pi^{2-\epsilon}}{(-k_1^2)^{\epsilon}} \int_{c-i\infty}^{c+i\infty} dz_1 \int_{d'-i\infty}^{d'+i\infty} dz_2 \left( \frac{-m_3^2+i0}{k_1^2} \right)^{z_1} \left( \frac{-m_2^2+i0}{k_1^2} \right)^{z_2} \frac{\Gamma(-z_1) \Gamma(-z_2)}{\Gamma(2-2\epsilon-z_1-z_2)} \nonumber \\ 
	& \hspace{6.5cm} \times \Gamma(1-\epsilon-z_1) \Gamma(1-z_2-\epsilon) \Gamma(\epsilon+z_1+z_2)
\end{align}
where the full set of constraints for the second term is $\text{Re}(z_2) \in  ]0,1[$, $\text{Re}(z_1) \in  ]-1,0[$, $\text{Re}(\epsilon+z_1+z_2)>0$, $\text{Re}(1-\epsilon-z_1) > 0$ and $\text{Re}(1-\epsilon-z_2) > 0$. These are satisfied for $\epsilon=0$.

The integral over $k_2$ in the first term converges if $\text{Re}(\epsilon+z_1)>0$. This condition, while simultaneously satisfiable with the constraint $\text{Re}(z_1) \in ]-1,0[$ for infinitesimally small values of $\epsilon$, cannot be satisfied for $\epsilon=0$. As previously, shifting the contour of the $z_1$ complex-plane integral to the right resolves the conflict. Extracting the corresponding residue one gets:
\begin{align}
	\left( 1 + \frac{-m_3^2+i0}{(k_1-k_2)^2} \right)^{-1} = 1 + \frac{1}{2\pi i} \int_{c'-i\infty}^{c'+i\infty}  dz_1 \left( \frac{-m_3^2+i0}{(k_1-k_2)^2} \right)^{z_1} \Gamma(-z_1) \Gamma(1+z_1) 
\label{Eq:MBsKaon2}
\end{align}
where $c' = \text{Re}(z_1) \in  ]0,1[$.
Redoing the calculation with Eq.~\ref{Eq:MBsKaon2} and Eq.~\ref{Eq:MBtKaon2} inserted into the expression for $I_{k_2}$ results in $I_{k_2}$ now reading:
\begin{align}
	I_{k_2} &= \int \frac{d^d k_2}{ k_2^2 (k_1-k_2)^2 } 
	+	\frac{1}{2\pi i} \int_{c'-i\infty}^{c'+i\infty} dz_1 \, \Gamma(-z_1) \Gamma(1+z_1) \left( -m_3^2+i0 \right)^{z_1} \int \frac{d^d k_2}{ k_2^2 [ (k_1-k_2)^2]^{1+z_1} } \nonumber \\
	& +	\frac{1}{2\pi i} \int_{d'-i\infty}^{d'+i\infty} dz_2 \, \Gamma(-z_2) \Gamma(1+z_2) \left( -m_2^2+i0 \right)^{z_2} \int \frac{d^d k_2}{ k_2^2 [ (k_1-k_2)^2]^{1+z_2} } \nonumber \\
	& + \frac{1}{(2\pi i)^2} \frac{i \pi^{2-\epsilon}}{(-k_1^2)^{\epsilon}} \int_{c-i\infty}^{c+i\infty} dz_1 \int_{d'-i\infty}^{d'+i\infty} dz_2  \left( \frac{-m_3^2+i0}{k_1^2} \right)^{z_1} \left( \frac{-m_2^2+i0}{k_1^2} \right)^{z_2}  \frac{\Gamma(-z_1) \Gamma(-z_2)}{\Gamma(2-2\epsilon-z_1-z_2)} \nonumber \\ 
	& \hspace{6.5cm} \times \Gamma(1-\epsilon-z_1) \Gamma(1-\epsilon-z_2) \Gamma(\epsilon+z_1+z_2)
	\label{Eq:Ik2kaon}
\end{align}

The first term is solved using Feynman parameterization to give:
\begin{align}
	\int \frac{d^d k_2}{k_2^2 (k_1-k_2)^2} 
	= \frac{i \pi^{2-\epsilon}}{(-k_1^2)^\epsilon} \frac{\Gamma(\epsilon) \Gamma^2(1-\epsilon)}{\Gamma(2-2\epsilon)}
\end{align}
subject to the constraints $\text{Re}(\epsilon)>0$ and $\text{Re}(1-\epsilon)>0$. It is obviously not possible to set $\epsilon=0$, but given the fact that the corresponding singularity does not involve an MB variable, one can evaluate this term for infinitesimally small but finite $\epsilon$ and take an expansion around $\epsilon=0$ at the end of the calculation, as it is usually done.

The second term of Eq.~\ref{Eq:Ik2kaon} evaluates to:
\begin{align}
	\frac{\Gamma(1-\epsilon)}{(-k_1^2)^{\epsilon}}  \frac{i \pi^{2-\epsilon}}{2\pi i} \int_{c'-i\infty}^{c'+i\infty} dz_1 \, \left( \frac{-m_3^2+i0}{k_1^2} \right)^{z_1} \frac{\Gamma(-z_1)  \Gamma(1-\epsilon-z_1) \Gamma(\epsilon+z_1)}{\Gamma(2-2\epsilon-z_1)}
\end{align}
subject to the constraints $\text{Re}(\epsilon+z_1)>0$, $\text{Re}(1-\epsilon-z_1)>0$, $\text{Re}(1-\epsilon)>0$ and $\text{Re}(z_1) \in  ]0,1[$, which can be satisfied simultaneously for both $\epsilon \neq 0$ as well as $\epsilon=0$. The third term of $I_{k_2}$ is symmetric to the second term, and can be found  by the exchanges $m_2^2 \leftrightarrow m_3^2$ and $z_1 \leftrightarrow z_2$.

The constraints of the fourth term: $c' = \text{Re}(z_1) \in  ]0,1[$, $d' = \text{Re}(z_2) \in  ]0,1[$, $\text{Re}(\epsilon+z_1+z_2)>0$, $\text{Re}(1-\epsilon-z_1) > 0$ and $\text{Re}(1-\epsilon-z_2) > 0$, can  be simultaneously satisfied for both zero and non-zero values of $\text{Re}(\epsilon)$.

The full integral now reads:
\begin{align}
	H_{\{1,1,1\}} &\left(m_1, m_2, m_3; m_1^2 \right)
	= \int \frac{d^d k_1}{[(k_1-p)^2 - m_1^2 + i 0]} \Bigg\{ \frac{i 4^\epsilon \pi^{5/2-\epsilon}}{2(-k_1^2)^\epsilon} \frac{\Gamma(\epsilon) \Gamma(1-\epsilon)}{\Gamma(\frac{3}{2}-\epsilon)} \nonumber \\
	& + \frac{\Gamma(1-\epsilon)}{(-k_1^2)^{\epsilon}}  \frac{i \pi^{2-\epsilon}}{2\pi i} \int_{c'-i\infty}^{c'+i\infty} dz_1 \, \left( \frac{-m_3^2+i0}{k_1^2} \right)^{z_1} \frac{\Gamma(-z_1)  \Gamma(1-\epsilon-z_1) \Gamma(\epsilon+z_1)}{\Gamma(2-2\epsilon-z_1)} \nonumber \\
	& + \frac{\Gamma(1-\epsilon)}{(-k_1^2)^{\epsilon}}  \frac{i \pi^{2-\epsilon}}{2\pi i} \int_{d'-i\infty}^{d'+i\infty} dz_2 \, \left( \frac{-m_2^2+i0}{k_1^2} \right)^{z_2} \frac{\Gamma(-z_2)  \Gamma(1-\epsilon-z_2) \Gamma(\epsilon+z_2)}{\Gamma(2-2\epsilon-z_2)}\nonumber \\
	& + \frac{1}{(2\pi i)^2} \frac{i \pi^{2-\epsilon}}{(-k_1^2)^{\epsilon}} \int_{c'-i\infty}^{c'+i\infty} dz_1 \int_{d'-i\infty}^{d'+i\infty} dz_2 \left( \frac{-m_3^2+i0}{k_1^2} \right)^{z_1} \left( \frac{-m_2^2+i0}{k_1^2} \right)^{z_2} \Gamma(-z_1) \Gamma(-z_2) \nonumber \\ 
	& \hspace{5.75cm} \times \frac{ \Gamma(1-\epsilon-z_1) \Gamma(1-\epsilon-z_2) \Gamma(\epsilon+z_1+z_2)}{\Gamma(2-2\epsilon-z_1-z_2)}  \Bigg\} \label{Eq:H111kaonMBmid}
\end{align}

When evaluated, the second term of Eq.~\ref{Eq:H111kaonMBmid} leads to the following constraints: $0<\text{Re}(z_1)<1$, $\text{Re}(\epsilon+z_1)>0$, $\text{Re}(2\epsilon+z_1)>0$, $\text{Re}(1-\epsilon-z_1)>0$ and $\text{Re}(1-2\epsilon-z_1)>0$. These can be satisfied for infinitesimally small values of $\epsilon$, but not for $\epsilon=0$. The problem arises due to the $\text{Re}(1-\epsilon-z_2)>0$ condition. 
The derivation of an MB representation of $H_{\{1,1,1\}}\left(m_1, m_2, m_3; m_1^2 \right)$ where we can set $\epsilon=0$ ab initio will require a completely different approach, and may have a form that varies significantly from the representation presented below.

Evaluating the $k_1$ integral in each term of Eq.~\ref{Eq:H111kaonMBmid} yields the following `symmetric' representation:
\begin{align}
	H_{\{1,1,1\}}\left(m_1, m_2, m_3; m_1^2 \right) = I_1+I_2+I_3+I_4
\end{align}
where
\begin{align}
	I_1 \equiv \frac{\pi^{4-2\epsilon}}{(m_1^2)^{2\epsilon-1}} \frac{\Gamma(\epsilon) \Gamma(1-\epsilon)^2 \Gamma(2\epsilon-1) \Gamma(3-4\epsilon)}{\Gamma(2-2\epsilon) \Gamma(3-3\epsilon)},
	\label{Eq:I1}
\end{align}
\begin{align}
	I_2 \equiv \frac{\pi^{4-2\epsilon}}{(m_1^2)^{2\epsilon-1}} \frac{\Gamma(1-\epsilon)}{2\pi i} \int_{c'-i\infty}^{c'-i\infty} dz_1 \left( \frac{m_3^2-i0}{m_1^2} \right)^{z_1} \frac{\Gamma(-z_1) \Gamma(1-\epsilon-z_1) \Gamma(\epsilon+z_1) \Gamma(2\epsilon-1+z_1) \Gamma(3-4\epsilon-2z_1)}{\Gamma(2-2\epsilon-z_1) \Gamma(3-3\epsilon-z_1)}
\end{align}
where $c'=\text{Re}(z_1) \in ]1-2\epsilon, 1-\epsilon[$,
\begin{align}
	I_3 \equiv \frac{\pi^{4-2\epsilon}}{(m_1^2)^{2\epsilon-1}} \frac{\Gamma(1-\epsilon)}{2\pi i} \int_{d'-i\infty}^{d'-i\infty} dz_2 \left( \frac{m_2^2-i0}{m_1^2} \right)^{z_2} \frac{\Gamma(-z_2) \Gamma(1-\epsilon-z_2) \Gamma(\epsilon+z_2) \Gamma(2\epsilon-1+z_2) \Gamma(3-4\epsilon-2z_2)}{\Gamma(2-2\epsilon-z_2) \Gamma(3-3\epsilon-z_2)}
\end{align} 
subject to the same constraints as $I_2$ but with $z_1 \leftrightarrow z_2$, and
\begin{multline}
	I_4 \equiv \frac{\pi^{4-2\epsilon}}{(m_1^2)^{2\epsilon-1}} \frac{1}{(2\pi i)^2} \int_{c'-i\infty}^{c'+i\infty} dz_1 \int_{d'-i\infty}^{d'+i\infty} dz_2 \left( \frac{m_3^2-i0}{m_1^2} \right)^{z_1} \left( \frac{m_2^2-i0}{m_1^2} \right)^{z_2} \Gamma(-z_1) \Gamma(-z_2)  \\
	 \times \frac{\Gamma(2\epsilon-1+z_1+z_2) \Gamma(\epsilon+z_1+z_2) \Gamma(1-\epsilon-z_1) \Gamma(1-\epsilon-z_2) \Gamma(3-4\epsilon-2z_1-2z_2)}{\Gamma(2-2\epsilon-z_1-z_2) \Gamma(3-3\epsilon-z_1-z_2)}
  \label{I4}
\end{multline}
which holds when $c'=\text{Re}(z_1) \in ]0,1[$, $d'=\text{Re}(z_2) \in ]0,1[$, $\text{Re}(\epsilon+z_1+z_2)>0$, $\text{Re}(2\epsilon-1+z_1+z_2)>0$, $\text{Re}(1-\epsilon-z_1)>0$, $\text{Re}(1-\epsilon-z_2)>0$ and $\text{Re}(3-4\epsilon-2z_1-2z_2)>0$.

Note that in getting to Eq.~\ref{Eq:Ik2kaon}, which features in the derivation of the `symmetric' representation of $H_{\{1,1,1\}}\left(m_1, m_2, m_3; m_1^2 \right)$ above, we shifted the contours on both propagator representations (i.e. Eqs.~\ref{Eq:MBsKaon2} and ~\ref{Eq:MBtKaon2}) in order to get a representation of $H_{\{1,1,1\}}\left(m_1, m_2, m_3; m_1^2 \right)$ where we could put $\epsilon=0$ from the beginning. In the case of $H_{\{1,1,1\}}\left(m_1, m_2, m_3; m_1^2 \right)$ the double shift of contours is unnecessary. A shift in a single one of either propagator allows us to get a `minimal' representation of $H_{\{1,1,1\}}\left(m_1, m_2, m_3; m_1^2 \right)$ consisting of three terms:
\begin{align}
	& H_{\{1,1,1\}} \left(m_1, m_2, m_3; m_1^2 \right) = \frac{\pi^{4-2\epsilon}}{(m_1^2)^{2\epsilon-1}} \Bigg\{ \frac{\Gamma(\epsilon) \Gamma^2(1-\epsilon) \Gamma(2\epsilon-1) \Gamma(3-4\epsilon)}{\Gamma(2-2\epsilon) \Gamma(3-3\epsilon)} \nonumber \\
	& \quad + \frac{\Gamma(1-\epsilon)}{2\pi i} \int_{d'-i\infty}^{d'-i\infty} dz_2 \left( \frac{m_2^2-i0}{m_1^2} \right)^{z_2} \frac{\Gamma(-z_2) \Gamma(1-\epsilon-z_2) \Gamma(\epsilon+z_2) \Gamma(2\epsilon-1+z_2) \Gamma(3-4\epsilon-2z_2)}{\Gamma(2-2\epsilon-z_2) \Gamma(3-3\epsilon-z_2)} \nonumber \\
	& \quad + \frac{1}{(2\pi i)^2} \int_{c'-i\infty}^{c'+i\infty} dz_1 \int_{d-i\infty}^{d+i\infty} dz_2 \left( \frac{m_3^2-i0}{m_1^2} \right)^{z_1} \left( \frac{m_2^2-i0}{m_1^2} \right)^{z_2} \Gamma(-z_1) \Gamma(-z_2) \Gamma(\epsilon+z_1+z_2) \nonumber \\
	& \qquad \qquad \times \frac{\Gamma(2\epsilon-1+z_1+z_2)  \Gamma(1-\epsilon-z_1) \Gamma(1-\epsilon-z_2) \Gamma(3-4\epsilon-2z_1-2z_2)}{\Gamma(2-2\epsilon-z_1-z_2) \Gamma(3-3\epsilon-z_1-z_2)} \Bigg\}
\end{align}
subject to similar constraints as the `symmetric' representation, but with $\text{Re}(z_2) \in ]-1,0[$ instead of $\text{Re}(z_2) \in ]0,1[$.

However, in the case of $H_{\{2,1,1\}}\left(m_1, m_2, m_3; m_1^2 \right)$, which follows a very similar derivation, this double shift of contours is necessary, as it is not possible to get a singularity resolved representation due to the constraint arising from the Beta function integral.

Although the `symmetric' MB representation is not the most compact one, we shall use it to evaluate $H_{\{1,1,1\}}\left(m_1, m_2, m_3; m_1^2 \right)$, as the symmetry in its treatment of the $z_1$ and $z_2$ MB integration variables simplifies the calculation, and as it parallels the MB representation of $H_{\{2,1,1\}}\left(m_1, m_2, m_3; m_1^2 \right)$.

\subsubsection*{Solving the MB integral}

In order to compute the MB representation of $H_{\{1,1,1\}}\left(m_1, m_2, m_3; m_1^2 \right)$ we use the same technique as in Sec.~\ref{SMBSCH111mM} for the $I_2$ integral, and the techniques of \cite{Tsikh:1998,Friot:2011ic} for $I_4$. The integral $I_3$ can be evaluated from $I_2$ using the replacement $m_3 \rightarrow m_2$.

\begin{figure}[h!]
\centering
\includegraphics[scale=0.7]{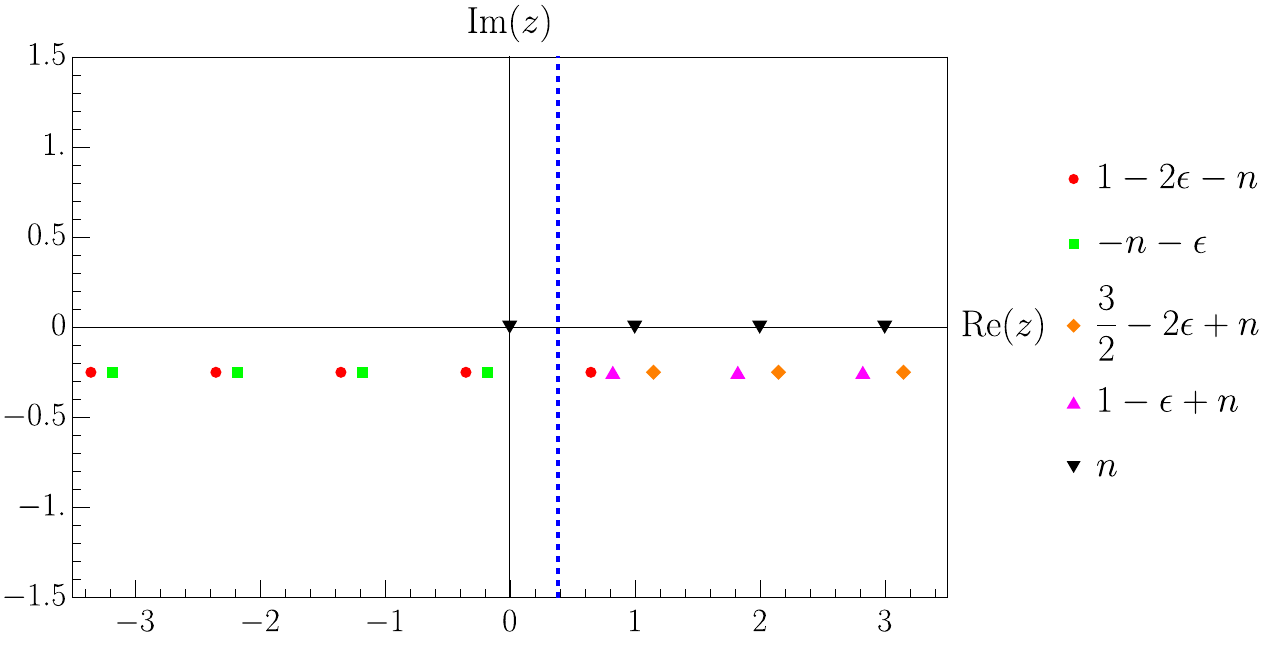}
\caption{Singularity structure of integral $I_2$}
\label{Fig:I2}
\end{figure}

The singularity structure of $I_2$ is shown in Fig.~\ref{Fig:I2}, and the correspondence between the singularities and their residues is given in Table~\ref{Table:H111k}. 

\begin{figure}[h]
\center
\begin{tabular}{lll}
\includegraphics[scale=0.35]{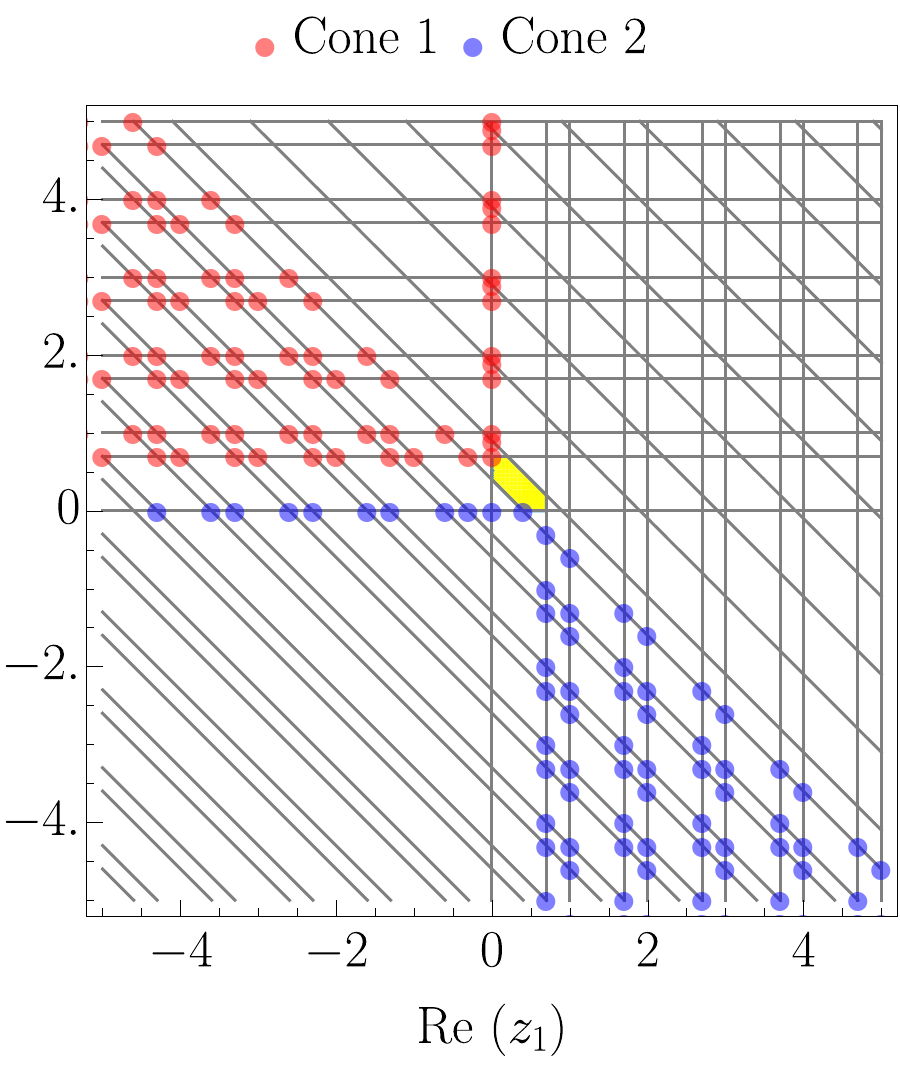}
&
\includegraphics[scale=0.35]{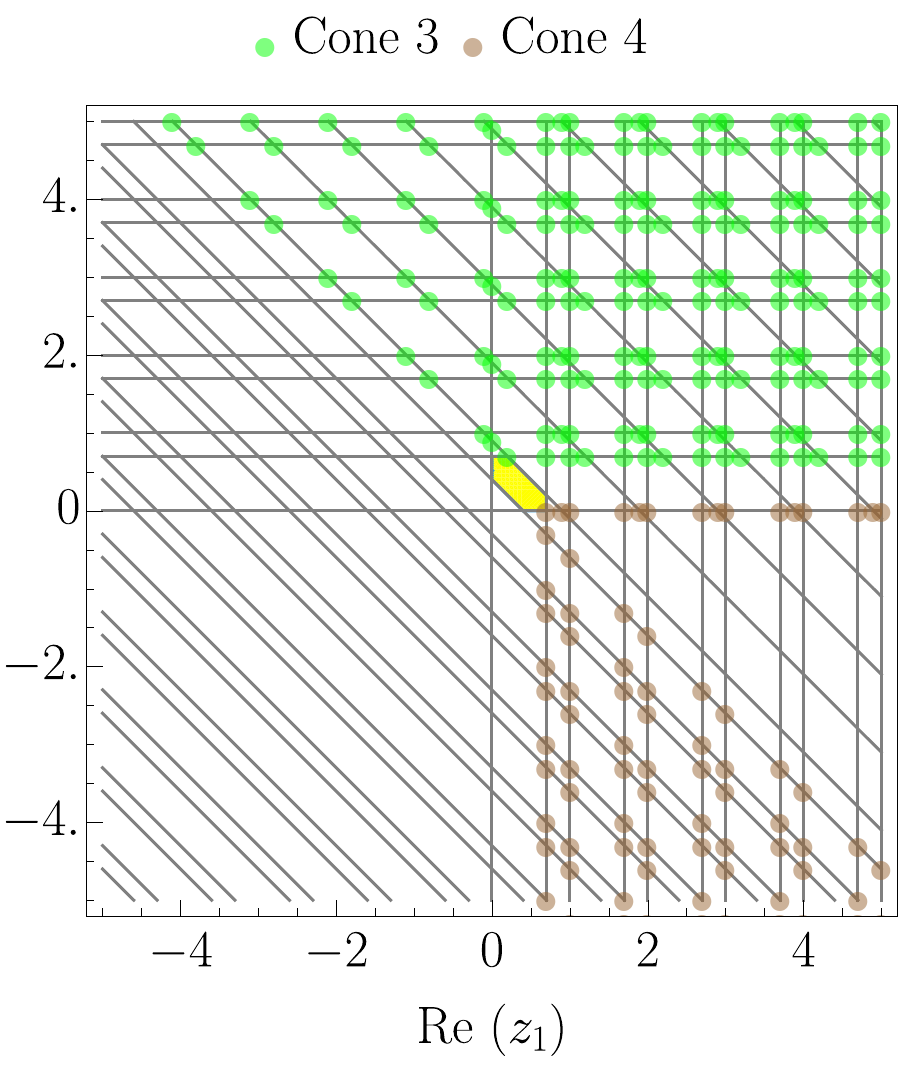}
&
\includegraphics[scale=0.35]{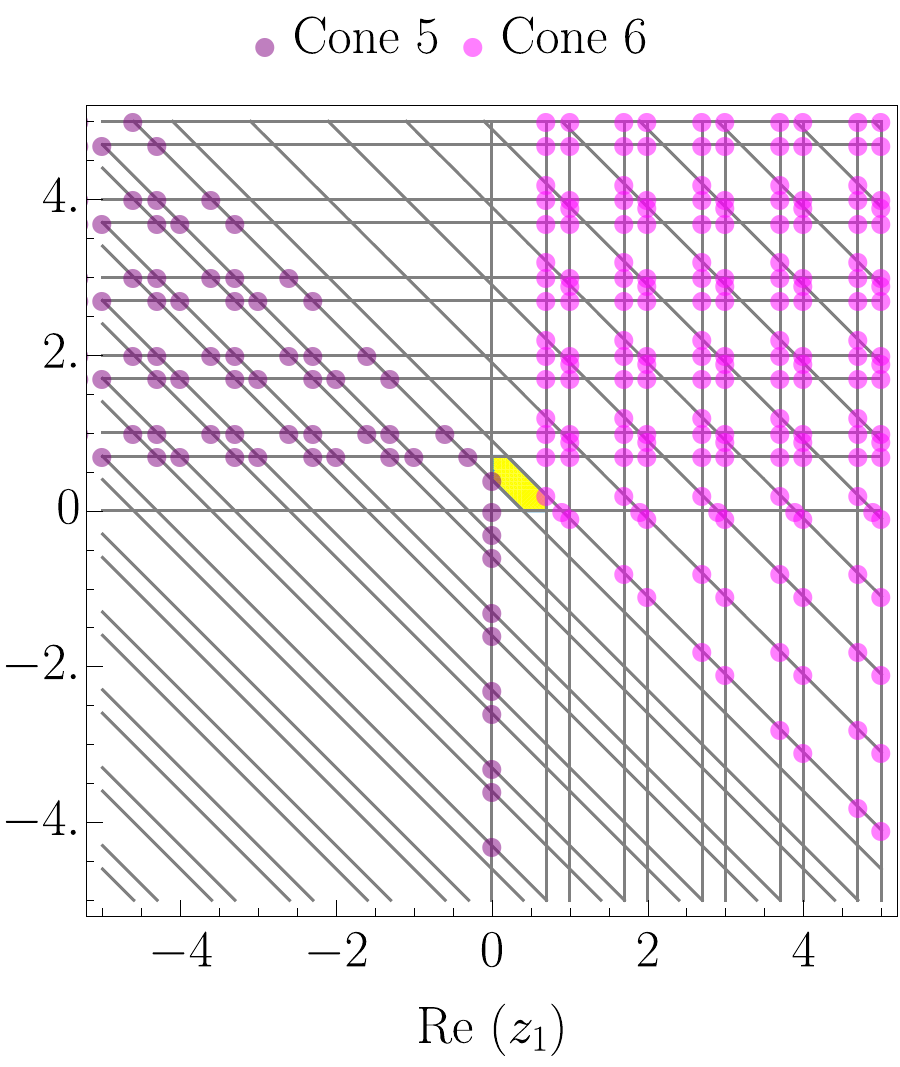}
\end{tabular}
\caption{The contributing singularities of the various cones of $I_4$}
\label{Fig:I4_cones}
\end{figure}

\begin{figure}[h]
\center
\includegraphics[scale=0.7]{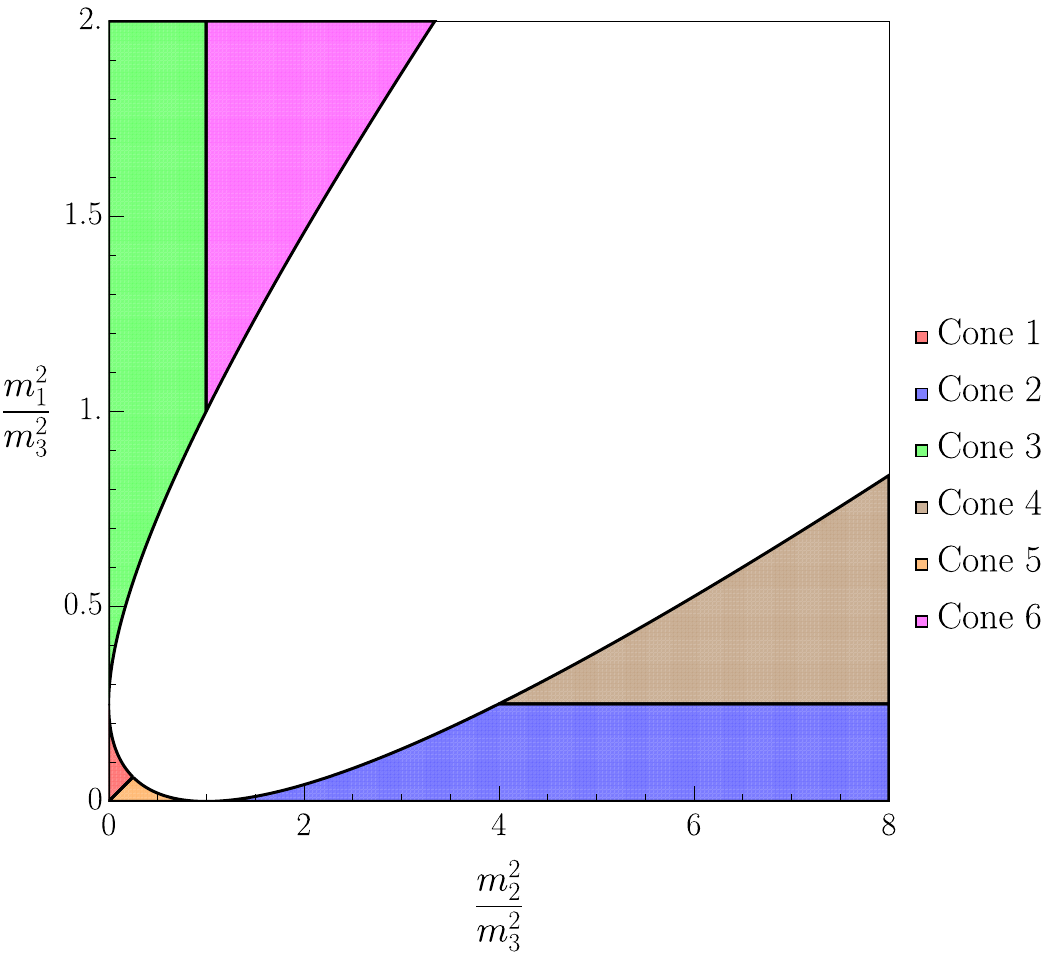}
\caption{The region of convergence of the various cones of $I_4$}
\label{Fig:I4_RoC}
\end{figure}

The integral $I_4$ has six cones, the regions of convergence of which are shown in Fig.~\ref{Fig:I4_RoC}. However, due to the $z_1 \leftrightarrow z_2$ symmetry of the integral, we need to explicitly compute only three of them. The residues corresponding to the various cones are shown in Fig.~\ref{Fig:I4_cones}, and given in Table~\ref{Table:H111k} and App.~\ref{Sec:H111kResults}.

\begin{table}[h!]
    \centering
    \begin{tabular}{|l|c||c|c|c|}
    \hline
    \rowcolor{lightgray}
    \multicolumn{2}{|c||}{} & \multicolumn{3}{c|}{\it Full} \\
    \hline
    \rowcolor{lightgray}
    \bf{Region of Convergence} & $C$ & $c$ & $\alpha$ & $\beta$ \\
    \hline
    $ (m_2 + m_3 < 2 m_1 ) \wedge (m_2^2 < m_3^2) $ & 1 & 7 & L & R \\
    \hline
    $ (2 m_1 + m_2 < m_3) \wedge (m_2^2 < 4 m_1^2) $ & 2 & 7 & L & L \\
    \hline
    $ (m_2 + m_3 < 2 m_1 ) \wedge (m_2^2 > m_3^2) $ & 3 & 9 & R & R \\
    \hline
    $ (2 m_1 + m_2 < m_3) \wedge (m_2^2 > 4 m_1^2) $ & 4 & 7 & L & R \\
    \hline
    $ (2 m_1 + m_3 < m_2) \wedge (4 m_1^2 < m_3^2) $ & 5 & 7 & L & L \\
    \hline
    $ (2 m_1 + m_3 < m_2) \wedge (4 m_1^2 > m_3^2) $ & 6 & 9 & R & R \\
    \hline
    \end{tabular}
    \caption{Table of correspondences}
    \label{Table:H111k_link}
\end{table}

The full solution for the integral $H_{\{1,1,1\}} (m_1, m_2, m_3; m_1^2)$ is given by:
\begin{align}
    H_{\{1,1,1\}} (m_1, m_2, m_3; m_1^2)
    & = I_1  
    + \sum_{i=1}^3 I_2^{\alpha,i}
    + \sum_{i=1}^3 I_3^{\beta,i} 
    + \sum_{i=1}^b I_4^{c,i}
\end{align}
The appropriate values of $\alpha$, $\beta$ and $c$ depend on the particular values of $m_1$, $m_2$ and $m_3$ being considered. Recourse must be had of Table~\ref{Table:H111k_link} to find the appropriate region of convergence which satisfies $m_1$, $m_2$ and $m_3$, and to then use the corresponding values of $\alpha$, $\beta$ and $c$ to obtain the appropriate residues from Table~\ref{Table:H111k}. 

For example, with $m_1=0.495$, $m_2=0.1395$ and $m_3=2.547$, we find that the cone 1 ($C=1$) region of convergence is satisfied. Therefore, for these mass values, we have:
\begin{align}
    H_{\{1,1,1\}} (m_1=0.495, m_2=0.1395, m_3=2.547; s=0.495^2)
    & = I_1 + \sum_{i=1}^3 I_2^{L,i} + \sum_{i=1}^3 I_3^{R,i} + \sum_{i=1}^7 I_4^{1,i}
\end{align}

Due to cancellation of various residues, the full expressions simplify significantly. For instance, in the above example, after such cancellations, we get:
\begin{align}
    H_{\{1,1,1\}} (m_1=0.495, m_2=0.1395, m_3=2.547; s=0.495^2)
    & = I_1 + \sum_{i=1}^3 I_2^{L,i} + \sum_{i=4}^7 I_4^{1,i}
\end{align}

We demonstrate the equivalence of these results with those of Sec.~\ref{Sec:H111Results} by means of another example. We see that for cone 2 ($C=2$), we have:
\begin{align}
    H_{\{1,1,1\}} (m_1, m_2, m_3; m_1^2)
     = I_1 
    + \sum_{i=1}^3 I_2^{L,i}
    + \sum_{i=1}^3 I_3^{L,i}+ \sum_{i=1}^7 I_4^{2,i}
    \label{Eq:H111k_example2}
\end{align}
which gives the exact same expression as Eq.~\ref{SeriesH111}.
 
Indeed, by inspecting the various contributions, one can see that $I_1+I_2^{L,1}=0$, $I_3^{L,1}+I_4^{2,7}=0$, $I_2^{L,2}+I_4^{2,6}=0$ and $I_2^{L,3}+I_4^{2,5}=0$ so that Eq.~\ref{Eq:H111k_example2} in fact reduces to
\begin{align}
    H_{\{1,1,1\}} (m_1, m_2, m_3; m_1^2) = I_3^{1,2}+I_3^{1,3}+\sum_{i=1}^4 I_4^{2,i}
\end{align}
Now, if one labels the four terms of Eq.~\ref{SeriesH111} as $T_i, (i=1,...,4)$ then one can prove that $I_3^{L,2}+I_4^{2,3}=T_1$, $I_3^{L,3}+I_4^{2,4}=T_2$, $I_4^{2,2}=T_3$ and $I_4^{2,1}=T_4$. So that at the end one gets Eq.~\ref{SeriesH111}.

\begin{table}[]
\centering
\noindent
\begin{minipage}{0.47\textwidth}
\centering
\begin{tabular}{|c|l|}
\hline
\rowcolor{lightgray}
\rule{0pt}{15pt}{$I_2$ \bf LHS} & $(\alpha=L)$ \\
\hline
$0$ & $I_2^{L,1}$ \\ 
\hline 
$-m-\epsilon$ & $I_2^{L,2}$ \\ 
\hline
$1-m-2\epsilon$ & $I_2^{L,3}$ \\ 
\hline
\hline
\rowcolor{lightgray}
\rule{0pt}{15pt}{$I_3$ \bf LHS} & $(\alpha=L)$ \\
\hline
$0$ & $I_3^{L,1} = I_2^{L,1}$ \\ 
\hline 
$-m-\epsilon$ & $I_3^{L,2} = I_2^{L,2} |_{m_3 \rightarrow m_2}$ \\ 
\hline
$1-m-2\epsilon$ & $I_3^{L,3} = I_2^{L,3} |_{m_3 \rightarrow m_2}$ \\ 
\hline
\end{tabular}
\end{minipage}%
\hfill
\begin{minipage}{0.47\textwidth}
\centering
\begin{tabular}{|c|l|}
\hline
\rowcolor{lightgray}
\rule{0pt}{15pt}{$I_2$ \bf RHS} & $(\alpha=R)$ \\
\hline
$1+m$ & $I_2^{R,1}$ \\ 
\hline 
$1-\epsilon+m$ & $I_2^{R,2}$ \\ 
\hline 
$\tfrac{3}{2}-2\epsilon+m$ & $I_2^{R,3}$ \\ 
\hline 
\hline
\rowcolor{lightgray}
\rule{0pt}{15pt}{$I_3$ \bf RHS} & $(\alpha=R)$ \\
\hline
$1+m$ & $I_3^{R,1} = I_2^{R,1}$ \\ 
\hline 
$1-\epsilon+m$ & $I_3^{R,2} = I_2^{R,2} |_{m_3 \rightarrow m_2}$ \\ 
\hline 
$\tfrac{3}{2}-2\epsilon+m$ & $I_3^{R,3} = I_2^{R,3} |_{m_3 \rightarrow m_2}$ \\ 
\hline 
\end{tabular}
\end{minipage}

\vspace{1cm}

\centering
\noindent
\begin{minipage}{0.47\textwidth}
\centering
\begin{tabular}{|c|l|}
\hline
\rowcolor{lightgray}
\bf{Cone 1} $(C=1)$ & $c=7$ \\
\hline
$(0,1-\epsilon+m)$ & $I_4^{1,1}$ \\ 
\hline 
$(0,1+m)$ & $I_4^{1,2}$ \\ 
\hline 
$(0,\tfrac{3}{2}-2\epsilon+m)$ & $I_4^{1,3}$ \\ 
\hline 
$(-\epsilon-m-n,1-\epsilon+m)$ & $I_4^{1,4}$ \\ 
\hline 
$(-1-m-n,1+m-\epsilon)$ & $I_4^{1,5}$ \\ 
\hline 
$(-1-\epsilon-m-n,1+m)$  & $I_4^{1,6}$ \\ 
\hline 
$(-2\epsilon-m-n,1+m)$  & $I_4^{1,7}$ \\ 
\hline 
\hline
\rowcolor{lightgray}
\bf{Cone 2} $(C=2)$ & $c=7$ \\
\hline 
$(0,0)$ & $I_4^{2,1}$ \\ 
\hline 
$(-\epsilon-m,0)$ & $I_4^{2,2}$ \\ 
\hline
$(1-2\epsilon-m,0)$ & $I_4^{2,3}$ \\ 
\hline 
$(1-\epsilon+m,-\epsilon-m-n)$ & $I_4^{2,4} = I_4^{4,4} $ \\ 
\hline 
$(1-\epsilon+m,-1-m-n)$ & $I_4^{2,5} = I_4^{4,5} $ \\ 
\hline 
$(1+m,-1-\epsilon-m-n)$ & $I_4^{2,6} = I_4^{4,6}$ \\ 
\hline 
$(1+m,-2\epsilon-m-n)$ & $I_4^{2,7} = I_4^{4,7}$ \\ 
\hline
\hline
\rowcolor{lightgray}
\bf{Cone 3} $(C=3)$ & $c=9$ \\
\hline
$(0,\tfrac{3}{2}-2\epsilon+m)$ & $I_4^{3,1}$ \\ 
\hline 
$(\tfrac{1}{2}-\epsilon+m,1-\epsilon+n)$ & $I_4^{3,2}$ \\ 
\hline 
$(\tfrac{1}{2}-2\epsilon+m,1+n)$ & $I_4^{3,3}$ \\ 
\hline 
$(1-\epsilon+m,1-\epsilon+n)$ & $I_4^{3,4}$ \\ 
\hline 
$(1-\epsilon+m,1+n)$ & $I_4^{3,5}$ \\ 
\hline 
$(1+m,1+n)$ & $I_4^{3,6}$ \\ 
\hline
$(1+m,1-\epsilon+n)$ & $I_4^{3,7}$ \\ 
\hline 
$(-\tfrac{1}{2}-\epsilon-m,2-\epsilon+m+n)$ & $I_4^{3,8}$ \\ 
\hline
$(-\tfrac{1}{2}-2\epsilon-m,2+m+n)$ & $I_4^{3,9}$ \\ 
\hline
\end{tabular}
\end{minipage}%
\hfill
\begin{minipage}{0.47\textwidth}
\centering
\begin{tabular}{|c|l|}
\hline 
\rowcolor{lightgray}
\bf{Cone 4} $(C=4)$ & $c=7$ \\
\hline
$(1-\epsilon+m,0)$ & $I_4^{4,1} = I_4^{1,1}|_{m_2 \leftrightarrow m_3}$ \\ 
\hline 
$(1+m,0)$ & $I_4^{4,2} = I_4^{1,2}|_{m_2 \leftrightarrow m_3}$ \\ 
\hline 
$(\tfrac{3}{2}-2\epsilon+m,0)$ & $I_4^{4,3} = I_4^{1,3}|_{m_2 \leftrightarrow m_3}$ \\ 
\hline 
$(1-\epsilon+m,-\epsilon-m-n)$ & $I_4^{4,4} = I_4^{1,4}|_{m_2 \leftrightarrow m_3}$ \\ 
\hline 
$(1+m-\epsilon,-1-m-n)$ & $I_4^{4,5} = I_4^{1,5}|_{m_2 \leftrightarrow m_3}$ \\ 
\hline 
$(1+m,-1-\epsilon-m-n)$  & $I_4^{4,6} = I_4^{1,6}|_{m_2 \leftrightarrow m_3}$ \\ 
\hline 
$(1+m,-2\epsilon-m-n)$  & $I_4^{4,7} = I_4^{1,7}|_{m_2 \leftrightarrow m_3}$ \\ 
\hline 
\hline
\rowcolor{lightgray}
\bf{Cone 5} $(C=5)$ & $c=7$ \\
\hline
$(0,0)$ & $I_4^{5,1}= I_4^{2,1}$ \\ 
\hline
$(0,-\epsilon-m)$ & $I_4^{5,2} = I_4^{2,2}|_{m_2 \leftrightarrow m_3}$ \\ 
\hline 
$(0,1-2\epsilon-m)$ & $I_4^{5,3}= I_4^{2,3}|_{m_2 \leftrightarrow m_3}$ \\ 
\hline 
$(-\epsilon-m-n,1-\epsilon+m)$ & $I_4^{5,4} = I_4^{1,4}$ \\ 
\hline 
$(-1-m-n,1-\epsilon+m)$ & $I_4^{5,5} = I_4^{1,5}$ \\ 
\hline 
$(-1-\epsilon-m-n,1+m)$ & $I_4^{5,6} = I_4^{1,6}$ \\ 
\hline 
$(-2\epsilon-m-n,1+m)$ & $I_4^{5,7} = I_4^{1,7}$ \\ 
\hline
\hline
\rowcolor{lightgray}
\bf{Cone 6} $(C=6)$ & $c=9$ \\
\hline
$(\tfrac{3}{2}-2\epsilon+m,0)$ & $I_4^{6,1}= I_4^{3,1}|_{m_2 \leftrightarrow m_3} $ \\ 
\hline 
$(1-\epsilon+m,\tfrac{1}{2}-\epsilon+n)$ & $I_4^{6,2}= I_4^{3,2}|_{m_2 \leftrightarrow m_3}$ \\ 
\hline 
$(1+m,\tfrac{1}{2}-2\epsilon+n)$ & $I_4^{6,3}= I_4^{3,3}|_{m_2 \leftrightarrow m_3}$ \\ 
\hline 
$(1-\epsilon+m,1-\epsilon+n)$ & $I_4^{6,4}= I_4^{3,4}|_{m_2 \leftrightarrow m_3}$ \\ 
\hline 
$(1+m,1-\epsilon+n)$ & $I_4^{6,5}= I_4^{3,5}|_{m_2 \leftrightarrow m_3}$ \\ 
\hline 
$(1+m,1+n)$ & $I_4^{6,6}= I_4^{3,6}|_{m_2 \leftrightarrow m_3}$ \\ 
\hline
$(1-\epsilon+m,1+n)$ & $I_4^{6,7}= I_4^{3,7}|_{m_2 \leftrightarrow m_3}$ \\ 
\hline 
$(2-\epsilon+m+n,-\tfrac{1}{2}-\epsilon-n)$ & $I_4^{6,8}= I_4^{3,8}|_{m_2 \leftrightarrow m_3}$ \\ 
\hline
$(2+m+n,-\tfrac{1}{2}-2\epsilon-n)$ & $I_4^{6,9}= I_4^{3,9}|_{m_2 \leftrightarrow m_3}$ \\ 
\hline
\end{tabular}
\end{minipage}

    \caption{Singularities of $I_2$, $I_3$ and $I_4$. Expressions for the residues are given in App.~\ref{Sec:H111kResults}.}
    \label{Table:H111k}
\end{table}

 \subsection{The integral $H_{\{2,1,1\}} (m_1, m_2, m_3; m_1^2)$ \label{Sec:H211MB}}

\begin{align}
	H_{\{2,1,1\}} \left(m_1, m_2, m_3; m_1^2 \right) \equiv
	\int \frac{d^d k_1 d^d k_2}{\left[ (k_1-p)^2 - m_1^2 + i 0 \right]^2 \left[ k_2^2 - m_2^2 +i 0 \right] \left[ (k_1-k_2)^2 - m_3^2 + i 0 \right]}
\label{Eq:H211kaon}
\end{align}
is similar to that of $H_{\{1,1,1\}}\left(m_1, m_2, m_3; m_1^2 \right)$, and follows the exact same steps until Eq.~\ref{Eq:Ik2kaon} and its simplification thereafter. Analogous to Eq.~\ref{Eq:H111kaonMBmid}, we have:
\begin{align}
	H_{\{2,1,1\}}&\left(m_1, m_2, m_3; m_1^2 \right)
	= \int \frac{d^d k_1}{[(k_1-p)^2 - m_1^2 + i 0]^2} \Bigg\{ \frac{i \pi^{2-\epsilon}}{(-k_1^2)^\epsilon} \frac{\Gamma(\epsilon) \Gamma^2(1-\epsilon)}{\Gamma(2-2\epsilon)} \nonumber \\
	& + \frac{\Gamma(1-\epsilon)}{(-k_1^2)^{\epsilon}}  \frac{i \pi^{2-\epsilon}}{2\pi i} \int_{c'-i\infty}^{c'+i\infty} dz_1 \, \left( \frac{-m_3^2+i0}{k_1^2} \right)^{z_1} \frac{\Gamma(-z_1)  \Gamma(1-\epsilon-z_1) \Gamma(\epsilon+z_1)}{\Gamma(2-2\epsilon-z_1)} \nonumber \\
	& + \frac{\Gamma(1-\epsilon)}{(-k_1^2)^{\epsilon}}  \frac{i \pi^{2-\epsilon}}{2\pi i} \int_{d'-i\infty}^{d'+i\infty} dz_2 \, \left( \frac{-m_2^2+i0}{k_1^2} \right)^{z_2} \frac{\Gamma(-z_2)  \Gamma(1-\epsilon-z_2) \Gamma(\epsilon+z_2)}{\Gamma(2-2\epsilon-z_2)} \nonumber \\
	& + \frac{1}{(2\pi i)^2} \frac{i \pi^{2-\epsilon}}{(-k_1^2)^{\epsilon}} \int_{c'-i\infty}^{c'+i\infty} dz_1 \int_{d'-i\infty}^{d'+i\infty} dz_2 \left( \frac{-m_3^2+i0}{k_1^2} \right)^{z_1}  \left( \frac{-m_2^2+i0}{k_1^2} \right)^{z_2} \Gamma(-z_1)  \nonumber \\ 
	& \hspace{4.5cm} \times 
	\frac{\Gamma(-z_2) \Gamma(1-\epsilon-z_1) \Gamma(1-\epsilon-z_2) \Gamma(\epsilon+z_1+z_2)}
	{\Gamma(2-2\epsilon-z_1-z_2)} \Bigg\} \label{Eq:H211kaonMBmid}
\end{align}
We now integrate over the $k_1$ loop in each term. The $k_1$ integral in the second term of the above expression is:
\begin{align}
	\int \frac{d^d k_1}{[(k_1-p)^2 - m_1^2 + i 0]^2 [-k_1^2]^{\epsilon+z_1}} = 
	\frac{i \pi^{2-\epsilon}}{(m_1^2)^{2\epsilon+z_1}} \frac{\Gamma(2\epsilon+z_1) \Gamma(2-4\epsilon-2z_1)}{\Gamma(2-3\epsilon-z_1)}
\end{align}
provided that $\text{Re}(2\epsilon+z_1)>0$, $\text{Re}(2-4\epsilon-2z_1)>0$ and $\text{Re}(\epsilon+z_1)>0$, and where we make use of the kinematic invariant $p^2 = m_1^2$.
The result of the $k_1$ integral of the first term can be obtained from the above by setting $z_1=0$, and the one of the fourth term by setting $z_1=z_1+z_2$.

Combining all the factors, we get the following MB representation for  $H_{\{2,1,1\}} \left(m_1, m_2, m_3; m_1^2 \right)$ with straight contours, and in which we can set $\epsilon=0$ at the integrand level of each of the involved MB integrals:
\begin{align}
	H_{\{2,1,1\}} \left(m_1, m_2, m_3; m_1^2 \right) = J_1 + J_2 + J_3 + J_4
\label{Eq:H211C}
\end{align}
where:
\begin{align}
	J_1 \equiv -\frac{\pi^{4-2\epsilon}}{(m_1^2)^{2\epsilon}} \frac{\Gamma(\epsilon)\Gamma(2\epsilon)\Gamma^2(1-\epsilon)\Gamma(2-4\epsilon)}{\Gamma(2-2\epsilon)\Gamma(2-3\epsilon)},
	\label{Eq:J1}
\end{align}
\begin{align}
	J_2 \equiv &
	-\frac{\pi^{4-2\epsilon}}{(m_1^2)^{2\epsilon}}
	\frac{\Gamma(1-\epsilon)}{2\pi i}
	\int_{c'-i\infty}^{c'-i\infty} dz_1 \left( \frac{m_3^2-i0}{m_1^2} \right)^{z_1} \frac{\Gamma(-z_1) \Gamma(1-\epsilon-z_1) \Gamma(\epsilon+z_1) \Gamma(2\epsilon+z_1) \Gamma(2-4\epsilon-2z_1)}{\Gamma(2-2\epsilon-z_1) \Gamma(2-3\epsilon-z_1)}
	\label{Eq:J2}
\end{align}
with $c' = \text{Re}(z_1) \in ]0,1-2\epsilon[$
$J_3$ is equivalent to $J_2$ but with $m_3 \rightarrow m_2$, and 
\begin{multline}
	J_4 \equiv 
	-\frac{\pi^{4-2\epsilon}}{(m_1^2)^{2\epsilon}}
	\frac{1}{(2\pi i)^2} \int_{c'-i\infty}^{c'+i\infty} dz_1 \int_{d'-i\infty}^{d'+i\infty} dz_2 \left( \frac{m_3^2-i0}{m_1^2} \right)^{z_1} \left( \frac{m_2^2-i0}{m_1^2} \right)^{z_2} \Gamma(-z_1) \Gamma(-z_2)  \\
	 \times \frac{\Gamma(2\epsilon+z_1+z_2) \Gamma(\epsilon+z_1+z_2) \Gamma(1-\epsilon-z_1) \Gamma(1-\epsilon-z_2) \Gamma(2-4\epsilon-2z_1-2z_2)}{\Gamma(2-2\epsilon-z_1-z_2) \Gamma(2-3\epsilon-z_1-z_2)} 
  \label{Eq:J4}
\end{multline}
which holds if $c' = \text{Re}(z_1) \in ]0,1[$, $d' = \text{Re}(z_2) \in ]0,1[$, $\text{Re}(\epsilon+z_1+z_2)>0$, $\text{Re}(2\epsilon+z_1+z_2)>0$, $\text{Re}(1-\epsilon-z_1)>0$, $\text{Re}(1-\epsilon-z_2)>0$ and $\text{Re}(1-2\epsilon-z_1-z_2)>0$.

\subsubsection*{Solving the integral}

\begin{figure}[hbtp]
\centering
\includegraphics[scale=0.7]{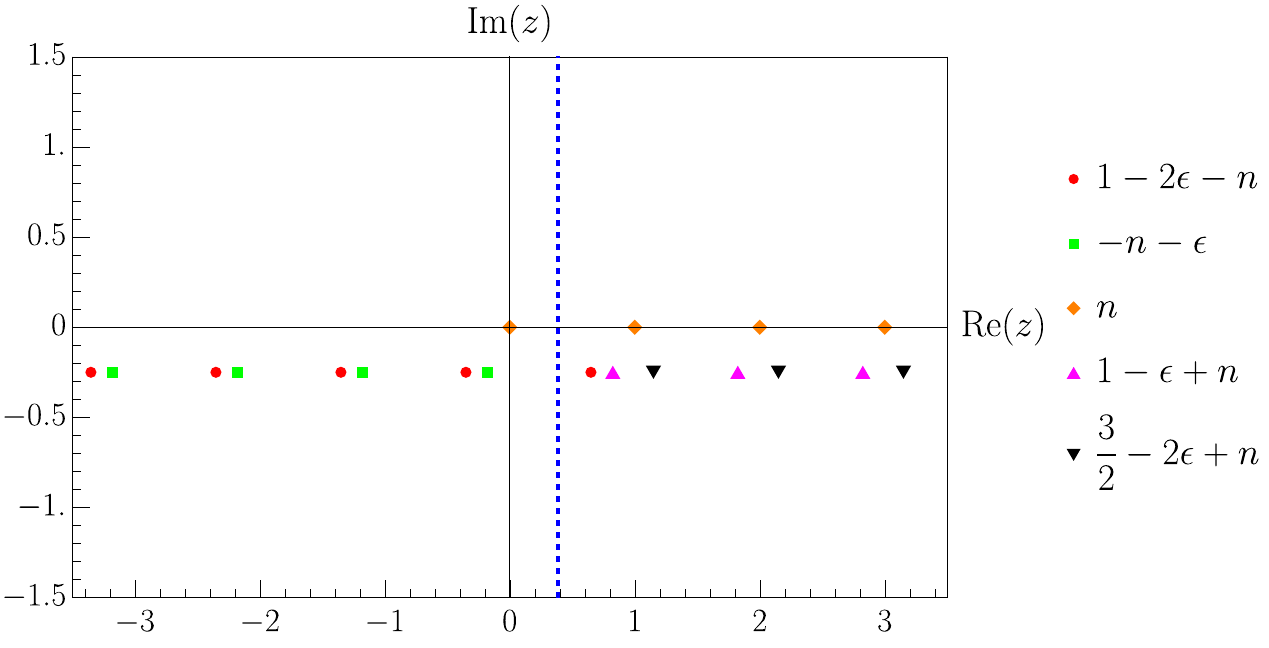}
\caption{Singularity structure of $J_2$ and $J_3$}
\label{Fig:J2Sing}
\end{figure}

\begin{figure}[h]
\center
\begin{tabular}{lll}
\includegraphics[scale=0.35]{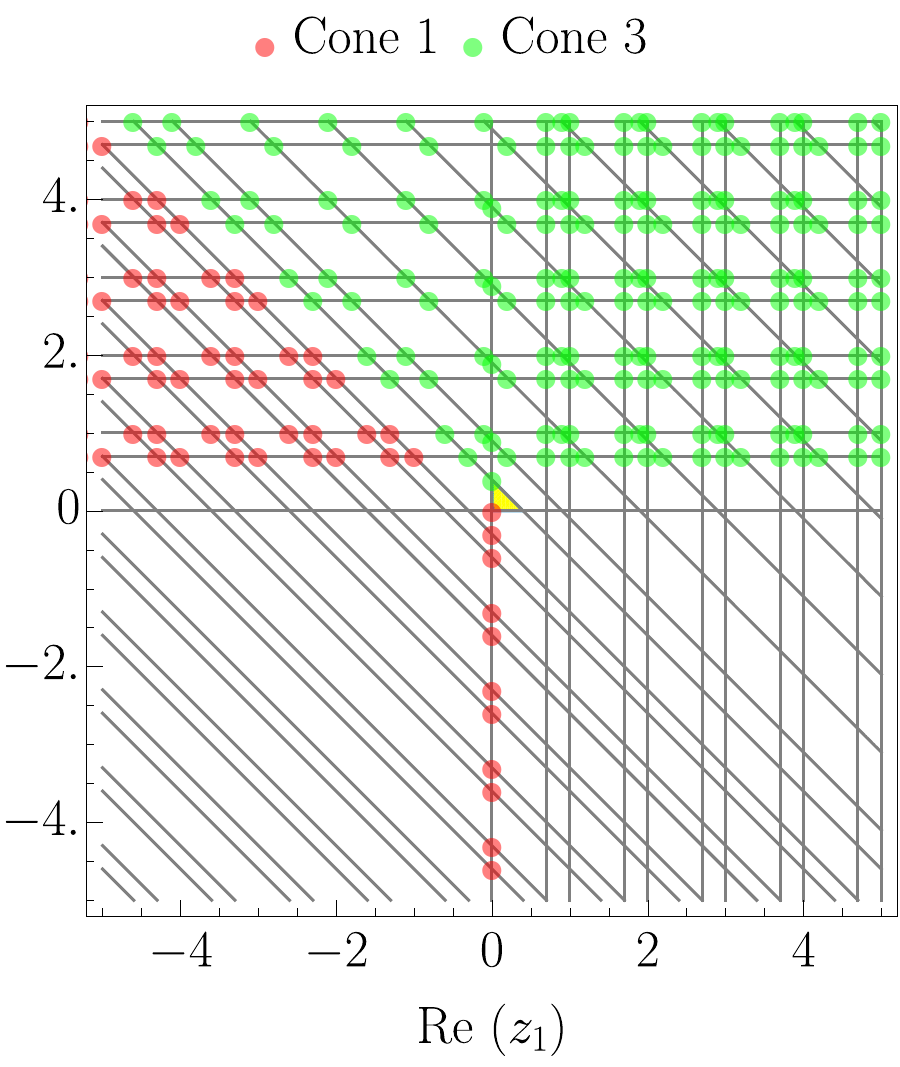}
&\hspace{-0.4cm}
\includegraphics[scale=0.35]{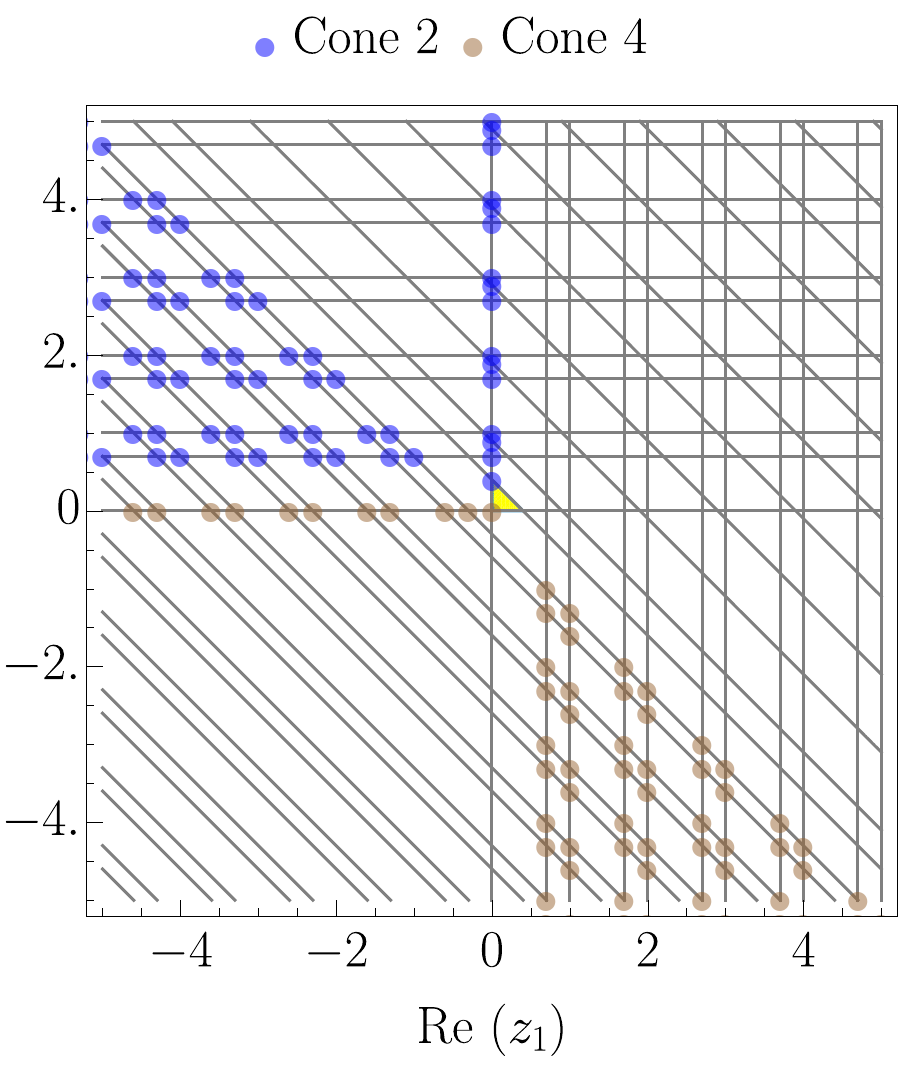}
&\hspace{-0.4cm}
\includegraphics[scale=0.35]{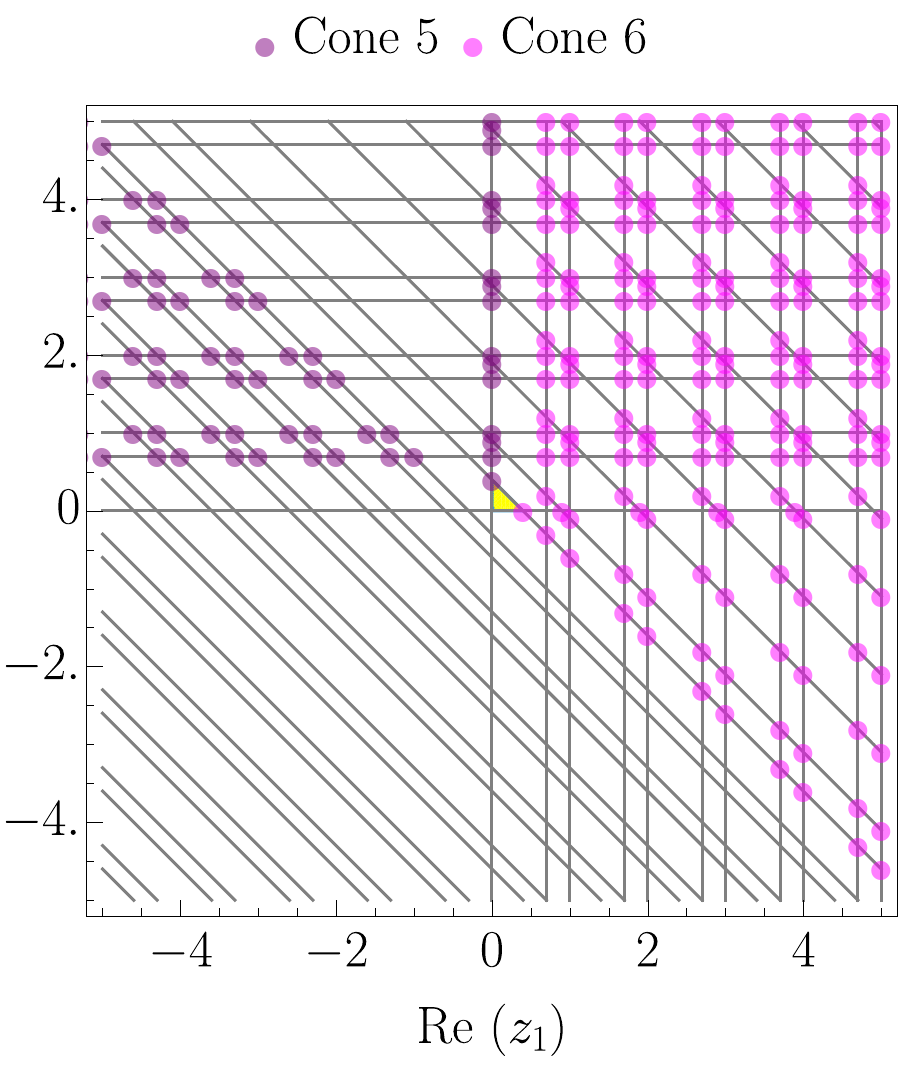}
\end{tabular}
\caption{The contributing singularities of the various cones of $J_4$}
\label{Fig:J4sing}
\end{figure}

\begin{figure}[h]
\center
\includegraphics[scale=0.7]{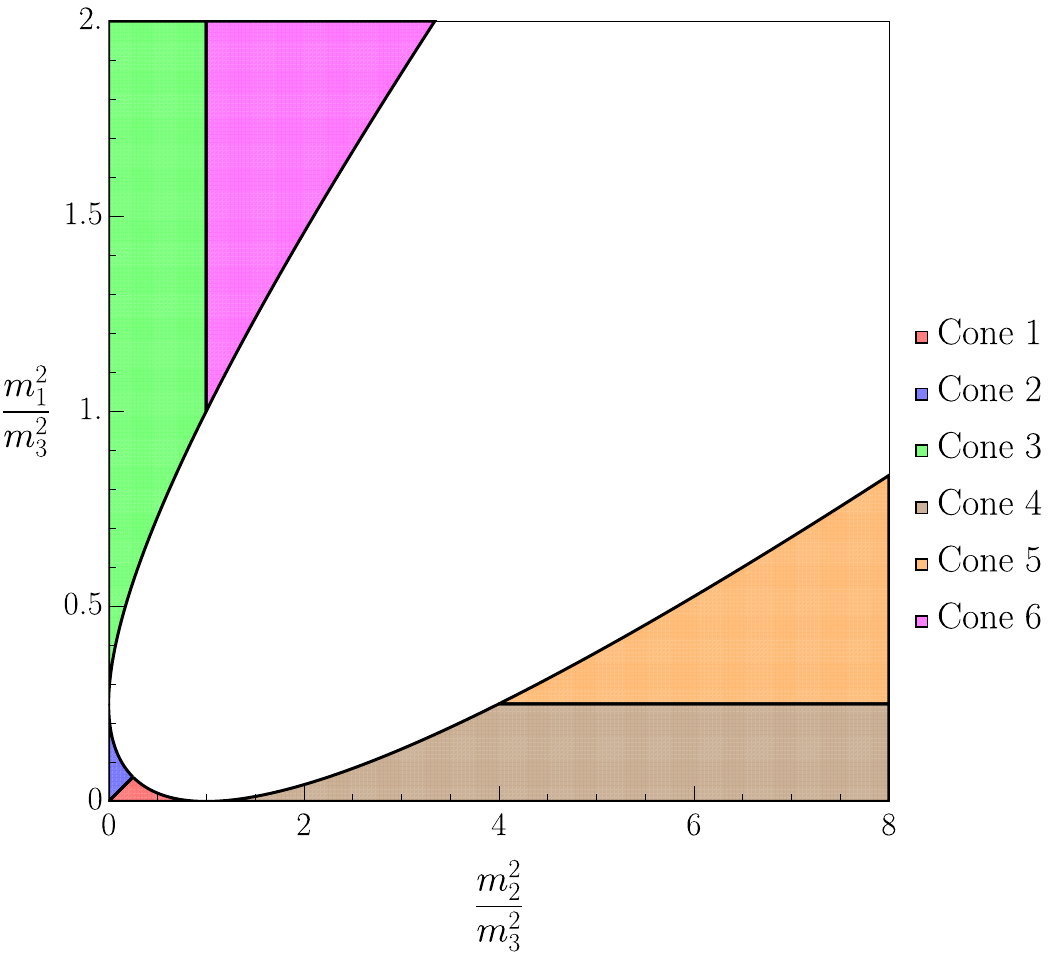}
\caption{The region of convergence of the various cones of $J_4$}
\label{Fig:J4_RoC}
\end{figure}

We solve the MB representation in a manner similar to the previous section.
The singularity structure of $J_2$ and $J_3$ is shown in Fig.~\ref{Fig:J2Sing}, and that of $J_4$ in Fig.~\ref{Fig:J4sing}. The regions of convergence corresponding to each cone of $J_4$ is shown in Fig.~\ref{Fig:J4_RoC}. The singularities and their corresponding residues are given in Table~\ref{Table:H211J4} and App.~\ref{Sec:H211Results}.

The full solution for the integral $H_{\{2,1,1\}} (m_1^2, m_2^2, m_3^2; m_1^2)$ is given by:
\begin{align}
    H_{\{2,1,1\}} (m_1, m_2, m_3; m_1^2)
    & = J_1  
    + \sum_{i=1}^a J_2^{\alpha,i}
    + \sum_{i=1}^b J_3^{\beta,i} 
    + \sum_{i=1}^c J_4^{C,i}
\end{align}
where the values of $\alpha$, $\beta$, $a$, $b$ and $c$ can be determined using Table~\ref{Table:H211_link} by finding the appropriate region of convergence which satisfies the particular values of $m_1$, $m_2$ and $m_3$ being considered.

\begin{table}
    \centering
    \begin{tabular}{|l|c|c||c|c||c|c|}
    \hline
    \rowcolor{lightgray}
    \bf{Region of Convergence} & $C$ & $c$ & $\alpha$ & a & $\beta$ & $b$\\
    \hline
    $ (2m_1 + m_2 < m_3 ) \wedge (4m_1^2 < m_2^2) $ & 1 & 7 & L & 3 & L & 3 \\
    \hline
    $ (2m_1 + m_2 < m_3 ) \wedge (4m_1^2 > m_2^2) $ & 2 & 8 & L & 3 & R & 4 \\
    \hline
    $ (m_2 + m_3 < 2 m_1 ) \wedge (m_2^2 < m_3^2) $ & 3 & 12 & R & 4 & R & 4 \\
    \hline
    $ (m_3 + 2m_1 < m_2 ) \wedge (4m_1^2 < m_3^2) $ & 4 & 7 & L & 3 & L & 3 \\
    \hline
    $ (2m_1 + m_3 < m_2 ) \wedge (4m_1^2 > m_3^2)  $ & 5 & 8 & R & 4 & L & 3 \\
    \hline
    $ (m_2 + m_3 < 2 m_1 ) \wedge (m_2^2 > m_3^2) $ & 6 & 12 & R & 4 & R & 4 \\
    \hline
    \end{tabular}
    \caption{Table of correspondences}
    \label{Table:H211_link}
\end{table}

\begin{table}[]
\centering
\noindent
\begin{minipage}{0.47\textwidth}
\centering
\begin{tabular}{|c|l|}
\hline

\rowcolor{lightgray}
\rule{0pt}{15pt}{$J_2$ \bf LHS} & $(\alpha=L, a=3)$ \\
\hline
$0$ & $J_2^{L,1} $ \\ 
\hline 
$-m-\epsilon$ & $J_2^{L,2}$ \\ 
\hline
$-m-2\epsilon$ & $J_2^{L,3}$ \\ 
\hline
\hline

\rowcolor{lightgray}
\rule{0pt}{15pt}{$J_2$ \bf RHS} & $(\alpha=R, a=4)$ \\
\hline
$1-2\epsilon$ & $J_2^{R,1}$ \\ 
\hline 
$m+1$ & $J_2^{R,2}$ \\ 
\hline
$m-\epsilon+1$ & $J_2^{R,3}$ \\ 
\hline
$m-2\epsilon+\tfrac{3}{2}$ & $J_2^{R,4}$ \\ 
\hline

\end{tabular}
\end{minipage}%
\hfill
\begin{minipage}{0.47\textwidth}
\centering
\begin{tabular}{|c|l|}
\hline

\rowcolor{lightgray}
\rule{0pt}{15pt}{$J_3$ \bf LHS} & $(\alpha=L, a=3)$ \\
\hline
$0$ & $J_3^{L,1} = J_2^{L,1} |_{m_3 \rightarrow m_2} $ \\ 
\hline 
$-m-\epsilon$ & $J_3^{L,2} = J_2^{L,2} |_{m_3 \rightarrow m_2} $ \\ 
\hline
$-m-2\epsilon$ & $J_3^{L,3} = J_2^{L,3} |_{m_3 \rightarrow m_2} $ \\ 
\hline
\hline

\rowcolor{lightgray}
\rule{0pt}{15pt}{$J_3$ \bf RHS} & $(\alpha=R, a=4)$ \\
\hline
$1-2\epsilon$ & $J_3^{R,1} = J_2^{R,1} |_{m_3 \rightarrow m_2}  $ \\ 
\hline 
$m+1$ & $J_3^{R,2} = J_2^{R,2} |_{m_3 \rightarrow m_2}$ \\ 
\hline
$m-\epsilon+1$ & $J_3^{R,3} = J_2^{R,3} |_{m_3 \rightarrow m_2}$ \\ 
\hline
$m-2\epsilon+\tfrac{3}{2}$ & $J_2^{R,4} = J_3^{R,4} |_{m_3 \rightarrow m_2}$ \\ 
\hline

\end{tabular}
\end{minipage}

\vspace{1cm}

    \centering
    \noindent
\begin{minipage}{0.47\textwidth}
\begin{tabular}{|c|l|}
\hline
\rowcolor{lightgray}
\bf{Cone 1} $(C=1)$ & $c=7$ \\
\hline
$(0,0)$ & $J_4^{1,1}$ \\ 
\hline 
$(0,-\epsilon-m)$ & $J_4^{1,2}$ \\ 
\hline 
 $(0,-2\epsilon-m)$ & $J_4^{1,3}$ \\ 
\hline 
$(-1-m-n,1-\epsilon+m)$ & $J_4^{1,4}$ \\ 
\hline 
$(-1-\epsilon-m-n,1-\epsilon+m)$ & $J_4^{1,5}$ \\ 
\hline 
$(-1-\epsilon-m-n,1+m)$ & $J_4^{1,6}$ \\ 
\hline 
$(-1-2\epsilon-m-n,1+m)$ & $J_4^{1,7}$ \\ 
\hline 
\hline

\rowcolor{lightgray}
\bf{Cone 2} $(C=2)$ & $c=8$ \\
\hline
$(0,1-2\epsilon)$ & $J_4^{2,1} = J_4^{3,1}$ \\ 
\hline 
$(0,1-\epsilon+m)$ & $J_4^{2,2}$ \\ 
\hline 
 $(0,1+m)$ & $J_4^{2,3}$ \\ 
\hline 
$(0,\tfrac{3}{2}-2\epsilon+m)$ & $J_4^{2,4} = J_4^{3,4}$ \\ 
\hline
$(-1-m-n,1-\epsilon+m)$ & $J_4^{2,5} = J_4^{1,4} $ \\ 
\hline
$(-1-\epsilon-m-n,1-\epsilon+m)$ & $J_4^{2,6} = J_4^{1,5}$ \\ 
\hline 
$(-1-\epsilon-m-n,1+m)$ & $J_4^{2,7} = J_4^{1,6}$ \\ 
\hline 
$(-1-2\epsilon-m-n,1+m)$ & $J_4^{2,8} = J_4^{1,7}$ \\ 
\hline 
\hline

\rowcolor{lightgray}
\bf{Cone 3} $(C=3)$ & $c=12$ \\
\hline
$(0,1-2\epsilon)$ & $J_4^{3,1}$ \\ 
\hline 
$(-\epsilon-m,1-\epsilon+m)$ & $J_4^{3,2}$ \\ 
\hline 
 $(-2\epsilon-m,1+m)$ & $J_4^{3,3}$ \\ 
\hline 
$(0,\tfrac{3}{2}-2\epsilon+m)$ & $J_4^{3,4}$ \\ 
\hline 
$(\tfrac{1}{2}-\epsilon+m,1-\epsilon+n)$ & $J_4^{3,5}$ \\ 
\hline 
$(\tfrac{1}{2}-2\epsilon+m,1+n)$ & $J_4^{3,6}$ \\ 
\hline 
$(1-\epsilon+m,1-\epsilon+n)$ & $J_4^{3,7}$ \\ 
\hline 
$(1-\epsilon+m,1+n)$ & $J_4^{3,8}$ \\ 
\hline 
$(1+m,1+n)$ & $J_4^{3,9}$ \\ 
\hline 
$(1+m,1-\epsilon+n)$ & $J_4^{3,10}$ \\ 
\hline 
$(-\tfrac{1}{2}-\epsilon-m,2-\epsilon+m+n)$ & $J_4^{3,11}$ \\ 
\hline 
$(-\tfrac{1}{2}-2\epsilon-m,2+m+n)$ & $J_4^{3,12}$ \\ 
\hline

\end{tabular}
\end{minipage}%
\hfill
\begin{minipage}{0.47\textwidth}
\centering
\begin{tabular}{|c|l|}
\hline

\rowcolor{lightgray}
\bf{Cone 4} $(C=4)$ & $c=7$ \\
\hline
$(0,0)$ & $J_4^{4,1} = J_4^{1,1} |_{m_2 \leftrightarrow m_3}$ \\ 
\hline 
$(-\epsilon-m,0)$ & $J_4^{4,2} = J_4^{1,2} |_{m_2 \leftrightarrow m_3}$ \\ 
\hline 
 $(-2\epsilon-m,0)$ & $J_4^{4,3} = J_4^{1,3} |_{m_2 \leftrightarrow m_3}$ \\ 
\hline 
$(1-\epsilon+m,-1-m-n)$ & $J_4^{4,4} = J_4^{1,4} |_{m_2 \leftrightarrow m_3}$ \\ 
\hline 
$(1-\epsilon+m,-1-\epsilon-m-n)$ & $J_4^{4,5} = J_4^{1,5} |_{m_2 \leftrightarrow m_3}$ \\ 
\hline 
$(1+m,-1-\epsilon-m-n)$ & $J_4^{4,6} = J_4^{1,6} |_{m_2 \leftrightarrow m_3}$ \\ 
\hline 
$(1+m,-1-2\epsilon-m-n)$ & $J_4^{4,7} = J_4^{1,7} |_{m_2 \leftrightarrow m_3}$ \\ 
\hline 
\hline

\rowcolor{lightgray}
\bf{Cone 5} $(C=5)$ & $c=10$ \\
\hline
$(1-2\epsilon,0)$ & $J_4^{5,1} = J_4^{3,1} |_{m_2 \leftrightarrow m_3}$  \\ 
\hline 
$(1-\epsilon+m,0)$ & $J_4^{5,2} = J_4^{2,2} |_{m_2 \leftrightarrow m_3}$ \\ 
\hline 
 $(1+m,0)$ & $J_4^{5,3} = J_4^{2,1} |_{m_2 \leftrightarrow m_3}$ \\ 
\hline 
$(\tfrac{3}{2}-2\epsilon+m,0)$ & $J_4^{5,4} = J_4^{3,4} |_{m_2 \leftrightarrow m_3}$ \\ 
\hline
$(1-\epsilon+m,-1-m-n)$ & $J_4^{5,5} = J_4^{1,4} |_{m_2 \leftrightarrow m_3}$ \\ 
\hline
$(1-\epsilon+m,-1-\epsilon-m-n)$ & $J_4^{5,6} = J_4^{2,1} |_{m_2 \leftrightarrow m_3}$  \\ 
\hline 
$(1+m,-1-\epsilon-m-n)$ & $J_4^{5,7} = J_4^{2,1} |_{m_2 \leftrightarrow m_3}$ \\ 
\hline 
$(1+m,-1-2\epsilon-m-n)$ & $J_4^{5,8} = J_4^{2,1} |_{m_2 \leftrightarrow m_3}$ \\ 
\hline 
\hline

\rowcolor{lightgray}
\bf{Cone 6} $(C=6)$ & $c=12$ \\
\hline
$(0,1-2\epsilon)$ & $J_4^{2,1} = J_4^{6,1} |_{m_2 \leftrightarrow m_3}$ \\ 
\hline 
$(-\epsilon-m,1-\epsilon+m)$ & $J_4^{2,2} = J_4^{6,2} |_{m_2 \leftrightarrow m_3}$ \\ 
\hline 
 $(-2\epsilon-m,1+m)$ & $J_4^{2,3} = J_4^{6,3} |_{m_2 \leftrightarrow m_3}$ \\ 
\hline 
$(0,\tfrac{3}{2}-2\epsilon+m)$ & $J_4^{2,4} = J_4^{6,4} |_{m_2 \leftrightarrow m_3}$ \\ 
\hline 
$(\tfrac{1}{2}-\epsilon+m,1-\epsilon+n)$ & $J_4^{6,5} = J_4^{3,5} |_{m_2 \leftrightarrow m_3}$ \\ 
\hline 
$(\tfrac{1}{2}-2\epsilon+m,1+n)$ & $J_4^{6,6} = J_4^{3,6} |_{m_2 \leftrightarrow m_3}$ \\ 
\hline 
$(1-\epsilon+m,1-\epsilon+n)$ & $J_4^{6,7} = J_4^{3,7} |_{m_2 \leftrightarrow m_3}$ \\ 
\hline 
$(1-\epsilon+m,1+n)$ & $J_4^{6,8} = J_4^{3,8} |_{m_2 \leftrightarrow m_3}$ \\ 
\hline 
$(1+m,1+n)$ & $J_4^{6,9} = J_4^{6,9} |_{m_2 \leftrightarrow m_3}$ \\ 
\hline 
$(1+m,1-\epsilon+n)$ & $J_4^{6,10} = J_4^{6,10} |_{m_2 \leftrightarrow m_3}$ \\ 
\hline 
$(-\tfrac{1}{2}-\epsilon-m,2-\epsilon+m+n)$ & $J_4^{6,11} = J_4^{3,11} |_{m_2 \leftrightarrow m_3}$ \\ 
\hline 
$(-\tfrac{1}{2}-2\epsilon-m,2+m+n)$ & $J_4^{6,12} = J_4^{3,12} |_{m_2 \leftrightarrow m_3}$  \\ 
\hline

\end{tabular}
\end{minipage}
    \caption{Cones, regions of convergence, and singularities of $J_2$, $J_3$ and $J_4$. Expressions for the residues can be found in App.~\ref{Sec:H211Results}.}
    \label{Table:H211J4}
\end{table}

\subsection{The integral $H_{\{1,1,1\}} \left(m_1, m_2, m_2; m_3^2 \right)$ \label{Sec:H111eMB}}

We now tackle the last of the three independent three mass sunset master integrals, namely the off-shell configuration:
\begin{align}
	 H_{\{1,1,1\}}(m_1, m_2, m_2; m_3^2) \equiv \int  \frac{d^dk_1 d^dk_2}{\left[ (k_1-p)^2 - m_1^2 + i 0 \right] \left[ k_2^2 - m_2^2 +i 0 \right] \left[ (k_1-k_2)^2 - m_2^2 + i 0 \right]}\ \bigg|_{p^2=m_3^2}
\label{Eq:H111_1}
\end{align}

We begin by considering the $k_2$ loop first, and applying the Feynman parameterization on the two corresponding propagators:
\begin{align}
	I_{k_2}\equiv \int \frac{d^dk_2}{\left[ k_2^2 - m_2^2 +i 0 \right] \left[ (k_1-k_2)^2 - m_2^2 + i 0 \right]} =i\pi^{2-\epsilon}\Gamma(\epsilon)\int_0^1 dx \frac{1}{\left[m_2^2-i0-x(1-x)k_1^2\right]^\epsilon}\ .
	\label{Eq:H111_2}
\end{align}
The last equality holds only if $2-\text{Re}(d/2)>0$, that is if Re$(\epsilon)>0$. 

Let us rewrite Eq.~\ref{Eq:H111_2} as
\begin{align}
	I_{k_2} =\frac{i\pi^{2-\epsilon}\Gamma(\epsilon)}{\left(m_2^2-i0\right)^\epsilon}\int_0^1 dx \frac{1}{\left[1-\frac{x(1-x)k_1^2}{m_2^2-i0}\right]^\epsilon}\ .
\label{Eq:subloop}
\end{align}
Since Re$(\epsilon)>0$ the first condition to apply Eq.~\ref{Eq:MBformula} in Eq.~\ref{Eq:subloop} is fulfilled and, thanks to the $i0$, the second condition is also always satisfied, therefore we may safely write
\begin{align}
	I_{k_2} =\frac{i\pi^{2-\epsilon}}{\left(m_2^2-i0\right)^\epsilon}\int_0^1 dx\  \frac{1}{2i\pi}\int_{c-i\infty}^{c+i\infty} dz_1 \left(-\frac{x(1-x)k_1^2}{m_2^2-i0}\right)^{z_1} \Gamma(-z_1)\Gamma(\epsilon+z_1)\ ,
	\label{Eq:H111_4}
\end{align}
where the straight contour parallel to the imaginary axis is given by $c=\text{Re}(z_1)\in]-\text{Re}(\epsilon),0[$.The Feynman parameter integral may be computed if $\text{Re}(1+z_1)>0$ (this is the beta function convergence condition) which is checked for any small value of $\epsilon$. 

We then finish with
\begin{align}
	I_{k_2} =\frac{i\pi^{2-\epsilon}}{\left(m_2^2-i0\right)^\epsilon} \frac{1}{2i\pi}\int_{c-i\infty}^{c+i\infty} dz_1 \left(-\frac{k_1^2}{m_2^2-i0}\right)^{z_1}\frac{\Gamma(1+z_1)^2\Gamma(-z_1)\Gamma(\epsilon+z_1)}{\Gamma(2+2z_1)}\ ,
	\label{Eq:H111_5}
\end{align}
with $\text{Re}(z_1)\in]-\text{Re}(\epsilon),0[$ and $\text{Re}(1+z_1)>0$. These constraints, obtained during the intermediate steps of the calculation, are in fact nothing but the requirement that the real parts of the arguments of all gamma functions in the numerator of Eq.~\ref{Eq:H111_5} are positive. Note that since we are interested in small values of $\epsilon$ the second constraint disappears, because it is weaker than the first one. Moreover, one may simplify Eq.~\ref{Eq:H111_5}, using the duplication formula for the gamma function and get
\begin{align}
	I_{k_2} =\frac{i\pi^{2-\epsilon}}{\left(m_2^2-i0\right)^\epsilon}\frac{\sqrt{\pi}}{2} \frac{1}{2i\pi}\int_{c-i\infty}^{c+i\infty} dz_1 \left(-\frac{k_1^2}{4(m_2^2-i0)}\right)^{z_1} \frac{\Gamma(1+z_1)\Gamma(-z_1)\Gamma(\epsilon+z_1)}{\Gamma\left(\frac{3}{2}+z_1
	\right)}\ .
	\label{Eq:H111_6}
\end{align}

Inserting this in Eq.~\ref{Eq:H111_1}, we may now compute the $k_1$ loop integral which looks like
\begin{align}
	I_{k_1} \equiv\int \frac{d^dk_1}{\left[ (k_1-p)^2 - m_1^2 + i 0 \right] \left(k_1^2\right)^{-z_1}}\ .
	\label{Eq:H111_7}
\end{align}
It is clear that this integral converges for $\text{Re}(1-z_1-d/2)>0$, that is for $\text{Re}(-1-z_1+\epsilon)>0$. From the constraint $\text{Re}(z_1)\in]-\text{Re}(\epsilon),0[$ above, we see that this cannot be satisfied for small values of $\epsilon$.
We need more negative values for $\text{Re}(z_1)$. This may be achieved by returning to the point of the calculation where Eq.~\ref{Eq:MBformula} has been introduced, which is the first place where a constraint on $\text{Re}(z_1)$ appeared, and by computing the first residue to the left. In our case, this is equivalent to going back to Eq.~\ref{Eq:H111_6} because the $I_{k_2}$ integral is a simple 1-fold MB integral. In more complicated situations where the subloop could be an $N$-fold MB integral with $N>2$, it would be easier to be back to the step(s) in the calculations where the fundamental MB representation Eq.~\ref{Eq:MBformula} has been introduced, because there, one is dealing with a 1-fold MB integral.

We therefore rewrite Eq.~\ref{Eq:H111_6} as
\begin{align}
	I_{k_2} & =i\pi^{2-\epsilon}\frac{\sqrt{\pi}}{2}\frac{4^\epsilon\Gamma(1-\epsilon)\Gamma(\epsilon)}{\left(-k_1^2\right)^\epsilon\Gamma\left(\frac{3}{2}-\epsilon\right)} \nonumber \\
	& +\frac{i\pi^{2-\epsilon}}{\left(m_2^2-i0\right)^\epsilon}\frac{\sqrt{\pi}}{2} \frac{1}{2i\pi}\int_{c'-i\infty}^{c'+i\infty} dz_1 \left(-\frac{k_1^2}{4(m_2^2-i0)}\right)^{z_1} \frac{\Gamma(1+z_1)\Gamma(-z_1)\Gamma(\epsilon+z_1)}{\Gamma\left(\frac{3}{2}+z_1 \right)}\ ,
	\label{Eq:H111_8}
\end{align}
where now $c'\in]-1,-\text{Re}(\epsilon)[$.

Injecting Eq.~\ref{Eq:H111_8} in Eq.~\ref{Eq:H111_1}, we see that the following two $k_1$ integrals will have to be evaluated:
\begin{align}
	I_{k_1}' \equiv\int \frac{d^dk_1}{\left[ (k_1-m_3)^2 - m_1^2 + i 0 \right] \left(k_1^2\right)^{\epsilon}} 
	\label{Eq:H111_9}
\end{align}
and
\begin{align}
	I_{k_1}'' \equiv\int d^dk_1 \frac{1}{\left[ (k_1-m_3)^2 - m_1^2 + i 0 \right] \left(k_1^2\right)^{-z_1}}
	\label{Eq:H111_9_prime}
\end{align}
where the second one has the same form as Eq.~\ref{Eq:H111_7}, but now with $\text{Re}(z_1)\in]-1,-\text{Re}(\epsilon)[$ instead of $\text{Re}(z_1)\in]-\text{Re}(\epsilon),0[$.

Let us first consider Eq.~\ref{Eq:H111_9}. This integral does not converge for small values of $\epsilon$. In fact, it converge only for $\text{Re}(-1+2\epsilon)>0$. If, nevertheless, we try to give the result of this integral as the analytic continuation, when $\epsilon$ goes to an infinitesimally small value, of the expression obtained for a chosen value of $\epsilon$ that satisfies this constraint, we will get a result on which we will want to apply Eq.~\ref{Eq:MBformula} in order to get an expansion in powers of $m_3^2/m_1^2$. This will not be possible because the first condition of Eq.~\ref{Eq:MBformula} will not be fulfilled. Therefore, the simplest alternative choice is to first apply the MB representation of Eq.~\ref{Eq:MBformula} on the massive propagator of Eq.~\ref{Eq:H111_9} and only then to perform the loop integral. Thanks to the presence of the $i0$ in the propagator, the second condition for applying Eq.~\ref{Eq:MBformula} is always satisfied, and we get
\begin{align}
	I_{k_1}' =-\frac{1}{m_1^2-i0} \int \frac{d^dk_1}{\left(k_1^2\right)^\epsilon}\ \frac{1}{2i\pi}\int_{d-i\infty}^{d+i\infty} dz_2 \left(-\frac{(k_1-m_3)^2}{m_1^2-i0}\right)^{z_2} \Gamma(-z_2) \Gamma(1+z_2)
	\label{Eq:H111_10}
\end{align}
where $\text{Re}(z_2)\in]-1,0[$. The convergence condition of the $k_1$ integral is $\text{Re}(-z_2+\epsilon-d/2)>0$, that is $\text{Re}(-2-z_2+2\epsilon)>0$. For small values of $\epsilon$ it is not possible to satisfy this constraint for the range of allowed $z_2$ values. A shift of the MB integral is therefore necessary. We then obtain
\begin{align}
	I_{k_1}' =\int \frac{d^dk_1}{(k_1-m_3)^2\left(k_1^2\right)^\epsilon} -\frac{1}{m_1^2-i0}\int \frac{d^dk_1}{\left(k_1^2\right)^\epsilon} \frac{1}{2i\pi}\int_{d'-i\infty}^{d'+i\infty} dz_2 \left(-\frac{(k_1-m_3)^2}{m_1^2-i0)}\right)^{z_2} \Gamma(-z_2)\Gamma(1+z_2)
	\label{Eq:H111_11}
\end{align}
where $d'\in]-2,-1[$.
The first term still does not converge for small values of $\epsilon$ because the corresponding convergence constraint is the same as the one of Eq.~\ref{Eq:H111_9}. However here we may safely compute the contribution of this term from an analytic continuation of the result obtained for a finite value of $\epsilon$ satisfying the constraint. Indeed, there is no dependency in the MB integration variables and we will not apply the MB representation Eq.~\ref{Eq:MBformula} to this term (this means that we cannot meet the problem mentioned for Eq.~\ref{Eq:H111_9}). As for the second term of Eq.~\ref{Eq:H111_11}, we know that the $k_1$ integral converges for $\text{Re}(-2-z_2+2\epsilon)>0$. But since now $d'=\text{Re}(z_2)\in]-2,-1[$, this constraint may be satisfied for infinitesimally small values of $\epsilon$.

Therefore, we obtain
\begin{align}
	I_{k_1}' &=
	(-1)^{-\epsilon}i\pi^{2-\epsilon}(m_3^2)^{1-2\epsilon}\frac{\Gamma(-1+2\epsilon)\Gamma(2-2\epsilon)\Gamma(1-\epsilon)}{\Gamma(\epsilon)\Gamma(3-3\epsilon)}
    -\frac{i\pi^{2-\epsilon} (m_3^2)^{2-2\epsilon}(-1)^{-\epsilon}}{m_1^2-i0}\frac{\Gamma(2-2\epsilon)}{\Gamma(\epsilon)} \nonumber \\
    & \qquad 
    \times \frac{1}{2i\pi}\int_{d'-i\infty}^{d'+i\infty} z_2 \left(-\frac{z_2^2}{m_1^2-i0}\right)^{z_2}
	\frac{\Gamma(1+z_2)\Gamma(-2-z_2+2\epsilon)\Gamma(2+z_2-\epsilon)}{\Gamma(4-3\epsilon+z_2)}
\label{Eq:H111_12}
\end{align}
The first equality holds if $\text{Re}(1-\epsilon)>0$ and, as we said just above, for $\text{Re}(-2-z_2+2\epsilon)>0$ while the second equality holds for $\text{Re}(2-2\epsilon)>0$ and $\text{Re}(2+z_2-\epsilon)>0$. It is therefore possible to find some values for $z_2$ in the allowed range $d'=\text{Re}(z_2)\in]-2,-1[$ so that all constraints are satisfied for infinitesimally small values of $\epsilon$.

Now, to finish with the derivation of our MB representation, we must consider the second term of Eq.~\ref{Eq:H111_8} which gave birth to the $k_1$ integral Eq.~\ref{Eq:H111_9_prime}, that is
\begin{align}
	I_{k_1}'' =\int \frac{d^dk_1}{\left[ (k_1-m_3)^2 - m_1^2 + i 0 \right] \left(k_1^2\right)^{-z_1}}\ 
	\label{Eq:H111_13}
\end{align}
with $\text{Re}(z_1)\in]-1,-\text{Re}(\epsilon)[$. The corresponding convergence condition $\text{Re}(-1-z_1+\epsilon)>0$ may thus be satisfied for infinitesimally small values of $\epsilon$ and we have
\begin{align}
	I_{k_1}'' = (-1)^{1-z_1}i\pi^{2-\epsilon}\frac{\Gamma(-1-z_1+\epsilon)}{\Gamma(-z_1)}\int_0^1 dx \frac{(1-x)^{-z_1-1}}{\left[(m_1^2-i0-(1-x)m_3^2)x\right]^{-1-z_1+\epsilon}}
	\label{Eq:H111_14}
\end{align}
Now we may apply the MB representation Eq.~\ref{Eq:MBformula} because the power of the denominator in the integrand of Eq.~\ref{Eq:H111_14} is positive. After performing the Feynman parameter integral, which converges for $\text{Re}(-z_1)>0$ and $\text{Re}(1+z_2)>0$ we find, since the latter two conditions are also fulfilled:
\begin{align}
	I_{k_1}'' = \frac{(-1)^{1-s}i\pi^{2-\epsilon}}{\left[(m_1^2-i0)x\right]^{-1-s+\epsilon}}\frac{1}{2i\pi}\int_{d-i\infty}^{d+\infty} dz_2 \left(-\frac{m_3^2}{m_1^2-i0}\right)^{z_2} \frac{\Gamma(z_2)\Gamma(-1-z_1+\epsilon+z_2)\Gamma(1+z_2)}{\Gamma(1+z_2-z_1)}\ ,
	\label{Eq:H111_15}
\end{align}
where $d=\text{Re}(z_2)\in]-\text{Re}(-1-z_1+\epsilon),0[$.

Taking into account the overall factors and putting everything together, our final MB representation for the sunset diagram is
\begin{align}
H_{\{1,1,1\}}(m_1^2, m_2^2, m_2^2; m_3^2) = K_1 + K_2 + K_3
	\label{Eq:MBrep}
\end{align}
where
\begin{align}
K_1\equiv \frac{\pi^{4-2\epsilon}}{\left(-m_3^2\right)^{2\epsilon-1}}\frac{\Gamma(1-\epsilon)^3\Gamma(-1+2\epsilon)}{\Gamma(3-3\epsilon)}
	\label{Eq:H111_17}
\end{align}
\begin{align}
	K_2\equiv \frac{\pi^{4-2\epsilon} \Gamma(1-\epsilon)^2}{(m_1^2-i0)\left(-m_3^2\right)^{2\epsilon-2}} \frac{1}{2i\pi}\int_{d'-i\infty}^{d'+i\infty} dz_2 \left(-\frac{m_3^2}{m_1^2-i0}\right)^{z_2}
	\frac{\Gamma(1+z_2)\Gamma(-2-z_2+2\epsilon)\Gamma(2+z_2-\epsilon)}{\Gamma(4-3\epsilon+z_2)}
	\label{Eq:H111_18}
\end{align}
where $d'=\text{Re}(z_2)\in]-2,-1[$, $\text{Re}(2-\epsilon+z_2)>0$ and $\text{Re}(-z_2-2+2\epsilon)>0$, and
\begin{align}
	K_3 \equiv &\frac{\pi^{4-2\epsilon}}{\left(m_2^2-i0\right)^\epsilon\left(m_1^2-i0\right)^{\epsilon-1}}\frac{\sqrt{\pi}}{2}\left(\frac{1}{2i\pi}\right)^2 \int_{c'-i\infty}^{c'+i\infty} dz_1 \int_{d-i\infty}^{d+i\infty} dz_2 \left(\frac{m_1^2-i0}{4(m_2^2-i0)}\right)^{z_1} \left(-\frac{m_3^2}{m_1^2-i0}\right)^{z_2} \nonumber\\
	& \qquad \times \frac{\Gamma(z_1+\epsilon)\Gamma(1+z_1)\Gamma(-z_2)\Gamma(-1-z_1+\epsilon+z_2)\Gamma(2+z_1-\epsilon)\Gamma(z_2-z_1)}{\Gamma \left(\frac{3}{2}+z_1\right)\Gamma(2+z_2-\epsilon)}
	\label{Eq:H111_19}
\end{align}
with $c'=\text{Re}(z_1)\in]-1,-\text{Re}(\epsilon)[$, $d=\text{Re}(z_2)\in]\text{Re}(1+z_1-\epsilon),0[$, $\text{Re}(z_2-z_1)>0$ and $\text{Re}(2+z_1-\epsilon)>0$.

\subsubsection*{Solving the MB integral}

\begin{figure}[hbtp]
\centering
\includegraphics[scale=0.7]{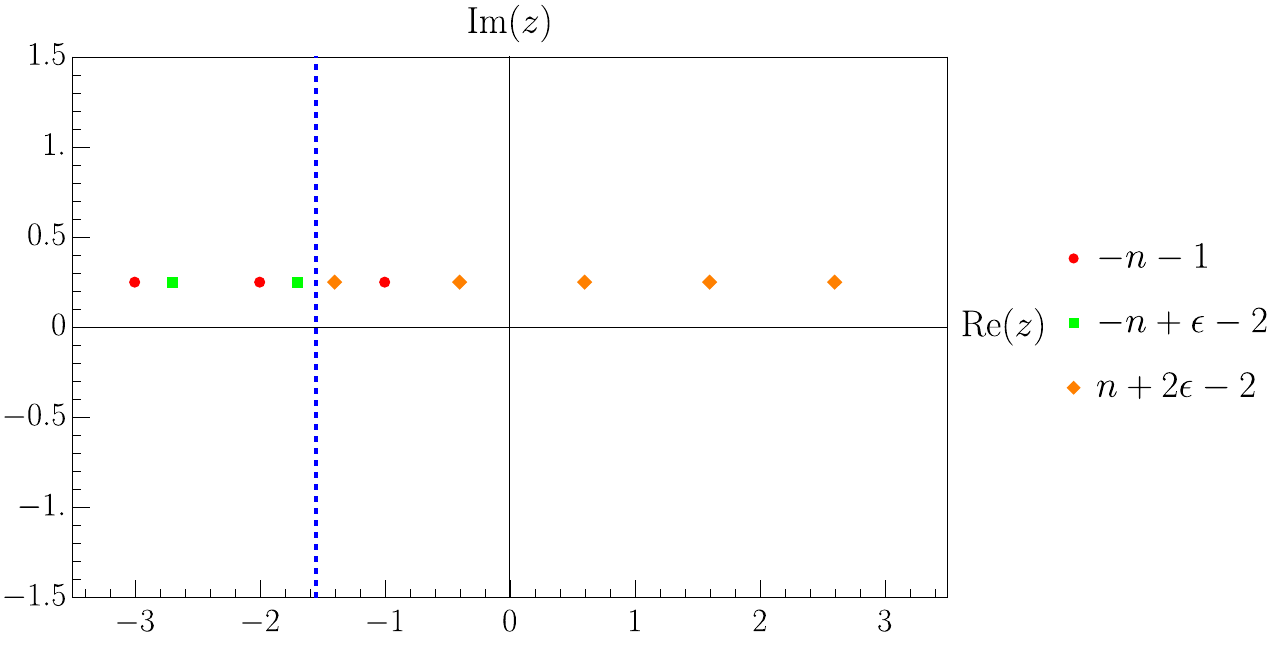}
\caption{Singularity structure of $K_2$}
\label{Fig:K2sing}
\end{figure}

The singularity structure for $K_2$ is shown in Fig.~\ref{Fig:K2sing}, and the correspondence between the singularities and their residues is given in Table~\ref{Table:H111e}.

\begin{figure}[h]
\center
\begin{tabular}{lll}
\includegraphics[scale=0.4]{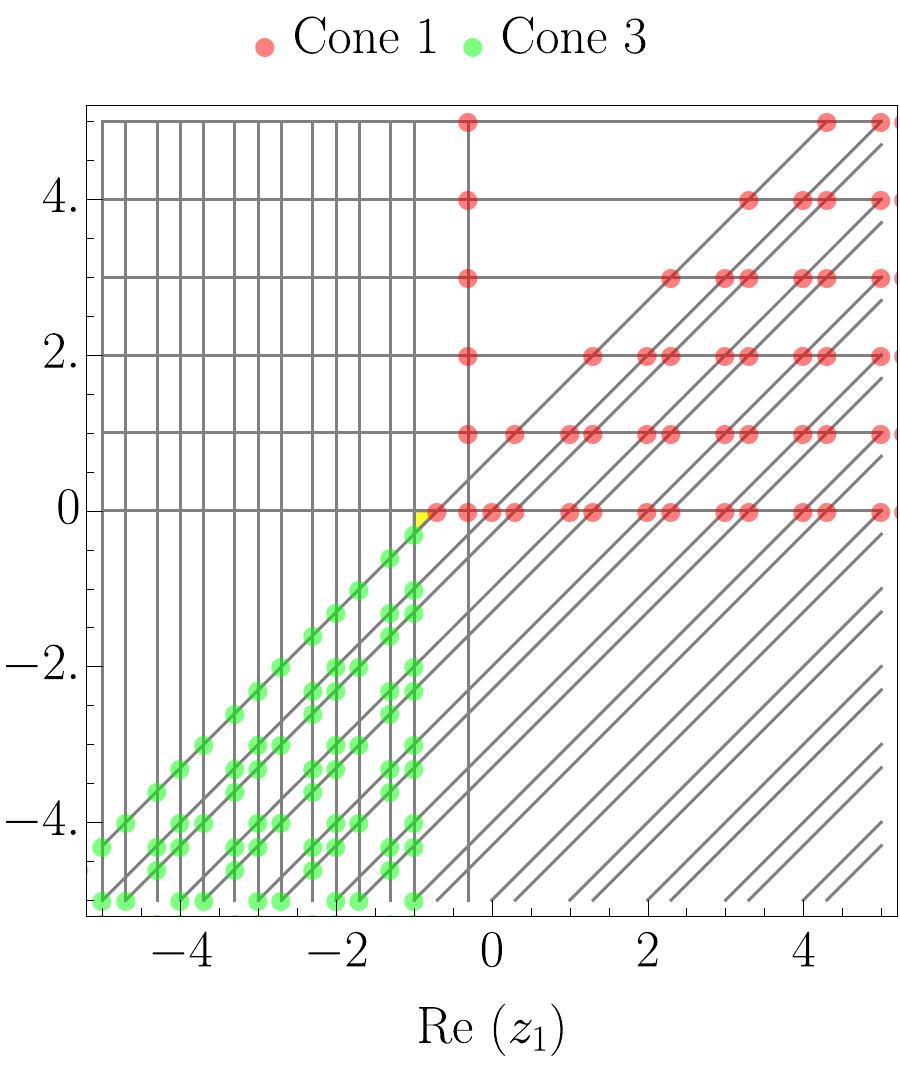}
& \hspace{2cm}
\includegraphics[scale=0.4]{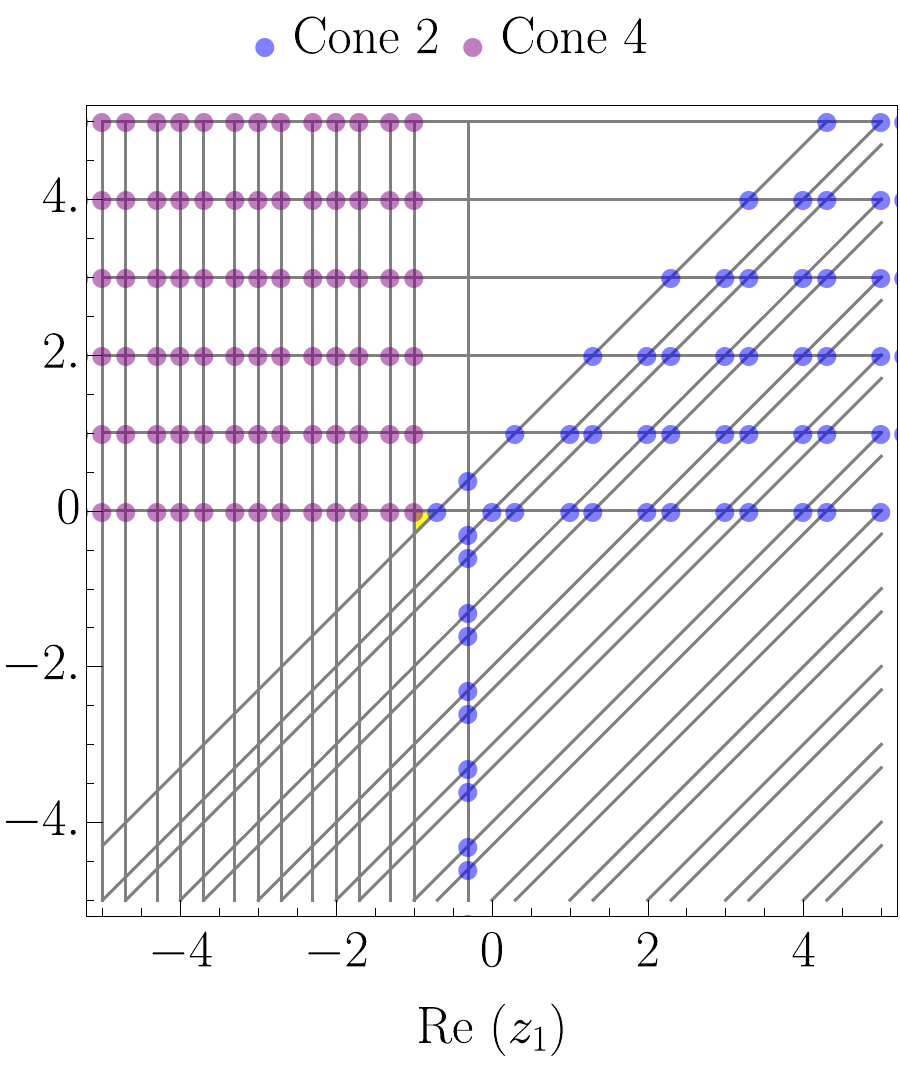}
\end{tabular}
\caption{The contributing singularities of the various cones of $K_3$}
\label{Fig:K3sing}
\end{figure}

\begin{figure}[h]
\center
\includegraphics[scale=0.7]{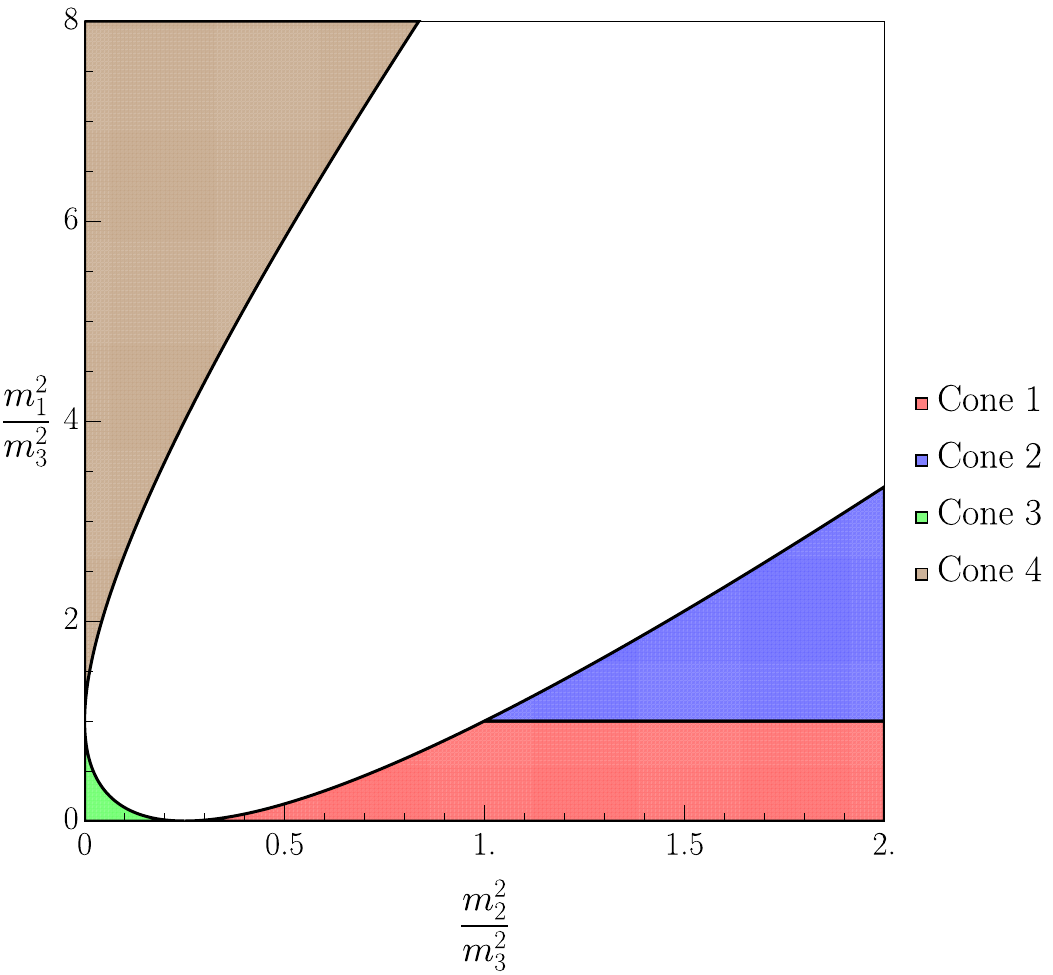}
\caption{The region of convergence of the various cones of $K_3$}
\label{Fig:K3RoC}
\end{figure}

The integral $K_3$ has four cones, the singularities of which are shown in Fig.~\ref{Fig:K3sing}, and whose regions of convergence are shown in Fig.~\ref{Fig:K3RoC}. The residues corresponding to the various cones are given in Table~\ref{Table:H111e}.

The full solution for the integral $H_{\{1,1,1\}} (m_1, m_2, m_2; m_3^2)$ is given by:
\begin{align}
    H_{\{1,1,1\}} (m_1, m_2, m_2; m_3^2) 
    = K_1 + \sum_{i=1}^2 K_2^{\alpha,i} + \sum_{i=1}^c K_3^{C,i}
\end{align}
where the values of $\alpha$ and $\beta$ appropriate for the mass-ratio being considered is given in Table~\ref{Table:H111e_link}.

\begin{table}
    \centering
    \begin{tabular}{|l|c|c||c|}
    \hline
    \rowcolor{lightgray}
    \bf{Region of Convergence} & $C$ & $c$ & $\alpha$ \\
    \hline
    $ (m_1 + m_3 < 2m_2 ) \wedge (m_3^2 > m_1^2) $ & 1 & 3 & R \\
    \hline
    $ (m_1 + m_3 < 2m_2 ) \wedge (m_3^2 < m_1^2) $ & 2 & 4 & L \\
    \hline
    $ (m_1 + 2m_2 < m_3 ) \wedge (m_3^2 > m_1^2) $ & 3 & 5 & L \\
    \hline
    $ (m_3 + 2m_2 < m_1 ) \wedge (m_3^2 < m_1^2) $ & 4 & 3 & R \\
    \hline
    \end{tabular}
    \caption{Table of correspondences}
    \label{Table:H111e_link}
\end{table}

\begin{table}[h!]
\centering
\noindent

\begin{minipage}{0.4\textwidth }
\centering
\begin{tabular}{|c|l|}
\hline
\rowcolor{lightgray}
$K_2$ {\bf LHS} & $\alpha=L$ \\
\hline
$-2-m$ & $K_2^{L,1}$ \\ 
\hline 
$-2+\epsilon-m$ & $K_2^{L,2}$ \\ 
\hline
\end{tabular}
\end{minipage}
\hfill
\begin{minipage}{0.4\textwidth }
\centering
\begin{tabular}{|c|l|}
\hline
\rowcolor{lightgray}
$K_2$ {\bf RHS} & $\alpha=R$ \\
\hline
$-1$ & $K_2^{R,1}$ \\ 
\hline 
$-2+2\epsilon+m$ & $K_2^{R,2}$ \\ 
\hline
\end{tabular}
\end{minipage}

\vspace{0.5cm}

\begin{minipage}[t]{0.47\textwidth}
\centering
\begin{tabular}{|c|l|}
\hline
\rowcolor{lightgray}
$K_3$ {\bf Cone 1} ($C=1$) & $c=7$ \\
\hline
$(-\epsilon,m)$ & $K_3^{1,1}$ \\ 
\hline 
$(-1+\epsilon+m+n,m)$ & $K_3^{1,2}$ \\ 
\hline 
$(m+n,n)$ & $K_3^{1,3}$ \\ 
\hline 
\hline
\rowcolor{lightgray}
$K_3$ {\bf Cone 2} ($C=2$) & $c=4$ \\
\hline
$(-\epsilon,1-2\epsilon-m)$ & $K_3^{2,1}$ \\ 
\hline 
$(-\epsilon,-\epsilon-m)$ & $K_3^{2,2}$ \\ 
\hline
$(-1+\epsilon+m+n,m)$ & $K_3^{2,3} = K_3^{1,2}$ \\ 
\hline 
$(m+n,n)$ &  $K_3^{2,4} = K_3^{1,3}$ \\ 
\hline 
\end{tabular}
\end{minipage}
\hfill
\begin{minipage}[t]{0.47\textwidth}
\centering
\begin{tabular}{|c|l|}
\hline
\rowcolor{lightgray}
$K_3$ {\bf Cone 3} ($C=3$) & $c=5$ \\
\hline
$(-1-m,-\epsilon-m-n)$ & $K_3^{3,1}$ \\ 
\hline 
$(-1-m,-1-m-n)$ & $K_3^{3,2}$ \\ 
\hline 
$(-1-\epsilon-m,-2\epsilon-m-n)$ & $K_3^{3,3}$ \\ 
\hline 
$(-1-\epsilon-m,-1-\epsilon-m-n)$ & $K_3^{3,4}$ \\ 
\hline 
$(-2+\epsilon-m,-1-m-n)$ & $K_3^{3,5}$ \\ 
\hline
\hline
\rowcolor{lightgray}
$K_3$ {\bf Cone 4} ($C=4$) & $c=3$ \\
\hline
$(-1-m,n)$ & $K_3^{4,1}$ \\ 
\hline 
$(-1-\epsilon-m,n)$ & $K_3^{4,2}$ \\ 
\hline 
$(-2+\epsilon-m,n)$ & $K_3^{4,3}$ \\ 
\hline 
\end{tabular}
\end{minipage}

    \caption{Cones, regions of convergence, and singularities of $K_2$ and $K_3$}
    \label{Table:H111e}
\end{table}

\section{On further analytic continuations}
\label{Sec:AC}
We begin by noting that the figures showing the regions of convergence of the cones of all the two-fold MB integrals considered here contain a white region. It is not possible, by a direct application of the method presented above, to obtain series solutions for the integrals for mass configurations falling into these white regions. For these, one possibility is to analytically continuing the series that make up the solution of any given cone. Another approach consists in performing transformations at the MB integral level \cite{Friot:2022dme}.

In this section, we outline the steps to obtain such analytic continuations following the first approach, while expressions for the analytic continuations into the white area of all the integrals discussed in this paper can be found in the appendices of the thesis~\cite{Ghosh:2021jsk}. 

There are standard methods that one can use in order to analyze, and continue bivariate series of this type, that begin often by summing over one of the indices, then introducing an analytic continuation for the resulting hypergeometric function of one variable, which could result in a sum of hypergeometric functions of one variables, not necessarily of the original variable per se, with pre-factors involving the parameters and the unsummed summation index. The new hypergeometric functions are then written out as series in the new region of convergence and the result re-expressed as hypergeometric functions of two-variables, or of Horn series in general.  One efficient method is the one due to Olsson, which has recently been automated~\cite{Ananthanarayan:2021yar}.

As an example, let us begin with a double series, with summation indices ($m,n$):
\begin{align}
    I = \sum_{m,n=0}^{\infty} \frac{\Gamma(a_1+m) ... \Gamma(a_i+m) \Gamma(b_1+n) ... \Gamma(b_j+n) \Gamma(c_1+m+n) ... \Gamma(c_k+m+n)}{\Gamma(a'_1+m) ... \Gamma(a'_{i'}+m) \Gamma(b'_1+n) ... \Gamma(b'_{j'}+n) \Gamma(c'_1+m+n) ... \Gamma(c'_{k'}+m+n)} \frac{x^m}{m!} \frac{y^n}{n!}
\end{align}

As mentioned in the preamble above, we
    express the series over one of the indices as a hypergeometric function, $_2F_1$:
    \begin{align}
        I \sim \sum_{n=0}^{\infty} \frac{\Gamma(a_1+n)...\Gamma(a_i+n)}{\Gamma(a'_1+n)...\Gamma(a'_{j'}+n)} y^n
        {}_2F_1 \left( \begin{array}{c}
\alpha_1(n), \alpha_2(n) \\
\alpha_3(n) \\
\end{array} \bigg| x \right)
    \end{align}
In order to proceed, we     
must now divide the calculation into two parts now: for $x<1$ and for $x>1$. For $x<1$, we continue to the next step. For $x>1$, we apply the following transformation to the $_2F_1$ function to convert its argument to a value $< 1$, so that the formula of the next step can validly be applied to it, before continuing to the next step:
    \begin{align}
{}_2F_1 \left( \begin{array}{c}
a, b \\
c \\
\end{array} \bigg| z \right) =& \frac{\Gamma(c)\Gamma(b-a)}{\Gamma(b)\Gamma(c-a)} (-z)^{-a}
{}_2F_1 \left( \begin{array}{c}
a,1-c+a \\
1-b+a \\
\end{array} \bigg| \frac{1}{z} \right) \nonumber \\
& + \frac{\Gamma(c)\Gamma(a-b)}{\Gamma(a)\Gamma(c-b)} (-z)^{-b}
{}_2F_1 \left( \begin{array}{c}
b,1-c+b \\
1-a+b \\
\end{array} \bigg| \frac{1}{z} \right) & \text{for} \; | arg(-z)| < \pi \label{EqAC3}
\end{align}
    We then apply the following transformation to each the $_2F_1$ functions appearing in the result of the previous step (in the case of $x<1$, there will be only one $_2F_1$ function, whilst for $x>1$ there shall be two):
    \begin{align}
    {}_2F_1 \left( \begin{array}{c}
    a, b \\
    a-b+1 \\
    \end{array} \bigg| z \right)
    &= (1-z)^{-a} {}_2F_1 \left( \begin{array}{c}
    \frac{a}{2}, \frac{a}{2}-b+\frac{1}{2} \\
    a-b+1 \\
    \end{array} \bigg| -\frac{4z}{(1-z)^2} \right) \label{EqAC1}
    \end{align}
    We then apply the following analytic continuation of the hypergeometric function (valid for $|1-\arg(z)|<\pi$) to each $_2F_1$:
\begin{align}
{}_2F_1 \left( \begin{array}{c}
a, b \\
c \\
\end{array} \bigg| z \right) &= \frac{\Gamma(c)\Gamma(b-a)}{\Gamma(b)\Gamma(c-a)} (1-z)^{-a}
{}_2F_1 \left( \begin{array}{c}
a, c-b \\
a-b+1 \\
\end{array} \bigg| \frac{1}{1-z} \right) \nonumber \\
& + \frac{\Gamma(c)\Gamma(a-b)}{\Gamma(a)\Gamma(c-b)} (1-z)^{-b} {}_2F_1 \left( \begin{array}{c}
b, c-a \\
b-a+1 \\
\end{array} \bigg| \frac{1}{1-z} \right)
\end{align}
    
At the end of this procedure, we observe that the result of the last step is an analytic continuation of $I$, consisting of two terms for $x<1$, and four terms for $x>1$. Express each $_2F_1$ function as sums, resulting in an analytic continuation of two terms (for $x<1$) or four terms (for $x>1$), each of which is a double series in ($m,n$). We now reexpress the non-$_2F_1$ sum in the double series as a generalised hypergeometric function $_pFq$. This will result in each term of the analytic continuation of $I$ being of the form
    \begin{align}
        I_1 \sim \sum_{n=0} x^n \frac{\Gamma(n+a_1)...\Gamma(n+a_j)}{\Gamma(n+b_1)...\Gamma(n+b_k)}
        {}_pF_q \left( \begin{array}{c}
c_1(n), ..., c_l(n) \\
d_1(n), ..., d_m(n) \\
\end{array} \bigg| y \right)
    \end{align}
Having obtained this expression, in order to proceed, each of the $_pF_q$ functions is then expressed as a Meijer-G function, $G_{p,q}^{m,n}$. The Meijer-G function is defined as an inverse Mellin transform:

The Meijer-G functions is a very general function, defined as an inverse Mellin transform:
\begin{align}
G_{p,q}^{m,n} & \left( x \bigg|
        \begin{array}{c}
         a_1, ... , a_p \\
         b_1, ... , b_q \\
        \end{array}
        \right) = \frac{1}{2\pi i} \int_L \frac{\prod_{j=1}^m \Gamma(b_j-s) \prod_{j=1}^n \Gamma(1-a_j+s)}{\prod_{j=m+1}^q \Gamma(1-b_j+s) \prod_{j=n+1}^p \Gamma(a_j-s)} z^s ds
\end{align}
which reduces to a variety of common simpler special functions for special configurations of its arguments. These functions have been studied in detail for almost a century, and we use these well-known properties to analytically continue the Meijer-G function obtained in the last step by using the following formula:
    \begin{align}
G_{p,q}^{m,n} & \left( x \bigg|
        \begin{array}{c}
         a_1, ... , a_p \\
         b_1, ... , b_q \\
        \end{array}
        \right) = \sum_{h=1}^{m} \frac{\prod_{j=1}^{'m} \Gamma(b_j-b_h) \prod_{j=1}^{n} \Gamma(1+b_h-a_j)}{ \prod_{j=m+1}^{q} \Gamma(1+b_h-b_j) \prod_{j=n+1}^{p} \Gamma(a_j-b_h)} x^{b_h} \nonumber \\
        & \times {}_p F_{q-1} \left( \begin{array}{c}
        1+b_h-a_1, ... , 1+b_h-a_p \\
        1+b_h-b_1, ... , * , ... , 1+b_h-b_q  \\
        \end{array} \bigg| (-1)^{p-m-n} x \right)
        \end{align}
    
Having obtained these expressions, as a final step, the $_pF_q$ functions can be written as sums, resulting in a three term analytic continuation, each of which is an infinite series in two indices $\sum_{m,n=0}^\infty$. In the thesis~\cite{Ghosh:2021jsk}, closed form expressions have been given in the appendix. It should be borne in mind that many of the expressions represent multiple-valued functions, and in order to employ them in applications, and the choice of the branch cut has to be borne in mind, especially for numerical evaluation.  A similar issue was studied in detail for the case the case of the Appell $F_2$ in~\cite{Ananthanarayan:2021bqz}.

It may be noted that while the sunset diagram is one of the simplest two-loop diagrams and have been studied for well over four decades now, they continue to be the playground for developments in the interface between the mathematical structures of hypergeometric function theory, with numerous surprises being presented at every stage. However, it has been difficult to completely exhaust all aspects of their properties. In the present work, we have shown how the recently developed method of conic hulls for MB integrals has both helped as well as derived inspiration from the study of sunset diagrams.

\section{Applications to chiral perturbation theory \label{Sec:Applications}}

As one of the simpler two-loop integrals, the sunset diagrams appear in a variety of settings in quantum field theory.  In this section, we discuss applications to one special case namely chiral perturbation theory ($\chi$-PT). The latter is a widely used effective field theory of Quantum Chromodynamics in the domain of low energies/momenta, below the so-called chiral symmetry breaking scale $\Lambda_\chi \simeq 1$~GeV. It has been first elucidated at one loop order for mesonic SU(3) pseudoscalars octet degrees of freedom, namely the pions, kaons and eta in~\cite{Gasser:1984gg} and was later extended to the baryonic sector~\cite{Gasser:1987rb}. In that work, however, no scheme was
given that allowed to consistently deal with the power-counting breaking due to the nucleon mass. This was first solved within the
heavy baryon approach~\cite{Jenkins:1990jv,Bernard:1992qa}. The interested reader can find more details on effective field theories 
in~\cite{Meissner:2022cbi} and references therein. Two-loop studies have  been developed later. They have a long history in the meson sector, see e.g.~\cite{Bijnens:2014gsa} for a numerical package and references therein. Another application is the $\omega$-meson self-energy, which has been considered in the complex mass renormalization scheme in~\cite{Djukanovic:2015gna}.

More recently two loop calculations have also been performed in the baryon sector. As stated before,  one has to deal with the fact that the nucleon mass is of the same size as the chiral symmetry breaking scale. Different regularization schemes have been proposed to solve this issue, for a review see for example~\cite{Bernard:2007zu}. The nucleon mass at two loop order has been first obtained in the so-called Infrared Renormalization scheme~\cite{Schindler:2007dr} while more recent calculations have been done in the Extended-on-Mass-Shell scheme~\cite{Conrad:2024phd,Chen:2024twu,Liang:2025cjd}. 
Applying the Feynman-Hellmann theorem~\cite{Schindler:2007dr} determined the $\sigma$-term, a much discussed quantity, see e.g.~\cite{Hoferichter:2023ptl} (and references therein), to order ${\cal O}(M^5)$ with $M$ the pion mass.  Another quantity of interest is the axial-vector coupling $g_A$, whose leading two-loop correction is given in~\cite{Bernard:2025gto} (for earlier related  work using renormalization group arguments, 
see~\cite{Bernard:2006te}).

In all these studies a tensor reduction of the integrals involved 
allows to write the quantity of interest in terms of a small set of  (scalar) master integrals. The sunset integrals which are the non-reducible  two loop integrals with the smallest number of propagators of different types belong to it. Note that the reduction can also be done
in such a way that these master integrals appear for higher d-dimensional Minkowski space, see
for example~\cite{Schindler:2007dr}. The Mathematica package Tarcer~\cite{Mertig:1998vk} automates the process of reducing such integrals to their constituent master integrals, and has been used significantly in our previous work.

In the baryon sector except for~\cite{Chen:2024twu}  analytical results have been obtained using the method of regions to evaluate the master integrals allowing to differentiate between the infrared and regular parts. An expansion in the pion mass is then performed. The results given here will be the first step  for a (partial) full $\epsilon$-dependent analytical determination to all order in the pion mass.  

The cases discussed so far in the baryon sector are the simplest one as they involve only two mass scales. Cases including more mass scales,
namely the calculation of the ground state octet baryon masses at two-loop order or the $q^2$ dependence of $g_A$ for example have not been attempted yet.

As just discussed, all the integrals presented in this work play a central role in the calculation  of  physical quantities at two-loop order in $\chi$-PT. Let us give here the expression of sunset integrals with four different masses needed
in the calculation of   
the masses and decay-constants of the pion, kaon and eta at two-loops and three flavours. Two masses sunsets appearing in the two flavor baryon sector are also given. 

In the case of quantum field theory one has to deal with divergences. One has to regularize and renormalize the theory leading to the introduction of a scale parameter $\mu$ known as the t'Hooft parameter.  Note that there exists various renormalization scheme depending on how one absorbs the infinities. The most used ones are the so called Minimal subtraction  (MS) scheme
or the modified minimal subtraction scheme or $\overline{\text{MS}}$ scheme. In applications in $\chi$-PT one uses dimensional regularization which means that one changes $d^4 p \to \mu^{4-d} d^d p $ so that the integrals have the same dimension for arbitrary $d$.   It is also customary to use a modified version of the $\overline{\text{MS}}$ scheme by changing $\mu$ by $\mu_{\chi}^2 \equiv \mu^2 e^{\gamma_E - 1}/(4\pi)$ where $\gamma_E$ is the Euler-Mascheroni constant. One also defines the integrals such that each momentum integral is divided by $(2 \pi)^d$. Note that when performing two loop calculations  one needs in principle to 
perform the calculation of the one loop integrals to order $\mathcal{O}(\epsilon^1)$
while for the two loop ones one only needs to consider them up to order 
$\mathcal{O}(\epsilon^0)$. We will denote such renormalized sunset untegrals by use of the subscript $\chi$, i.e.
\begin{align}
	H^{\chi}_{\{ \alpha,\beta,\gamma \}} (m_1, m_2, m_3; p^2) \equiv \frac{(\mu_{\chi}^2)^{4-d}}{(4 \pi^2)^d} H_{\{ \alpha,\beta,\gamma \}} (m_1, m_2, m_3; p^2)
\end{align}

The inclusion of factor $\mu$ raised to a power of the dimension $d$ introduces terms involving chiral logarithms, i.e.
\begin{align}
	l_i \equiv \frac{1}{2(4\pi)^2} \log \left[ \frac{m_i^2}{\mu^2} \right]
\end{align}

In the $\chi$-PT applications of the works~\cite{Ananthanarayan:2016pos, Ananthanarayan:2017qmx, Ananthanarayan:2018irl}, all terms containing chiral logarithms are grouped together, and $H^{\log}$ is defined as those terms of the sunset integral containing chiral logarithms.
\begin{align}
	\overline{H}_{ \{ \alpha \beta \gamma \}} \equiv H^{\chi}_{\alpha \beta \gamma} - H^{\log}_{ \alpha \beta \gamma}
\end{align}
The chiral log terms may be explicitly fixed as:
\begin{align}
	& H^{\log}_{\{1,1,1\}} (m_1,m_2,m_3;m_4^2) = 4 p^2 (l_1)^2 + 4 m_2^2 (l_2)^2 + 4 m_3^2 (l_3)^2 - \frac{m_1^2}{8\pi^2} l_1 - \frac{m_2^2}{8\pi^2} l_2 -\frac{m_3^2}{8\pi^2} l_3 + \frac{p^2}{16\pi^2} l_{4} \nonumber \\
	& H^{\log}_{\{2,1,1\}} (m_1,m_2,m_3;m_4^2) = 4 (l_1)^2 + \frac{1}{8\pi^2} l_1 \nonumber \\
	& H^{\log}_{\{1,2,1\}} (m_1,m_2,m_3;m_4^2) = 4 (l_2)^2 + \frac{1}{8\pi^2} l_2 \nonumber \\
	& H^{\log}_{\{1,1,2\}} (m_1,m_2,m_3;m_4^2) = 4 (l_3)^2 + \frac{1}{8\pi^2} l_3
	\label{Eq:ChiralLogs}
\end{align}

Having fixed the notation, we now give the results of the sunset integrals presented in this work at order $\mathcal{O}(\epsilon^0)$ after chiral subtraction.

The one mass sunset, $H(m^2) \equiv H_{1,1,1}(m,m,m;m^2)$, is given by:
\begin{align}
	H^{\chi}(m^2) = \overline{H}(m^2) + H^{\log}(m^2)
\end{align} 
with $H^{\log}(m^2)$ calculable from Eq.~\ref{Eq:ChiralLogs} and $\overline{H}$ given by:
\begin{align}
	\overline{H} (m^2)= \frac{m^2}{512 \pi^4} \left(\frac{15}{4}+\frac{\pi ^2}{2}\right)
\end{align}

The two mass sunsets are given by:
\begin{align}
	& H^{\chi}_{\{1,1,1\}}(m,M,M;m^2) = \overline{H}_{\{1,1,1\}}(\bullet) + H^{\log}_{\{1,1,1\}}(\bullet) \nonumber \\
	& H^{\chi}_{\{2,1,1\}}(m,M,M;m^2) = \overline{H}_{\{2,1,1\}}(\bullet) + H^{\log}_{\{2,1,1\}}(\bullet)
\end{align} 
with $H^{\log}_{\{1,1,1\}}$ and $H^{\log}_{\{2,1,1\}}$ given by Eq.~\ref{Eq:ChiralLogs}, and:
\begin{multline}
\overline{H}_{\{1,1,1\}}(m,M,M;m^2) = -\frac{M^2}{512\pi^4} \Bigg\{ 4 + \frac{\pi^2}{3} + \frac{m^2}{M^2} \left(\frac{\pi ^2}{6}-\frac{1}{4}\right) - \frac{m^2}{M^2} \log^2 \left[ \frac{m^2}{M^2} \right]  \\
 + 2 \log \left[ \frac{m^2}{M^2} \right] + 2 \left( \frac{M^2}{m^2} + \frac{m^2}{M^2}-2 \right) \left(\text{Li}_2\left[\frac{m^2}{M^2}\right] + \log \left[1-\frac{m^2}{M^2}\right] \log \left[\frac{m^2}{M^2}\right] \right) \Bigg\}
\end{multline}
\begin{align}
\overline{H}_{\{2,1,1\}}(m,M,M;m^2) &= -\frac{1}{512\pi^4} \Bigg\{ \frac{\pi^2}{6} - 1 - \log^2 \left[ \frac{m^2}{M^2} \right] \nonumber \\
& + 2 \left(1-\frac{M^2}{m^2}\right) \left( \text{Li}_2 \left[ \frac{m^2}{M^2} \right] + \log \left[ \frac{m^2}{M^2} \right] \log \left[ 1 - \frac{m^2}{M^2} \right] \right) \Bigg\}
\end{align}


From these two last equations and using Eq.~\ref{eq:h211mMMm}  one can determine 
 $H_{\{1,2,1\}}(m,M,M;m^2)$ which  appears in the calculation of the nucleon mass and $g_A$. Note that, as mentioned previously, these expressions contain power counting violating terms which have to be subtracted  when dealing with baryon physical quantities.

In the chiral one- and two-mass sunset results above, we are able to obtain closed form expressions which can be used for all values of the mass arguments. For the three mass sunsets, the solutions are in the form of several infinite series, which are analytic continuations of each other, and which together cover all the possible values of the mass configurations, but none of which individually do. For the three mass sunset results therefore, we choose those series that converge for physical mass values of the meson masses (i.e. $m_\pi$, $m_K$ and $m_\eta$), and chirally subtract and expand those. This gives:
\begin{align}
&   H^{\chi}_{\{1,1,1\}}(m_1,m_2,m_3;m_1^2) = \overline{H}_{\{1,1,1\}}(\bullet) + H^{\log}_{\{1,1,1\}}(\bullet) \nonumber \\
& H^{\chi}_{\{2,1,1\}}(m_1,m_2,m_3;m_1^2) = \overline{H}_{\{2,1,1\}}(\bullet) + H^{\log}_{\{2,1,1\}}(\bullet) \nonumber \\
&	H^{\chi}_{\{1,1,1\}}(m_1,m_2,m_2;m_3^2) = \overline{H}_{\{1,1,1\}}(\bullet) + H^{\log}_{\{1,1,1\}}(\bullet)
\end{align}
where
\begin{align}
	& \overline{H}_{\{1,1,1\}} (m_1,m_2,m_2;m_3^2) = \frac{m_1^2}{512 \pi^4} \Bigg\{ \frac{\pi^2}{6} - 5 + 4 \log \left[\frac{m_1^2}{m_2^2}\right] - \log ^2 \left[\frac{m_1^2}{m_2^2}\right] + \frac{m_3^2}{m_1^2} \left( \log \left[ \frac{m_2^2}{m_3^2} \right] + \frac{5}{4} \right) \nonumber \\
	& \qquad + \frac{m_2^2}{m_1^2} \left(6 + \frac{\pi ^2}{3}\right) 
	- \frac{1}{18} \frac{m_3^2}{m_2^2} \frac{m_3^2}{m_1^2}  {}_3F_2 \bigg[ \begin{array}{c}
		1,1,2 \\
		\frac{5}{2},4 \\
	\end{array}	\bigg| \frac{m_3^2}{4m_2^2} \bigg] - \frac{1}{3} \frac{m_1^2}{m_2^2} \log \left[ \frac{m_1^2}{4 m_2^2} \right] {}_2F_1 \bigg[ \begin{array}{c}
		1,1 \\
		\frac{5}{2} \\
	\end{array}	\bigg| \frac{m_1^2}{4 m_2^2} \bigg] \nonumber \\
	& \qquad - \sqrt{\pi} \sum_{m=0}^{\infty} \frac{\Gamma (m+1)}{\Gamma \left(m+\frac{5}{2}\right)} \left(\frac{m_1^2}{4 m_2^2}\right)^{m+1} \Bigg( \psi(m+1) - \psi\left( m+\frac{5}{2} \right) \Bigg) \nonumber \\
	& \qquad - \sqrt{\pi} \sum_{m,n=0}^{\infty} \frac{\Gamma (m+n+1) \Gamma (m+n+2) \Gamma (m+n+3)}{\Gamma (m+2) \Gamma (m+3) \Gamma (n+1) \Gamma (n+2) \Gamma \left(m+n+\frac{5}{2}\right)} \left(\frac{m_3^2}{4 m_2^2}\right)^{m+1} \left(\frac{m_1^2}{4 m_2^2}\right)^n \nonumber \\
	& \qquad \times \Bigg( \log \left[\frac{m_1^2}{4 m_2^2}\right]-\psi(n+1)-\psi(n+2) + \psi(m+n+1) + \psi(m+n+2) +\psi(m+n+3) \nonumber \\
	& \qquad \quad  - \psi\left( m+n+\frac{5}{2} \right) \Bigg) \Bigg\}
\end{align}

\begin{align}
	& \overline{H}_{\{2,1,1\}} (m_1,m_2,m_3;m_1^2) = \frac{1}{512\pi ^4} \Bigg\{ -\frac{m_3^2}{m_1^2} \bigg( 1 + \frac{\pi^2}{3} + \frac{1}{2} \log ^2 \left[\frac{m_1^2}{m_3^2}\right] + \log \left[\frac{m_1^2}{m_3^2}\right] + \text{Li}_2 \left[ 1-\frac{m_2^2}{m_3^2} \right] \bigg)  \nonumber \\
	& \qquad + \frac{5\pi^2}{6} -1 -\frac{m_2^2}{m_1^2} \bigg( 1 + \frac{\pi^2}{3} - \log \left[\frac{m_2^2}{m_1^2}\right] - \frac{1}{2} \log^2 \left[ \frac{m_1^2}{m_3^2} \right] - \log \left[\frac{m_1^2}{m_3^2}\right] \log \left[\frac{m_2^2}{m_1^2}\right] - \text{Li}_2 \left[1-\frac{m_2^2}{m_3^2}\right] \bigg) \nonumber \\
	& \qquad - \frac{m_2^4}{4 m_1^4}
	{}_3F_2 \bigg[ \begin{array}{c}
		1,1,1 \\
		\frac{3}{2},3 \\
	\end{array}	\bigg| \frac{m_\pi^2}{4m_K^2} \bigg]
	 - \frac{m_3^4}{4 m_1^4} 
	 {}_3F_2 \bigg[ \begin{array}{c}
		1,1,1 \\
		\frac{3}{2},3 \\
	\end{array}	\bigg| \frac{m_\eta^2}{4m_K^2} \bigg] 
	+ \frac{2 \pi}{3} \left(\frac{m_3^2}{m_1^2}\right)^{3/2}
	{}_2F_1 \bigg[ \begin{array}{c}
		\frac{1}{2},\frac{1}{2} \\
		\frac{5}{2} \\
	\end{array}	\bigg| \frac{m_\eta^2}{4m_K^2} \bigg] \nonumber \\
	& \qquad + 4 \sqrt{\pi} \sum_{m,n=0}^{\infty} \frac{\Gamma \left(m+n+\frac{1}{2}\right)^2 \Gamma \left(m+n+\frac{3}{2}\right)}{\Gamma \left(m+\frac{1}{2}\right) \Gamma \left(m+\frac{3}{2}\right) \Gamma (n+1) \Gamma (n+2) \Gamma (m+n+1)} \left(\frac{m_3^2}{4m_1^2}\right)^{\frac{1}{2}+m} \left(\frac{m_2^2}{4m_1^2}\right)^{1+n} \nonumber \\
	& \qquad \times \left( \log \left[\frac{m_\pi^2}{m_\eta^2}\right] + \psi\left(m+\frac{1}{2}\right) + \psi\left(m+\frac{3}{2}\right) - \psi(n+1) - \psi(n+2) \right) \nonumber \\
	& \qquad - 4 \sqrt{\pi} \sum_{m,n=0}^{\infty}  \frac{ \Gamma (m+n+1)^2 \Gamma (m+n+2) }{\Gamma (m+1) \Gamma (m+2) \Gamma (n+1) \Gamma (n+2) \Gamma \left(m+n+\frac{3}{2} \right) } \left(\frac{m_3^2}{4m_1^2}\right)^{1+m} \left(\frac{m_2^2}{4m_1^2}\right)^{1+n} \nonumber \\
	& \qquad \times \bigg( \log \left[\frac{m_2^2}{4 m_1^2}\right] + \log \left[\frac{m_3^2}{4 m_1^2}\right] -\psi(m+1)-\psi(m+2) -\psi(n+1)-\psi(n+2) \nonumber \\
	& \qquad \quad + 4 \psi(m+n+1)+2 \psi(m+n+2)-2 \psi\left(m+n+\frac{3}{2}\right) \bigg) \nonumber \\
	& \qquad - \frac{4}{\pi^{3/2}} \sum_{m,n=0}^{\infty} \frac{\Gamma \left(m+\frac{1}{2}\right) \Gamma \left(m+\frac{3}{2}\right) \Gamma \left(n+\frac{1}{2}\right)^2 \Gamma \left(n+\frac{3}{2}\right)}{\Gamma (n+1) \Gamma (m+n+2) \Gamma (m+n+3)} \left(\frac{m_2^2}{m_3^2}\right)^{m+\tfrac{1}{2}} \left(\frac{m_2^2}{4m_1^2}\right)^{n+\tfrac{3}{2}} \nonumber \\
	& \qquad \times \left( \log \left[\frac{m_\pi^2}{m_\eta^2}\right] + \psi\left(m+\frac{1}{2}\right) + \psi\left(m+\frac{3}{2}\right) - \psi(m+n+2) - \psi(m+n+3) \right) \Bigg\}
\end{align}

\begin{align}
	& \overline{H}_{\{1,1,1\}} (m_1,m_2,m_3;m_1^2) = \frac{m_1^2}{512\pi ^4} \Bigg\{ -\frac{1}{4}+\frac{5 \pi ^2}{6}-\frac{7}{4}\left(\frac{m_3^4}{m_1^4}+\frac{m_2^4}{m_1^4}\right) + \left(1-\frac{\pi^2}{2}\right)\left(\frac{m_3^2}{m_1^2}+\frac{m_2^2}{m_1^2}\right)  \nonumber \\
	& \qquad +\frac{m_2^2}{m_1^2} \frac{m_3^2}{m_1^2} \left(7+\frac{2 \pi^2}{3}-2 \log\left[\frac{m_3^2}{m_1^2}\right]-2 \log\left[\frac{m_2^2}{m_1^2}\right]+\log\left[\frac{m_3^2}{m_1^2}\right] \log\left[\frac{m_2^2}{m_1^2}\right]\right) +\frac{m_3^4}{2 m_1^4} \log\left[\frac{m_3^2}{m_1^2}\right] \nonumber \\
	& \qquad -\frac{m_2^2}{m_1^2} \log\left[\frac{m_2^2}{m_1^2}\right]^2-\frac{m_3^2}{m_1^2} \log\left[\frac{m_3^2}{m_1^2}\right]^2 +\frac{8 \pi }{3}\left(\frac{m_3^2}{m_1^2}\right)^{3/2} 
	{}_2F_1 \bigg[ \begin{array}{c}
		\frac{1}{2},-\frac{1}{2} \\
		\frac{5}{2} \\
	\end{array}	\bigg| \frac{m_\eta^2}{4m_K^2} \bigg] \nonumber \\
	& \qquad + \frac{1}{36}\frac{m_3^6}{m_1^6}
	{}_3F_2 \bigg[ \begin{array}{c}
		1,1,2 \\
		\frac{5}{2},4 \\
	\end{array}	\bigg| \frac{m_\eta^2}{4m_K^2} \bigg]
	 + \frac{1}{36} \frac{m_2^6}{m_1^6}
	{}_3F_2 \bigg[ \begin{array}{c}
		1,1,2 \\
		\frac{5}{2},4 \\
	\end{array}	\bigg| \frac{m_\pi^2}{4m_K^2} \bigg] 
	+\frac{m_2^4}{2 m_1^4} \log\left[\frac{m_2^2}{m_1^2}\right] \nonumber \\
	& \qquad + \frac{1}{6} \frac{m_3^4}{m_K^4} \frac{m_2^2}{m_K^2}
	 \left( 2\gamma_E - 1 + \log \left[\frac{m_3^2}{4 m_K^2}\right] + \log \left[\frac{m_2^2}{4 m_K^2}\right] \right) {}_2F_1 \bigg[ \begin{array}{c}
		1,1 \\
		\frac{5}{2} \\
	\end{array}	\bigg| \frac{m_\pi^2}{4m_K^2} \bigg] \nonumber \\
	& \qquad + 2 \sqrt{\pi} \frac{m_3^2}{m_1^2} \sum_{m=0}^{\infty} \frac{\Gamma(1+m)}{ \Gamma(\frac{5}{2}+m)} \left(\frac{m_2^2}{4m_1^2}\right)^{m+2}  \bigg( 2 \psi(m+1)+\psi(m+2)+\psi(m+3)-2 \psi\left(m+\frac{5}{2}\right) \bigg) \nonumber \\
	& + 8 \sqrt{\pi} \sum_{m,n=0}^{\infty} \frac{ \Gamma (m+n+1) \Gamma (m+n+2) \Gamma (m+n+3)}{\Gamma (m+2) \Gamma (m+3) \Gamma (n+1) \Gamma (n+2) \Gamma \left(m+n+\frac{5}{2}\right)}\left(\frac{m_3^2}{4 m_1^2}\right)^{m+2}\left(\frac{m_2^2}{4 m_1^2}\right)^{n+1} \nonumber \\
	& \qquad \times \bigg( \log \left[\frac{m_3^2}{4 m_1^2}\right]+\log \left[\frac{m_2^2}{4 m_1^2}\right] -\psi(m+2)-\psi(m+3)-\psi(n+1)-\psi(n+2)\nonumber \\
	& \qquad \quad + 2 \psi(m+n+1) +2 \psi(m+n+2)+2 \psi(m+n+3) -2 \psi\left(m+n+\frac{5}{2}\right) \bigg) 
\label{Ch_Kaon:Eq:Hkpe}
\end{align}

\begin{align}
	& + \frac{32}{\pi^{3/2}} \left(\frac{m_1^2}{4m_3^2}\right)^{1/2} \sum_{m,n=0}^{\infty} \frac{\Gamma \left(m+\frac{1}{2}\right) \Gamma \left(m+\frac{3}{2}\right) \Gamma \left(n-\frac{1}{2}\right) \Gamma \left(n+\frac{1}{2}\right) \Gamma \left(n+\frac{3}{2}\right)}{\Gamma (n+1) \Gamma (m+n+2) \Gamma (m+n+3)} \left(\frac{m_2^2}{m_3^2}\right)^m\left(\frac{m_2^2}{4m_1^2}\right)^{n+2} \nonumber \\
	& \qquad \times \left( \log \left[\frac{m_\pi^2}{m_\eta^2}\right] + \psi\left(m+\frac{1}{2}\right) + \psi\left(m+\frac{3}{2}\right) - \psi(m+n+2) - \psi(m+n+3) \right) \nonumber \\
	& - 8 \sqrt{\pi} \sum_{m,n=0}^{\infty} \frac{\Gamma \left(m+n-\frac{1}{2}\right) \Gamma \left(m+n+\frac{1}{2}\right) \Gamma \left(m+n+\frac{3}{2}\right)}{\Gamma \left(m+\frac{1}{2}\right) \Gamma \left(m+\frac{3}{2}\right) \Gamma (n+1) \Gamma (n+2) \Gamma (m+n+1)} \left(\frac{m_3^2}{4m_1^2}\right)^{\frac{1}{2}+m} \left(\frac{m_2^2}{4m_1^2}\right)^{1+n} \nonumber 
	\\
	& \qquad \times \left( \log \left[\frac{m_\pi^2}{m_\eta^2}\right] + \psi\left(m+\frac{1}{2}\right) + \psi\left(m+\frac{3}{2}\right) - \psi(n+1) - \psi(n+2) \right) \Bigg\}
\nonumber
\end{align}

The other master integrals required are obtained by differentiating with respect to the appropriate mass.

\section{Conclusion}

We now give a discussion and summary of the main results presented in this paper.  We have considered here the two-loop self-energy or sunset diagrams, which are generic and which also play a role in chiral perturbation theory, in the context of the masses and decay constants of the pseudo-scalar octet, which contains three different masses, namely those of the pion, kaon, and the eta in the iso-spin conserving limit.  Here we confine ourselves to the case of two mass ratios. For the most general case, one may look into the paper of Berends et al.~\cite{Berends:1997vk} and the recent discussion~\cite{Ananthanarayan:2019icl}. The divergent part was first given by Tarasov~\cite{Tarasov:1997kx} a long time ago. The most general case concerns solutions expressed in terms of Lauricella functions with three variables, while here we are restricted to two variables, as in the case of chiral perturbation theory. Our aim is to present a new strategy to derive MB representations that allows us to resolve singularities in $\epsilon$ and retain straight-line contours throughout the derivation process. We then analyze the MB representations of the sunsets obtained using this strategy, and compare them with those obtained using the different strategies advocated by Smirnov~\cite{Smirnov:2012gma} in a series of investigations.

Our results are presented in Sections~\ref {Sec:TwoMassScale} and~\ref{Sec:ThreeMassScale}. In order to derive them, we have employed the method given by Friot and Greynat~\cite{Friot:2011ic}, which developed the method studied by de Rafael, Greynat and Aguillar~\cite{Aguilar:2008qj}. We also demonstrate how to use analytic continuation methods to obtain series solutions for the MB representations for all possible real values of its masses. This is shown in Section~\ref{Sec:AC}. It is also necessary to point out that a general problem concerning the series representations of $N$-fold MB representations was solved using the method of conic hulls.  That said, the present work is of stand-alone importance.  In order to check the validity of the results and their relationship to the conic hull method, we have considered a couple of simple examples, and we find general consistency. A complete one-to-one map is a complicated issue and could be taken up in the future.  Since the basic MBs are the same, either method would give the same solution.

\section*{Acknowledgments}
We would like to thank Sudeepan Datta for his help with Appendix B. SB is supported by a UZH Postdoc Grant (Grant No. [FK-24-100]) and Swiss National Science Foundation (Grant No.~PP00P21 76884). The work of UGM  was supported in part by the European
Research Council (ERC) under the European Union's Horizon 2020 research
and innovation programme (grant agreement No. 101018170), by the CAS President's International
Fellowship Initiative (PIFI) (Grant No.~2025PD0022), by the MKW NRW
under the funding code NW21-024-A and by  the Deutsche Forschungsgemeinschaft (DFG,
German Research Foundation) as part of the CRC 1639 NuMeriQS – project
no. 511713970.



\bigskip

\appendix

\begin{appendices}

\section{MB representations using the modified Method of Brackets \label{MoB}}

\renewcommand{\theequation}{A-\arabic{equation}}
  \setcounter{equation}{0}  

In this appendix, we derive the MB representation of the sunset Feynman integral following the procedure described in~\cite{Prausa:2017frh}. Our starting point is the
Schwinger representation of the sunset integral
\begin{equation}
\displaystyle{I(a_1,a_2,a_3)=\frac{1}{\Gamma(a_1)\Gamma(a_2)\Gamma(a_3)}\int\limits_0^\infty{dx_1}\,{x_1}^{a_1-1}\cdots\int\limits_0^\infty {dx_3}\, {x_3}^{a_3-1} \frac{e^{-\frac{F}{U}-x_1 m^2_1 - x_2 m^2_2 - x_3 m^2_3}}{U^{D/2}}}
\label{eq:Schwinger_2}
\end{equation}
where $F=p^2 x_1x_2x_3$ and $U=x_1 x_2 + x_2 x_3 + x_1 x_3$ are the (first and second) Symanzik polynomials.

We first apply Rule~A of~\cite{Prausa:2017frh} to expand the exponential function in terms of multiple MB integrals,
\begin{align}\label{exp_expand}
   e^{-\frac{F}{U}-x_1 m^2_1 - x_2 m^2_2 - x_3 m^2_3} 
   =  \int_{-i \infty}^{+i \infty} \frac{d z_1}{2 \pi i} \cdots \int  \frac{d z_4}{2 \pi i} \,
   \Gamma(-z_1) \cdots \Gamma(-z_4) \left( \frac{F}{U} \right)^{z_4} (x_1 m^2_1)^{z_1} (x_2 m^2_2)^{z_2} (x_3 m^2_3)^{z_3} .
\end{align}
We then expand the polynomial $U$ using Rule~B of~\cite{Prausa:2017frh} to obtain
\begin{align}\label{U_expand}
   \frac{1}{U^{D/2+z_4}} 
   & =  \frac{1}{\Gamma(D/2+z_4)} \int_{-i \infty}^{+i \infty} \frac{d z_5}{2 \pi i} \cdots \int  \frac{d z_7}{2 \pi i} 
   \Gamma(-z_5) \cdots \Gamma(-z_7)\,
   (x_1 x_2)^{z_5} (x_2 x_3)^{z_6} (x_1 x_3)^{z_7} \nonumber \\ & \times
   \left\langle z_5 + z_6 + z_7 + z_4 + D/2 \right\rangle ,
\end{align}
where the bracket $\left\langle \cdot \right\rangle$ is defined as $\left\langle a \right\rangle = \int_0^\infty dx \, x^{a-1}$. Inserting the above expansions into the Schwinger representation in Eq.~\ref{eq:Schwinger_2} and expressing it in terms of brackets yields
\begin{align}\label{sunset_mb_bracket}
    & I(a_1,a_2,a_3) =\frac{1}{\Gamma(a_1)\Gamma(a_2)\Gamma(a_3)} \int_{-i \infty}^{+i \infty} \frac{d z_1}{2 \pi i} \cdots \int  \frac{d z_7}{2 \pi i} \frac{\Gamma(-z_1) \cdots \Gamma(-z_7)}{\Gamma(D/2+z_4)}
    (m^2_1)^{z_1} (m^2_2)^{z_2} (m^2_3)^{z_3} (p^2)^{z_4}
    \nonumber \\ &
    \, \times 
    \left\langle z_1 + z_4 + z_5 + z_7 + a_1 \right\rangle
\left\langle z_2 + z_4 + z_5 + z_6 + a_2 \right\rangle
\left\langle z_3 + z_4 + z_6 + z_7 + a_3 \right\rangle
    \left\langle z_5 + z_6 + z_7 + z_4 + D/2 \right\rangle
\end{align}

Thus, we have a seven-fold MB integral with four brackets. 
By Rule~D of~\cite{Prausa:2017frh}, there are $\binom{7}{3}=\binom{7}{4}=35$ possible choices of three free MB variables, and hence at most 35 equivalent MB representations. We choose the case where $z_1, z_2 ,z_3$ are the free variables and solve the bracket equations
\begin{align}
    z_1 + z_4 + z_5 + z_7 + a_1 & = 0 \nonumber \\ 
    z_2 + z_4 + z_5 + z_6 + a_2 & = 0 \nonumber \\ 
    z_3 + z_4 + z_6 + z_7 + a_3 & = 0 \nonumber \\
    z_5 + z_6 + z_7 + z_4 + \frac{D}{2} & = 0 \nonumber 
\end{align}
to write the dependent variables in terms of the free variables,
\begin{align}
    z_4 & = D - a_1 - a_2 - a_3 - z_1 - z_2 - z_3 \nonumber \\ 
    z_5 & = -\frac{D}{2}+a_3+z_3 \nonumber \\ 
    z_6 & = -\frac{D}{2}+a_1+z_1 \nonumber \\
    z_7 & = -\frac{D}{2}+a_2+z_2 \nonumber .
\end{align}

Substituting these into Eq.~\ref{sunset_mb_bracket} and applying Rule~D of~\cite{Prausa:2017frh} gives
\begin{align}\label{sunset_mb_final_1}
I(a_1,a_2,a_3)&=\frac{ (p^2)^{D-a_1-a_2-a_3} }{\Gamma(a_1)\Gamma(a_2)\Gamma(a_3)}
 \int\limits_{-i \infty}^{+i \infty} \frac{ d z_1}{2 \pi i}
  \int\limits_{-i \infty}^{+i \infty} \frac{ d z_2}{2 \pi i}
  \int\limits_{-i \infty}^{+i \infty}\frac{ d z_3}{2 \pi i}
\left( \frac{m^2_1}{p^2} \right)^{z_1} \left( \frac{m^2_2}{p^2} \right)^{z_2} \left( \frac{m^2_3}{p^2} \right)^{z_3} \nonumber \\
& \times \frac{\Gamma(-z_1)\Gamma(-z_2)\Gamma(-z_3)\, \Gamma\!\left(\frac{D}{2}-a_1-z_1\right)\Gamma\!\left(\frac{D}{2}-a_2-z_2\right)\Gamma\!\left(\frac{D}{2}-a_3-z_3\right)}{\Gamma\!\left(\frac{3D}{2}-a_1-a_2-a_3-z_1-z_2-z_3\right)} \nonumber \\ 
& \times \Gamma\!\left(-D+a_1+a_2+a_3+z_1+z_2+z_3\right) .
\end{align}

One can similarly choose another set of free variables to obtain a different MB integral. However, these MB integrals are all equivalent up to a change of variables. For example, if we take $z_1, z_2 , z_5$ as the free variables, we obtain
\begin{align}\label{sunset_mb_final_2}
I(a_1,a_2,a_3)&=\frac{ (p^2)^{\frac{D}{2}-a_1-a_2} \,(m^2_3)^{-\!a_3+\frac{D}{2}} }{\Gamma(a_1)\Gamma(a_2)\Gamma(a_3)}
 \int\limits_{-i \infty}^{+i \infty} \frac{ d z_1}{2 \pi i}
  \int\limits_{-i \infty}^{+i \infty} \frac{ d z_2}{2 \pi i}
  \int\limits_{-i \infty}^{+i \infty}\frac{ d z_5}{2 \pi i}
\left( \frac{m^2_1}{p^2} \right)^{z_1} \left( \frac{m^2_2}{p^2} \right)^{z_2} \left( \frac{m^2_3}{p^2} \right)^{z_5} \nonumber \\
& \times \frac{\Gamma(-z_1)\Gamma(-z_2)\Gamma(-z_5)\, \Gamma\!\left(\frac{D}{2}-a_1-z_1\right)\Gamma\!\left(\frac{D}{2}-a_2-z_2\right)\Gamma\!\left(-\frac{D}{2}+a_3-z_5\right)}{\Gamma\!\left(D-a_1-a_2-z_1-z_2-z_5\right)} \nonumber \\ 
& \times \Gamma\!\left(-\frac{D}{2}+a_1+a_2+z_1+z_2+z_5\right) ,
\end{align}
which is the same as the MB representation in Eq.~\ref{sunset_mb_final_1} after the change of variables $z_5 \mapsto -\tfrac{D}{2} + a_3 + z_3$.

\section{Sunsets' hypergeometric representations\label{SunsetHyperRep2}}

\renewcommand{\theequation}{B-\arabic{equation}}
  \setcounter{equation}{0}  

In this appendix, we provide explicit series solutions of various MB representations in the main text, which are associated with different configurations of the sunset integral. These solutions were obtained using \texttt{MBConicHulls} and expressed in terms of known hypergeometric and Kamp\'e de Feri\'et functions using \texttt{Olsson}.

The Kamp\'e de Feri\'et function is defined as:
\begin{align}
    F^{p,q,r}_{p',q',r'} & \left[ \begin{array}{c}
    a_1, a_2, ... , a_p: b_1, b_2, ... , b_q: c_1, c_2, ..., c_r \\
    a'_1, a'_2, ... , a'_{p'}: b'_1, b'_2, ... , b'_{q'}: c'_1, c'_2, ..., c'_{r'} 
\end{array}	\bigg| x, y  \right] \nonumber \\
& = \sum_{m,n=0}^{\infty}
\frac{(a_1)_m (a_2)_m ... (a_p)_m}{(a'_1)_m (a'_2)_m ... (a'_{p'})_m} 
\frac{(b_1)_n (b_2)_n ... (b_q)_n}{(b'_1)_n (b'_2)_n ... (b'_{q'})_n} 
\frac{(c_1)_{m+n} (c_2)_{m+n} ... (c_r)_{m+n} }{(c'_1)_{m+n} (c'_2)_{m+n} ... (c'_{r'})_{m+n}}
\frac{x^m}{m!} \frac{y^n}{n!}
\end{align}

For the MB integral in Eq.~\ref{MBrepH211mM}, we have two series solutions.
\begin{flalign}
   & H^{\text{Sol1}}_{\{2,1,1\}}(m,M,M;m^2) = \frac{(m^2 M^2)^{-\epsilon} \pi^{\frac{7}{2}-2\epsilon}}{\Gamma(2-2\epsilon) \Gamma(2-\epsilon) \Gamma\left(\tfrac32+\epsilon\right)} \Bigg\{ \nonumber \\
    & \bigg(\frac{m^2}{M^2}\bigg)^{\epsilon} \pi
  \Gamma(2-2\epsilon) \Gamma(-\epsilon) \Gamma(2\epsilon) \Gamma(1+\epsilon)
  {}_3F_2 \left( \begin{array}{c}
	   1, \tfrac{3}{2}, 2\epsilon \\
	   2-\epsilon, \tfrac{3}{2}+\epsilon  \\
	\end{array}	\bigg| \frac{m^2}{M^2} \right) \nonumber \\
    +& 4 \Gamma(1-\epsilon) \Gamma(\tfrac{3}{2}-\epsilon) \Gamma(2-\epsilon) \Gamma(\epsilon)^2 \Gamma\bigg(\tfrac32+\epsilon\bigg)
  {}_3F_2 \left( \begin{array}{c}
	1, \tfrac{3}{2}-\epsilon, \epsilon \\
	\tfrac{3}{2}, 2-2\epsilon \\
	\end{array}	\bigg| \frac{m^2}{M^2} \right) 
\Bigg\}
&& \end{flalign}
which converges for $\left|\frac{m^2}{M^2}\right|<1$.

\begin{flalign}
& H^{\text{Sol2}}_{\{2,1,1\}}(m,M,M;m^2) = 4^{-\epsilon}(m^2M^2)^{-\epsilon} \pi^{-2\epsilon} \Bigg\{ \nonumber \\
& \pi^5 \bigg(\frac{m^2}{M^2}\bigg)^{\epsilon-\tfrac{3}{2}} 
\frac{\Gamma(\tfrac32-\epsilon) \Gamma(-\tfrac{1}{2}+\epsilon) \Gamma(-\tfrac{3}{2}+2\epsilon)}{\Gamma(\tfrac12-\epsilon) \Gamma(\epsilon)}
{}_2F_1 \left( \begin{array}{c}
	1-\epsilon,\tfrac12+\epsilon \\
	\tfrac52-2\epsilon \\
\end{array}	\bigg| \frac{m^2}{M^2} \right) \nonumber \\
-& \pi^4 \bigg(\frac{m^2}{M^2}\bigg)^{-\epsilon} \frac{\Gamma(1-2\epsilon) \Gamma(\tfrac32-2\epsilon) \Gamma(1-\epsilon) \Gamma(\epsilon) \Gamma(2\epsilon)}{\Gamma(2-3\epsilon) \Gamma(\tfrac32-\epsilon)}
{}_2F_1 \left( \begin{array}{c}
	-\tfrac12+\epsilon, -1+3\epsilon \\
	-\tfrac12+2\epsilon \\
\end{array}	\bigg| \frac{m^2}{M^2} \right) \nonumber \\
-& \pi^{\tfrac{7}{2}} \bigg( \frac{m^2}{M^2} \bigg)^{-1} \frac{ \Gamma(\tfrac12-\epsilon) \Gamma(-1+\epsilon) \Gamma(-\epsilon) \Gamma(1+\epsilon)}{\Gamma(1-2\epsilon)}
{}_3F_2 \left( \begin{array}{c}
	 \tfrac{1}{2}, 1, 2\epsilon \\
	 2-\epsilon, \tfrac{1}{2}+\epsilon \\
\end{array}	\bigg| \frac{m^2}{M^2} \right) \nonumber \\
-& \pi^{\tfrac{9}{2}} \bigg(\frac{m^2}{M^2}\bigg)^{-1+\epsilon} 
\frac{\Gamma(\epsilon) \Gamma(-1+2\epsilon)}{\Gamma(\tfrac{1}{2}+\epsilon)}
{}_3F_2 \left( \begin{array}{c}
	 1, \tfrac{1}{2}-\epsilon, \epsilon \\
	 \tfrac{1}{2}, 2-2\epsilon \\
\end{array}	\bigg| \frac{m^2}{M^2} \right)
\Bigg\}
&& \end{flalign}
which converges for $\left|\frac{m^2}{M^2}\right|>1$.


For the MB integral in Eq.~\ref{MBrepH111v2}, we have four series solutions.

\begin{flalign}
& H^{\text{Sol1}}_{\{1,1,1\}}(m_1,m_2,m_3; m_1^2) \nonumber \\
& = \pi^{4-2\epsilon} (m_2^2)^{-\epsilon} (m_3^2)^{1-\epsilon}
\frac{\Gamma(1-\epsilon) \Gamma(\epsilon-1) \Gamma(\epsilon)}{\Gamma(2-\epsilon)}
F^{2:1:0}_{0:2:1} \left[ \begin{array}{c}
    1,\epsilon : \tfrac{1}{2} : - \\
    - : 2-\epsilon,\epsilon : 2-\epsilon
\end{array}	\bigg| \frac{4 m_1^2}{m_2^2}, \frac{m_3^2}{m_2^2} \right] \nonumber \\
& + \pi^{4-2\epsilon} m_1^2 (m_2^2)^{-2\epsilon} \frac{\Gamma(1-\epsilon)^2 \Gamma(\epsilon) \Gamma(2\epsilon-1)}{\Gamma(2-\epsilon)} 
F^{2:0:1}_{0:1:2} \left[ \begin{array}{c}
	\epsilon, 2\epsilon-1 : - : \tfrac{1}{2} \\
	- : \epsilon : 2-\epsilon, \epsilon 
\end{array}	\bigg| \frac{m_3^2}{m_2^2}, \frac{4 m_1^2}{m_2^2} \right] \nonumber \\
& + \pi^{4-2\epsilon} (m_1^2)^{1-\epsilon} (m_2^2)^{-\epsilon}
\frac{\Gamma(1-\epsilon) \Gamma(\epsilon-1) \Gamma(\epsilon)}{\Gamma(2-\epsilon)}
F^{2:0:1}_{0:1:2} \left[ \begin{array}{c}
    1,\epsilon: - : \tfrac{3}{2}-\epsilon \\
    - : \epsilon : 3-2\epsilon, 2-\epsilon
\end{array}	\bigg| \frac{m_3^2}{m_2^2}, \frac{4 m_1^2}{m_2^2} \right] \nonumber \\
& + \pi^{4-2\epsilon} (m_1^2)^{1-\epsilon} (m_2^2)^{-\epsilon} \Gamma(\epsilon-1)^2
F^{2:1:0}_{0:2:1} \left[ \begin{array}{c}
    1,2-\epsilon : \tfrac32-\epsilon : - \\
    - : 3-2\epsilon,2-\epsilon : 2-\epsilon
\end{array}	\bigg| \frac{m_3^2}{m_2^2}, \frac{4 m_1^2}{m_2^2} \right]
&& \end{flalign}

\begin{flalign}
& H^{\text{Sol3}}_{\{1,1,1\}}(m_1,m_2,m_3; m_1^2) \nonumber \\
& = \pi^{4-2\epsilon} (m_1^2)^{-\tfrac{1}{2}}(m_2^2)^{\tfrac{3}{2}-2\epsilon} \frac{\Gamma(3-4\epsilon) \Gamma(1-\epsilon)^2 \Gamma(\epsilon) \Gamma(2\epsilon-1)}{\Gamma(3-3\epsilon) \Gamma(2-2\epsilon)}
F^{3:0:0}_{1:1:1} \left[ \begin{array}{c}
    \epsilon, 2\epsilon-1, 3\epsilon-2 : - : - \\
    2\epsilon-\tfrac{1}{2} : \epsilon : \epsilon
\end{array}	\bigg| \frac{m_3^2}{4 m_1^2}, \frac{m_2^2}{4 m_1^2} \right] \nonumber \\
& + \frac{1}{2} \pi^{4-2\epsilon} (m_1^2)^{-\tfrac{1}{2}}(m_2^2)^{\tfrac{3}{2}-2\epsilon} \Gamma(1-\epsilon) \Gamma (\epsilon-\tfrac{1}{2}) \Gamma(2\epsilon-\tfrac{3}{2})
F^{3:0:0}_{0:1:0} \left[ \begin{array}{c}
    \tfrac{1}{2}+\epsilon, \tfrac{3}{2}-\epsilon, \epsilon-\tfrac{1}{2} : - : - \\
    - : \epsilon : -
\end{array}	\bigg| \frac{m_3^2}{m_2^2}, \frac{m_2^2}{4 m_1^2} \right] \nonumber \\
& + \pi^{4-2\epsilon} (m_1^2)^{1-\epsilon} (m_2^2)^{-\epsilon}
\frac{\Gamma(1-2\epsilon) \Gamma(\epsilon-1) \Gamma(\epsilon)}{\Gamma(2-2\epsilon)}
F^{3:0:0}_{0:1:0} \left[ \begin{array}{c}
    1,\epsilon,-1+2\epsilon : - : - \\
    \tfrac{1}{2}+\epsilon : \epsilon : 2-\epsilon
\end{array}	\bigg| \frac{m_3^2}{4m_1^2}, \frac{m_2^2}{4m_1^2} \right] \nonumber \\
& + \pi^{4-2\epsilon} (m_1^2)^{-1} (m_2^2)^{1-\epsilon} (m_3^2)^{1-\epsilon} \frac{\Gamma(1-2\epsilon) \Gamma(\epsilon-1) \Gamma(\epsilon)}{\Gamma(2-2\epsilon)}
F^{3:0:0}_{1:1:1} \left[ \begin{array}{c}
    1,\epsilon,-1+2\epsilon : - : - \\
    \tfrac{1}{2}+\epsilon : \epsilon : 2-\epsilon
\end{array}	\bigg| \frac{m_2^2}{4 m_1^2}, \frac{m_3^2}{4 m_1^2} \right] \nonumber \\
& - \frac{1}{2} \pi^{\tfrac{9}{2}-2\epsilon} (m_1^2)^{-\epsilon} (m_2^2)^{1-\epsilon} \Gamma(\epsilon) \Gamma(2\epsilon-1)
\widetilde F\!\Bigl[
  \{\tfrac12,\,-\tfrac12+\epsilon\},\{\},\{\tfrac12,\,2-\epsilon\}, \{\},\{\},
  \bigl(\tfrac{m_{3}^{2}}{m_{2}^{2}},\,\tfrac{m_{2}^{2}}{m_{1}^{2}}\bigr)
\Bigr] \nonumber \\
& - \frac{1}{2} \pi^{4-2\epsilon} (m_2^2)^{1-2\epsilon} \frac{\Gamma(2-\epsilon) \Gamma(\epsilon-1)^2}{\Gamma(1-\epsilon)}
F^{3:0:0}_{1:1:1} \left[ \begin{array}{c}
    1,2-\epsilon,\epsilon : - : - \\
    \tfrac{3}{2} : 2-\epsilon : 2-\epsilon
\end{array}	\bigg| \frac{m_3^2}{4m_1^2}, \frac{m_2^2}{4m_1^2} \right]
&& \end{flalign}

The second and fourth series solutions can be obtained by performing $m_2 \leftrightarrow m_3$ and $m_1 \leftrightarrow m_3$ on the first and third solutions, respectively.

For the MB integral in Eq.~\ref{MBrepH211v2}, we have four series solutions.

\begin{flalign}
& H^{\text{Sol1}}_{\{2,1,1\}}(m_1,m_2,m_3; m_3^2) \nonumber \\
& = \pi^{4-2\epsilon} (m_2^2)^{-2\epsilon} \frac{\Gamma(1-\epsilon) \Gamma(-\epsilon) \Gamma(2\epsilon) \Gamma(1+\epsilon)}{\Gamma(2-\epsilon)}
F^{2:0:1}_{0:1:3} \left[ \begin{array}{c}
    2\epsilon, 1+\epsilon : - : \tfrac{3}{2} \\
    - : \epsilon : 2,2-\epsilon,1+\epsilon
\end{array}	\bigg| \frac{m_3^2}{m_2^2}, \frac{4 m_2^2}{m_2^2} \right] \nonumber \\
& - \pi^{4-2\epsilon} (m_2^2)^{-1-\epsilon} (m_3^2)^{1-\epsilon}
\frac{\Gamma(\epsilon-1) \Gamma(-\epsilon) \Gamma(1+\epsilon)}{\Gamma(2-\epsilon)}
F^{2:1:0}_{0:3:1} \left[ \begin{array}{c}
    2,1+\epsilon : \tfrac{3}{2} : - \\
    - : 2, 2-\epsilon, 1+\epsilon : 2-\epsilon
\end{array}	\bigg| \frac{4m_2^2}{m_2^2}, \frac{m_3^2}{m_2^2} \right] \nonumber \\
& + \pi^{4-2\epsilon} (m_1^2)^{-\epsilon} (m_2^2)^{-\epsilon}
\frac{\Gamma(1-\epsilon)\Gamma(\epsilon)^2}{\Gamma(2-\epsilon)}
F^{2:0:1}_{0:1:2} \left[ \begin{array}{c}
    1,\epsilon : - : \tfrac{3}{2}-\epsilon \\
    - : \epsilon : 2-2\epsilon,2-\epsilon
\end{array}	\bigg| \frac{m_3^2}{m_2^2}, \frac{4 m_2^2}{m_2^2} \right] \nonumber \\
& + \pi^{4-2\epsilon} (m_1^2)^{-\epsilon} (m_2^2)^{-\epsilon} \Gamma(\epsilon-1) \Gamma(\epsilon)
F^{2:1:0}_{0:2:1} \left[ \begin{array}{c}
    1, 2-\epsilon : \tfrac{3}{2}-\epsilon : - \\
    - : 2-2\epsilon, 2-\epsilon : 2-\epsilon
\end{array}	\bigg| \frac{4m_2^2}{m_2^2}, \frac{m_3^2}{m_2^2} \right] 
&& \end{flalign}

\begin{flalign}
&H^{\text{Sol3}}_{\{2,1,1\}}(m_1,m_2,m_3;m_3^2) \nonumber \\
& = \pi^{4-2\epsilon} (m_2^2)^{-2\epsilon}
\frac{\Gamma(2-4\epsilon) \Gamma(1-\epsilon)^2 \Gamma(\epsilon) \Gamma(2\epsilon)}{\Gamma(2-3\epsilon) \Gamma(2-2\epsilon)}
F^{3:0:0}_{1:1:1} \left[ \begin{array}{c}
    \epsilon, 2\epsilon-1, 3\epsilon-1 : - : - \\
    2\epsilon-\tfrac{1}{2} : \epsilon : \epsilon
\end{array}	\bigg| \frac{m_3^2}{4m_1^2}, \frac{m_2^2}{4m_1^2} \right] \nonumber \\
& - \pi^{4-2\epsilon} (m_2^2)^{-1-\epsilon} (m_3^2)^{1-\epsilon}
\frac{\Gamma(\epsilon-1) \Gamma(-\epsilon) \Gamma(1+\epsilon)}{\Gamma(2-\epsilon)}
F^{2:1:0}_{0:3:1} \left[ \begin{array}{c}
    2,\,1+\epsilon : \tfrac{3}{2} : - \\
    - : 2,\,2-\epsilon,\,1+\epsilon : 2-\epsilon
\end{array}	\bigg| \frac{4m_2^2}{m_1^2}, \frac{m_3^2}{m_1^2} \right] \nonumber \\
& + \frac{1}{2} \pi^{\tfrac{9}{2}-2\epsilon}
(m_1^2)^{-2} (m_2^2)^{1-\epsilon} (m_3^2)^{1-\epsilon}
\Gamma(\epsilon) \Gamma(2\epsilon-1)
\;\widetilde F\!\Bigl[
  \{\tfrac12,\,-\tfrac12+\epsilon\},\{\},\{\tfrac12,\,2-\epsilon\},\{\},\{\}\;
  \big|\;\tfrac{m_{3}^{2}}{m_{2}^{2}},\,\tfrac{m_{2}^{2}}{m_{1}^{2}}
\Bigr] \nonumber \\
& - \frac{1}{4} \pi^{4-2\epsilon} (m_1^2)^{-1} (m_2^2)^{1-2\epsilon} (m_3^2)^{1-\epsilon}
\frac{\Gamma(2-\epsilon) \Gamma(\epsilon-1)^2} {\Gamma(-\epsilon)}
F^{3:0:0}_{1:1:1} \left[ \begin{array}{c}
    1, 2-\epsilon, 1+\epsilon : - : - \\
    \tfrac{3}{2} : 2-\epsilon : 2-\epsilon
\end{array}	\bigg| \frac{m_3^2}{4m_1^2}, \frac{m_2^2}{4m_1^2} \right] \nonumber \\
& + \pi^{4-2\epsilon} (m_1^2)^{-\epsilon} (m_2^2)^{-\epsilon}
\frac{\Gamma(1-\epsilon) \Gamma(\epsilon)^2}{\Gamma(2-\epsilon)}
F^{2:0:1}_{0:1:2} \left[ \begin{array}{c}
    1, \epsilon : - : \tfrac{3}{2}-\epsilon \\
    - : \epsilon : 2-2\epsilon, 2-\epsilon
\end{array}	\bigg| \frac{m_3^2}{m_2^2}, \frac{4m_1^2}{m_2^2} \right] \nonumber \\
& - \pi^{4-2\epsilon} (m_1^2)^{-\epsilon} (m_2^2)^{-\epsilon} \Gamma(\epsilon-1) \Gamma(\epsilon)
F^{2:1:0}_{0:2:1} \left[ \begin{array}{c}
    1,\,2-\epsilon : \tfrac32-\epsilon : - \\
    - : 2-2\epsilon, 2-\epsilon : 2-\epsilon
\end{array}	\bigg| \frac{4m_1^2}{m_2^2}, \frac{m_3^2}{m_2^2} \right]
&& \end{flalign}

The second and fourth series solutions can be obtained by performing $m_2 \leftrightarrow m_3$ and $m_1 \leftrightarrow m_3$ on the first and third solutions, respectively.

For the MB integral in Eq.~\ref{MBrepH111m1m2m2m3v2}, we have three series solutions.

\begin{flalign}
& H^{\text{Sol1}}_{\{1,1,1\}}(m_1,m_2,m_2; m_3^2) \nonumber \\
& = 2^{1-2\epsilon} \pi^{\tfrac{9}{2}-2\epsilon} (m_1^2)^{-2\epsilon} (m_2^2)
\frac{\Gamma(1-\epsilon) \Gamma(\epsilon) \Gamma(2\epsilon-1)}{\Gamma(2-\epsilon) \Gamma(\tfrac{1}{2}+\epsilon)}
F^{3:0:0}_{1:1:1} \left[ \begin{array}{c}
    1, \epsilon, 2\epsilon-1 : - : - \\
    \tfrac{1}{2}+\epsilon : 2-\epsilon, \epsilon : \epsilon
\end{array}	\bigg| \frac{m_3^2}{4m_2^2}, \frac{m_1^2}{4m_2^2} \right] \nonumber \\
& + \pi^{4-2\epsilon} (m_1^2)^{1-\epsilon} (m_2^2)^{-\epsilon} \bigg(\frac{m_2^2}{m_1^2}\bigg)^{\epsilon}
\Gamma(\epsilon-1) \Gamma(\epsilon) 
F^{3:0:0}_{1:2:0} \left[ \begin{array}{c}
    1, 2-\epsilon, \epsilon : - : - \\
    \tfrac{3}{2} : 2-\epsilon,2-\epsilon : -
\end{array}	\bigg| \frac{m_3^2}{4m_2^2}, \frac{m_1^2}{4m_2^2} \right] 
&& \end{flalign}

\begin{flalign}
& H^{\text{Sol2}}_{\{1,1,1\}}(m_1,m_2,m_2; m_3^2) \nonumber \\ 
& = \pi^{\tfrac{9}{2}-2\epsilon} 2^{2\epsilon-1}
(m_1^2)^{1-2\epsilon} (m_2^2)^{3\epsilon}
\frac{\Gamma(2-2\epsilon) \Gamma(1-\epsilon) \Gamma(\epsilon) \Gamma(2\epsilon-1)}
     {\Gamma(\tfrac{3}{2}-\epsilon) \Gamma(2-\epsilon)}
F^{2:1:0}_{0:1:2} \left[ \begin{array}{c}
    \epsilon, 2\epsilon-1 : \epsilon-\tfrac{1}{2} : - \\
    - : 2-\epsilon : \epsilon, 2\epsilon-1
\end{array}	\bigg| \frac{m_3^2}{m_2^2}, \frac{4m_2^2}{m_2^2} \right] \nonumber \\
& + \pi^{\tfrac{9}{2}-2\epsilon} 2^{1-2\epsilon}  (m_1^2)^{-2\epsilon} (m_2^2)
\frac{ \Gamma(1-\epsilon) \Gamma(\epsilon) \Gamma(2\epsilon-1)}
     {\Gamma(2-\epsilon) \Gamma(\tfrac{1}{2}+\epsilon)}
F^{3:0:0}_{1:2:1} \left[ \begin{array}{c}
    1, \epsilon, 2\epsilon-1 : - : - \\
    \tfrac{1}{2}+\epsilon : 2-\epsilon,\epsilon : \epsilon
\end{array}	\bigg| \frac{m_3^2}{4m_1^2}, \frac{m_2^2}{4m_1^2} \right] \nonumber \\
& + \pi^{4-2\epsilon} (m_2^2)^{-2\epsilon}
\frac{\Gamma(1-\epsilon) \Gamma(\epsilon)^2}{\Gamma(2-\epsilon)}
F^{2:0:1}_{0:1:2} \left[ \begin{array}{c}
    1, \epsilon : - : \tfrac{1}{2} \\
    - : 2-\epsilon : 2-\epsilon, \epsilon
\end{array}	\bigg| \frac{m_3^2}{m_2^2}, \frac{4m_2^2}{m_2^2} \right] 
&& \end{flalign}

\begin{align}
&H^{\text{Sol3}}_{\{2,1,1\}}(m_1,m_2,m_2;m_3^2) \nonumber \\
& = \pi^{\tfrac{9}{2}-2\epsilon} 2^{3-2\epsilon} 
(m_2^2)^{1-3\epsilon} (m_3^2)^{\epsilon-1}
\frac{\Gamma(\epsilon-1)\,\Gamma(2-1+\epsilon)}{\Gamma(\epsilon-\tfrac{1}{2})}
F^{2:1:0}_{0:2:1} \left[ \begin{array}{c}
    1, \epsilon : \tfrac{3}{2}-\epsilon : - \\
    - : 3-2\epsilon, 2-\epsilon : \epsilon
\end{array}	\bigg| \frac{4m_2^2}{m_3^2}, \frac{m_1^2}{m_3^2} \right] \nonumber \\
& + \pi^{4-2\epsilon} 2^{4-2\epsilon} (m_1^2)^{-3\epsilon} (m_3^2)^{\epsilon-1}
\frac{\Gamma(1-2\epsilon) \Gamma(\epsilon-1)^2 \Gamma(\epsilon)}{\Gamma(2-2\epsilon)}
\;\widetilde F\!\Bigl[
  \{\;1,\epsilon\},\{\tfrac12\},\{\},\{\},\{2-\epsilon\},\{2-\epsilon\};
  \tfrac{4m_2^2}{m_3^2}, \tfrac{m_1^2}{m_3^2}
\Bigr] \nonumber \\
& + 2^{2\epsilon-1} \pi^{\tfrac{9}{2}-2\epsilon} (m_1^2)^{-3\epsilon} (m_3^2)^{\epsilon-1}
\frac{\Gamma(2-2\epsilon) \Gamma(1-\epsilon)^2 \Gamma(2\epsilon-1)}{\Gamma(\tfrac32-\epsilon) m_3^2}
F^{2:1:0}_{0:1:2} \left[ \begin{array}{c}
    \epsilon, 2\epsilon-1 : \epsilon-\tfrac{1}{2} : - \\
    - : 2-\epsilon : \epsilon, 2\epsilon-1
\end{array}	\bigg| \frac{4m_2^2}{m_3^2}, \frac{m_1^2}{m_3^2} \right] \nonumber \\
& + \pi^{\tfrac{9}{2}-2\epsilon} 2^{2\epsilon-1} (m_1^2)^{-3\epsilon} (m_3^2)^{\epsilon-1}
\frac{\Gamma(1-\epsilon) \Gamma(\epsilon-1) \Gamma(\epsilon)}{\Gamma(2-2\epsilon) \Gamma(\tfrac{3}{2}-\epsilon)}
\;\widetilde F\!\Bigl[
  \{\tfrac12,-\tfrac12+\epsilon\},\{\},\{\tfrac12,2-\epsilon\},\{\},\{\epsilon\},\{-1+2\epsilon\};
  \tfrac{4m_2^2}{m_3^2},\,\tfrac{m_1^2}{m_3^2}
\Bigr] \nonumber \\
& + \pi^{\tfrac{9}{2}-2\epsilon} 2^{2\epsilon-1}
(m_1^2)^{-1} (m_2^2)^{1-3\epsilon} (m_3^2)^{\epsilon-1}
\frac{\Gamma(1-\epsilon) \Gamma(\epsilon-1) \Gamma(\epsilon)}
     {\Gamma(2-2\epsilon) \Gamma(\tfrac{3}{2}-\epsilon)}
F^{2:1:0}_{0:2:1} \left[ \begin{array}{c}
    \epsilon, 2\epsilon-1 : \tfrac12 : - \\
    - : \epsilon, 2\epsilon-1 : 2-\epsilon
\end{array}	\bigg| \frac{4m_2^2}{m_3^2}, \frac{m_1^2}{m_3^2} \right]
&& \end{align}

\section{Residues of $H_{\{1,1,1\}} \left(m_1, m_2, m_3; m_1^2 \right)$ \label{Sec:H111kResults}}

\renewcommand{\theequation}{C-\arabic{equation}}
  \setcounter{equation}{0}  

\subsection{Residues of integral $I_1$}

\begin{flalign}
    I_1 = (m_1^2)^{1-2\epsilon} \pi^{4-2\epsilon} \frac{\Gamma(3-4\epsilon) \Gamma^2(1-\epsilon) \Gamma(\epsilon)  \Gamma(2\epsilon-1)}{\Gamma(3-3\epsilon) \Gamma(2-2\epsilon)} &&
\end{flalign}

\subsection{Residues of integral $I_2$}

\subsubsection*{LHS}

\begin{flalign}
    I_2^{L,1} = -(m_1^2)^{1-2\epsilon} \pi^{4-2\epsilon} \frac{\Gamma(3-4\epsilon) \Gamma^2(1-\epsilon) \Gamma(\epsilon)  \Gamma(2\epsilon-1)}{\Gamma(3-3\epsilon) \Gamma(2-2\epsilon)} &&
\end{flalign}

\begin{flalign}
	I_2^{L,2} = (m_1^2)^{1-2\epsilon} \pi^{4-2\epsilon} \left( \frac{m_3^2}{m_1^2}  \right)^{-\epsilon} \frac{\Gamma(1-\epsilon) \Gamma(\epsilon-1) \Gamma(\epsilon)}{\Gamma(2-\epsilon)}
    {}_3F_2 \left( \begin{array}{c}
		1, \frac{3}{2}-\epsilon, \epsilon \\
		3-2\epsilon, 2-\epsilon \\
	\end{array}	\bigg| \frac{4m_1^2}{m_3^2} \right) &&
\end{flalign}

\begin{flalign}
	I_2^{L,3} = -2 m_3^2 (m_1^2)^{-2\epsilon} 
    \pi^{4-2\epsilon} \left( \frac{m_3^2}{m_1^2} \right)^{-2\epsilon} 
    \Gamma(\epsilon) \Gamma(1-\epsilon) \Gamma(2\epsilon-2)
    {}_2F_1 \left( \begin{array}{c}
		\frac{1}{2}, 2\epsilon-1 \\
		2-\epsilon \\
	\end{array}	\bigg| \frac{4m_1^2}{m_3^2} \right) &&
\end{flalign}

\subsubsection*{RHS}

\begin{flalign}
	I_2^{R,1} = m_3^2 (4 m_1^2)^{-2\epsilon} \pi^{\tfrac{7}{2}-2\epsilon} 
    \frac{\Gamma\left(\tfrac{1}{2}-2\epsilon\right) \Gamma^2(1-\epsilon) \Gamma(\epsilon) \Gamma(2\epsilon)}{\Gamma(2-3\epsilon)}
    {}_3F_2 \left( \begin{array}{c}
		2\epsilon, 3\epsilon-1, 1 \\
		2, \frac{1}{2}+2\epsilon \\
	\end{array}	\bigg| \frac{m_3^2}{4 m_1^2} \right) &&
\end{flalign}

\begin{flalign}
    I_2^{R,2} = 2m_3^2 (2m_1^2)^{-2\epsilon} \pi^{\tfrac{9}{2}-2\epsilon}
    \left(\frac{m_3^2}{m_1^2}\right)^{-\epsilon}
    \frac{\Gamma(1-2\epsilon) \Gamma(\epsilon-1) \Gamma(2\epsilon)}{\Gamma(2-2\epsilon) \Gamma\left(\tfrac{1}{2}+\epsilon\right)}
    {}_3F_2 \left( \begin{array}{c}
		\epsilon, 2\epsilon-1, 1 \\
		2-\epsilon, \frac{1}{2}+\epsilon \\
	\end{array}	\bigg| \frac{m_3^2}{4 m_1^2} \right) &&
\end{flalign}

\begin{flalign}
	I_2^{R,3} = \frac{1}{2} m_3^2 (m_1^2)^{-2\epsilon} \pi^{4-2\epsilon}
    \left(\frac{m_3^2}{m_1^2}\right)^{\tfrac{1}{2}-2\epsilon}
    \Gamma(1-\epsilon) \Gamma\left(\epsilon-\tfrac{1}{2}\right) \Gamma\left(2\epsilon-\tfrac{3}{2}\right) 
    {}_2F_1 \left( \begin{array}{c}
		\frac{1}{2}, \epsilon-\frac{1}{2} \\
		\frac{5}{2}-2\epsilon \\
	\end{array}	\bigg| \frac{m_3^2}{4 m_1^2} \right)
\end{flalign}

\subsection{Residues of integral $I_4$}

The residues for the integral $I_4$ are given below in terms of a multiplicative prefactor $c_{I} = \pi^{\frac{3}{2}} (4 m_1^2)^{1-2\epsilon} \pi^{\tfrac{7}{2}-2\epsilon}$. 

\subsubsection*{Cone 1}

\begin{flalign}
	I_4^{1,1} = c_{I} \sqrt{\pi } 2^{2 \epsilon -1} \left(\frac{m_2^2}{4 m_1^2}\right)^{1-\epsilon } \frac{\Gamma (\epsilon -1) \Gamma \left(\epsilon -\frac{1}{2}\right) \Gamma (\epsilon )}{\Gamma \left(\epsilon +\frac{1}{2}\right)}  {}_3F_2 \left( \begin{array}{c}
		1,\epsilon ,2 \epsilon -1 \\
		2-\epsilon ,\epsilon +\frac{1}{2} \\
	\end{array}	\bigg| \frac{m_2^2}{4 m_1^2} \right) &&
\end{flalign}

\begin{flalign}
	I_4^{1,2} = - c_{I} \left( \frac{m_2^2}{4 m_1^2} \right) \frac{\Gamma \left(\frac{1}{2}-2 \epsilon \right) \Gamma (2 \epsilon ) \Gamma (1-\epsilon )^2 \Gamma (\epsilon )}{\Gamma (2-3 \epsilon )}  {}_3F_2 \left( \begin{array}{c}
		2 \epsilon ,3 \epsilon -1,1 \\
		2,2 \epsilon +\frac{1}{2} \\
	\end{array}	\bigg| \frac{m_2^2}{4 m_1^2} \right) &&
\end{flalign}

\begin{flalign}
	I_4^{1,3} = - c_{I} \sqrt{\pi } \left(\frac{m_2^2}{4 m_1^2}\right)^{\frac{3}{2}-2 \epsilon} \Gamma (1-\epsilon ) \Gamma \left(\epsilon -\frac{1}{2}\right) \Gamma \left(2 \epsilon -\frac{3}{2}\right) {}_2F_1 \left( \begin{array}{c}
		\frac{1}{2},\epsilon -\frac{1}{2} \\
		\frac{5}{2}-2 \epsilon \\
	\end{array}	\bigg| \frac{m_2^2}{4 m_1^2} \right) 
&& \end{flalign}

\begin{flalign}
	I_4^{1,4} &= - c_{I} \left(\frac{m_2^2}{4 m_1^2}\right)^{1-\epsilon } \left(\frac{m_3^2}{4 m_1^2}\right)^{-\epsilon } \Gamma (\epsilon )^2 \Gamma (1-\epsilon )^2 \nonumber \\
	& \times \sum_{m,n=0}^{\infty} \left(\frac{m_2^2}{m_3^2}\right)^m \left(\frac{4 m_1^2}{m_3^2}\right)^n \frac{\Gamma \left(n+\frac{1}{2}\right) \Gamma (m+n+1) \Gamma (m+n+\epsilon )}{\Gamma (m+1) \Gamma (n+1) \Gamma (m-\epsilon +2) \Gamma (n-\epsilon +2) \Gamma (n+\epsilon )} &&
\end{flalign}

\begin{flalign}
	I_4^{1,5} & = c_{I} \left( \frac{m_2^2}{m_3^2} \right) \left(\frac{m_2^2}{4 m_1^2}\right)^{-\epsilon} \Gamma (\epsilon )^2 \Gamma (1-\epsilon )^2 \nonumber \\
	\times & \sum_{m,n=0}^{\infty} \left(\frac{m_2^2}{m_3^2}\right)^m \left(\frac{4 m_1^2}{m_3^2}\right)^n \frac{\Gamma (m+n+1) \Gamma \left(n-\epsilon +\frac{3}{2}\right) \Gamma (m+n-\epsilon +2)}{\Gamma (m+1) \Gamma (n+1) \Gamma (m-\epsilon +2) \Gamma (n-2 \epsilon +3) \Gamma (n-\epsilon +2)}
&& \end{flalign}

\begin{flalign}
	I_4^{1,6} &= -c_{I} \left( \frac{m_2^2}{m_3^2} \right) \left(\frac{m_3^2}{4 m_1^2}\right)^{-\epsilon} \Gamma (\epsilon )^2 \Gamma (1-\epsilon )^2 \nonumber \\
	\times & \sum_{m,n=0}^{\infty} \left(\frac{m_2^2}{m_3^2}\right)^m \left(\frac{4 m_1^2}{m_3^2}\right)^n \frac{\Gamma (m+n+2) \Gamma \left(n-\epsilon +\frac{3}{2}\right) \Gamma (m+n+\epsilon +1)}{\Gamma (m+2) \Gamma (n+1) \Gamma (m+\epsilon +1) \Gamma (n-2 \epsilon +3) \Gamma (n-\epsilon +2)} \label{EqC16}
&& \end{flalign}

\begin{flalign}
	I_4^{1,7} &= c_{I} \left( \frac{m_2^2}{4 m_1^2} \right) \left(\frac{m_3^2}{4 m_1^2}\right)^{-2\epsilon} \Gamma (\epsilon )^2 \Gamma (1-\epsilon )^2 \nonumber \\
	& \times  \sum_{m,n=0}^{\infty} \left(\frac{m_2^2}{m_3^2}\right)^m \left(\frac{4 m_1^2}{m_3^2}\right)^n \frac{\Gamma \left(n+\frac{1}{2}\right) \Gamma (m+n+\epsilon +1) \Gamma (m+n+2 \epsilon )}{\Gamma (m+2) \Gamma (n+1) \Gamma (m+\epsilon +1) \Gamma (n-\epsilon +2) \Gamma (n+\epsilon )} \label{EqC17}
&& \end{flalign}

\subsubsection*{Cone 2}

\begin{flalign}
    I_4^{3,1} = c_{I} (m_1^2)^{1-2\epsilon} \pi^{4-2\epsilon} \frac{\Gamma(3-4\epsilon) \Gamma^2(1-\epsilon) \Gamma(\epsilon)  \Gamma(2\epsilon-1)}{\Gamma(3-3\epsilon) \Gamma(2-2\epsilon)} &&
\end{flalign}

\begin{flalign}
    I_4^{3,2} = c_{I} 2 m_3^2 (m_1^2)^{-2\epsilon} \pi^{4-2\epsilon} \left( \frac{m_3^2}{m_1^2}  \right)^{-\epsilon} \frac{\Gamma(1-\epsilon) \Gamma(\epsilon-1) \Gamma(\epsilon)}{\Gamma(2-\epsilon)}
    {}_3F_2 \left( \begin{array}{c}
		1, \frac{3}{2}-\epsilon, \epsilon \\
		3-2\epsilon, 2-\epsilon \\
	\end{array}	\bigg| \frac{4m_1^2}{m_3^2} \right) &&
\end{flalign}

\begin{flalign}
    I_4^{3,3} = c_{I} (m_1^2)^{1-2\epsilon} \pi^{4-2\epsilon} \left( \frac{m_3^2}{m_1^2} \right)^{-\epsilon} \Gamma(\epsilon) \Gamma(1-\epsilon) \Gamma(2\epsilon-2) \Gamma(2-\epsilon)
    {}_2F_1 \left( \begin{array}{c}
		\frac{1}{2}, 2\epsilon-1 \\
		2-\epsilon \\
	\end{array}	\bigg| \frac{4m_1^2}{m_3^2} \right) &&
\end{flalign}

\subsubsection*{Cone 3}

\begin{flalign}
	I_4^{2,1} &= - c_{I} \sqrt{\pi} \left(\frac{m_2^2}{4 m_1^2}\right)^{\frac{3}{2}-2 \epsilon} \Gamma (1-\epsilon ) \Gamma \left(\epsilon -\frac{1}{2}\right) \Gamma \left(2 \epsilon -\frac{3}{2}\right)  {}_2F_1 \left( \begin{array}{c}
		\frac{1}{2},\epsilon -\frac{1}{2} \\
		\frac{5}{2}-2 \epsilon \\
	\end{array}	\bigg| \frac{m_2^2}{4 m_1^2} \right) 
&& \end{flalign}

\begin{flalign}
	& I_4^{2,2} = - c_{I} \pi \left(\frac{m_2^2}{4 m_1^2}\right)^{1-\epsilon} \left(\frac{m_3^2}{4 m_1^2}\right)^{\frac{1}{2}-\epsilon } \Gamma (\epsilon ) \Gamma (1-\epsilon ) \nonumber \\
	& \times \sum_{m,n=0}^{\infty} \left(\frac{m_3^2}{4 m_1^2}\right)^m \left(\frac{m_2^2}{4 m_1^2}\right)^n \frac{\Gamma \left(m+n+\frac{1}{2}\right) \Gamma \left(m+n-\epsilon +\frac{3}{2}\right) \Gamma \left(m+n+\epsilon -\frac{1}{2}\right)}{\Gamma \left(m+\frac{1}{2}\right) \Gamma (n+1) \Gamma (m+n+1) \Gamma \left(m-\epsilon +\frac{3}{2}\right) \Gamma (n-\epsilon +2)}	
&& \end{flalign}

\begin{flalign}
	& I_4^{2,3} = - c_{I} \left( \frac{m_2^2}{4 m_1^2} \right) \left(\frac{m_3^2}{4 m_1^2}\right)^{\frac{1}{2}-2 \epsilon } \Gamma \left(\frac{3}{2}-2 \epsilon \right) \Gamma \left(2 \epsilon -\frac{1}{2}\right) \Gamma (\epsilon ) \Gamma (1-\epsilon ) \nonumber \\
	& \times \sum_{m,n=0}^{\infty} \left(\frac{m_3^2}{4 m_1^2}\right)^m \left(\frac{m_2^2}{4 m_1^2}\right)^n \frac{\Gamma \left(m+n+\frac{1}{2}\right) \Gamma \left(m+n-\epsilon +\frac{3}{2}\right) \Gamma \left(m+n+\epsilon -\frac{1}{2}\right)}{\Gamma (n+2) \Gamma (m+n+1) \Gamma \left(m-2 \epsilon +\frac{3}{2}\right) \Gamma \left(m-\epsilon +\frac{1}{2}\right) \Gamma (n+\epsilon +1)}
&& \end{flalign}

\begin{flalign}
	& I_4^{2,4}= - c_{I} \pi \left(\frac{m_2^2}{4 m_1^2}\right)^{1-\epsilon} \left(\frac{m_3^2}{4 m_1^2}\right)^{1-\epsilon } \Gamma (\epsilon ) \Gamma (1-\epsilon ) \nonumber \\
	& \times \sum_{m,n=0}^{\infty} \left(\frac{m_3^2}{4 m_1^2}\right)^m \left(\frac{m_2^2}{4 m_1^2}\right)^n \frac{\Gamma (m+n+1) \Gamma (m+n-\epsilon +2) \Gamma (m+n+\epsilon )}{\Gamma (m+1) \Gamma (n+1) \Gamma \left(m+n+\frac{3}{2}\right) \Gamma (m-\epsilon +2) \Gamma (n-\epsilon +2)}
&& \end{flalign}

\begin{flalign}
	& I_4^{2,5} = c_{I} 2 \pi   \left(\frac{m_2^2}{4 m_1^2}\right) \left(\frac{m_3^2}{4 m_1^2}\right)^{1-\epsilon } \Gamma (\epsilon ) \Gamma (1-\epsilon ) \nonumber \\
	& \times \sum_{m,n=0}^{\infty} \left(\frac{m_3^2}{4 m_1^2}\right)^m \left(\frac{m_2^2}{4 m_1^2}\right)^n \frac{\Gamma (m+n+2) \Gamma (m+n+\epsilon +1) \Gamma (m+n+2 \epsilon )}{\Gamma (m+1) \Gamma (n+2) \Gamma (m-\epsilon +2) \Gamma (n+\epsilon +1) \Gamma \left(m+n+\epsilon +\frac{3}{2}\right)}
&& \end{flalign}

\begin{flalign}
	& I_4^{2,6} = c_{I} \left(\frac{m_2^2}{4 m_1^2}\right) \left(\frac{m_3^2}{4 m_1^2}\right) \frac{\Gamma \left(-2 \epsilon -\frac{1}{2}\right) \Gamma (1-\epsilon )^2 \Gamma (\epsilon )^2 \Gamma \left(2 \epsilon +\frac{3}{2}\right)}{\Gamma (1-3 \epsilon ) \Gamma (3 \epsilon )} \nonumber \\
	& \times \sum_{m,n=0}^{\infty} \left(\frac{m_3^2}{4 m_1^2}\right)^m \left(\frac{m_2^2}{4 m_1^2}\right)^n \frac{\Gamma (m+n+\epsilon +2) \Gamma (m+n+2 \epsilon +1) \Gamma (m+n+3 \epsilon )}{\Gamma (m+2) \Gamma (n+2) \Gamma (m+\epsilon +1) \Gamma (n+\epsilon +1) \Gamma \left(m+n+2 \epsilon +\frac{3}{2}\right)}
&& \end{flalign}

\begin{flalign}
	& I_4^{2,7} = c_{I}  2 \pi  \left(\frac{m_2^2}{4 m_1^2}\right)^{1-\epsilon } \left(\frac{m_3^2}{4 m_1^2}\right) \Gamma (\epsilon ) \Gamma (1-\epsilon ) \nonumber \\
	& \times \sum_{m,n=0}^{\infty} \left(\frac{m_3^2}{4 m_1^2}\right)^m \left(\frac{m_2^2}{4 m_1^2}\right)^n \frac{\Gamma (m+n+2) \Gamma (m+n+\epsilon +1) \Gamma (m+n+2 \epsilon )}{\Gamma (m+2) \Gamma (n+1) \Gamma (m+\epsilon +1) \Gamma (n-\epsilon +2) \Gamma \left(m+n+\epsilon +\frac{3}{2}\right)}
&& \end{flalign}

\begin{flalign}
	& I_4^{2,8} = - c_{I} \left(\frac{m_2^2}{4 m_1^2}\right)^{2-\epsilon } \left(\frac{m_3^2}{4 m_1^2}\right)^{-\epsilon -\frac{1}{2}} \frac{\Gamma (1-\epsilon ) \Gamma (\epsilon )} {\Gamma \left(\frac{3}{2}-\epsilon \right) \Gamma \left(\epsilon -\frac{1}{2}\right)} \nonumber \\
	& \times \sum_{m,n=0}^{\infty} \left(\frac{m_2^2}{m_3^2}\right)^m \left(\frac{m_2^2}{4 m_1^2}\right)^n \frac{\Gamma \left(m+\frac{3}{2}\right) \Gamma \left(n+\frac{1}{2}\right) \Gamma \left(m+\epsilon +\frac{1}{2}\right) \Gamma \left(n-\epsilon +\frac{3}{2}\right) \Gamma \left(n+\epsilon -\frac{1}{2}\right)}{\Gamma (n+1) \Gamma (m+n+2) \Gamma (m+n-\epsilon +3)}
&& \end{flalign}

\begin{flalign}
	& I_4^{2,9} = c_{I} \left(\frac{m_2^2}{4 m_1^2}\right)^{2} \left(\frac{m_3^2}{4 m_1^2}\right)^{-2\epsilon -\frac{1}{2}} \frac{\Gamma (1-\epsilon ) \Gamma (\epsilon )} {\Gamma \left(\frac{3}{2}-\epsilon \right) \Gamma \left(\epsilon -\frac{1}{2}\right)} \nonumber \\
	& \times \sum_{m,n=0}^{\infty} \left(\frac{m_2^2}{m_3^2}\right)^m \left(\frac{m_2^2}{4 m_1^2}\right)^n \frac{\Gamma \left(n+\frac{1}{2}\right) \Gamma \left(m+\epsilon +\frac{3}{2}\right) \Gamma \left(m+2 \epsilon +\frac{1}{2}\right) \Gamma \left(n-\epsilon +\frac{3}{2}\right) \Gamma \left(n+\epsilon -\frac{1}{2}\right)}{\Gamma (n+1) \Gamma (m+n+3) \Gamma (m+n+\epsilon +2)}
&& \end{flalign}

\section{Residues of $H_{\{2,1,1\}} \left(m_1^2, m_2^2, m_3^2; m_1^2 \right)$ \label{Sec:H211Result}}

\renewcommand{\theequation}{D-\arabic{equation}}
  \setcounter{equation}{0}  

\subsection{Residues of integral $J_1$}

\begin{flalign}
    J_1 = -\frac{\pi^{4-2\epsilon}}{(m_1^2)^{2\epsilon}} \frac{\Gamma(\epsilon)\Gamma(2\epsilon)\Gamma^2(1-\epsilon)\Gamma(2-4\epsilon)}{\Gamma(2-2\epsilon)\Gamma(2-3\epsilon)} &&
\end{flalign}

\subsection{Residues of integral $J_2$}

\subsubsection*{LHS}

\begin{flalign}
    J_2^{L,1} = 2^{1-4\epsilon} \pi^{\frac{7}{2}-2\epsilon} \left(m_1^2\right)^{-2\epsilon}  \frac{\Gamma (1-2\epsilon) \Gamma \left(\frac{3}{2}-2 \epsilon \right) \Gamma (1-\epsilon)^2 \Gamma (\epsilon ) \Gamma (2\epsilon)}{\Gamma (2-3 \epsilon) \Gamma (2-2 \epsilon)} &&
\end{flalign}

\begin{flalign}
    J_2^{L,2} = \pi ^{4-2 \epsilon} \left(m_1^2\right)^{-2 \epsilon} \Gamma (\epsilon-1) \Gamma (\epsilon) \left(\frac{m_3^2}{m_1^2}\right)^{-\epsilon}
    {}_3F_2 \left( \begin{array}{c}
	   \frac{3}{2}-\epsilon ,\epsilon ,1 \\
	   2-2 \epsilon ,2-\epsilon \\
	\end{array}	\bigg| \frac{4 m_1^2}{m_3^2} \right)
&& \end{flalign}

\begin{flalign}
    J_2^{L,3} = \pi ^{4-2 \epsilon} \left(m_3^2\right)^{-2 \epsilon} \frac{\left(\Gamma (1-\epsilon )^2 \Gamma (\epsilon ) \Gamma (2 \epsilon )\right)}{\Gamma (2-\epsilon )}
    {}_3F_2 \left( \begin{array}{c}
	   \frac{3}{2},2 \epsilon ,1  \\
	   2,2-\epsilon \\
	\end{array}	\bigg| \frac{4 m_1^2}{m_3^2} \right)
&& \end{flalign}

\subsubsection*{RHS}

\begin{flalign}
    J_2^{R,1} = -\frac{1}{2} \pi ^{4-2 \epsilon } \left(m_1^2\right)^{-2 \epsilon } \Gamma (1-\epsilon ) \Gamma (\epsilon ) \Gamma (2 \epsilon -1) \left(\frac{m_3^2}{m_1^2}\right)^{1-2 \epsilon}  &&
\end{flalign}

\begin{flalign}
    J_2^{R,2} =  4 \pi ^{\frac{9}{2}-2 \epsilon} m_3^2  \left(4 m_1^2\right)^{-1-2\epsilon} 
    \frac{\Gamma (1-4 \epsilon) \Gamma (1-\epsilon)^2 \Gamma (\epsilon ) \Gamma (4 \epsilon)}{ \left(\Gamma (1-3 \epsilon ) \Gamma (1-2 \epsilon ) \Gamma \left(2 \epsilon +\frac{1}{2}\right)\right)}
    {}_3F_2 \left( \begin{array}{c}
	   2\epsilon, 3\epsilon, 1 \\
	   2, 2\epsilon+\frac{1}{2}\\
	\end{array}	\bigg| \frac{m_3^2}{4m_1^2} \right)    
    &&
\end{flalign}

\begin{flalign}
    J_2^{R,3} = -\pi ^{\frac{9}{2}-2 \epsilon } \left(2 m_1^2\right)^{-2 \epsilon }
    \left(\frac{m_3^2}{m_1^2}\right)^{1-\epsilon }
    \frac{\Gamma (-2 \epsilon ) \Gamma (-\epsilon ) \Gamma (\epsilon +1) \Gamma (2 \epsilon +1) }{\Gamma (1-2 \epsilon ) \Gamma (2-\epsilon ) \Gamma \left(\epsilon +\frac{1}{2}\right)}
    {}_3F_2 \left( \begin{array}{c}
		\epsilon, 2\epsilon, 1  \\
		2-\epsilon, \epsilon+\frac{1}{2} \\
	\end{array}	\bigg| \frac{m_3^2}{4m_1^2} \right)
    &&
\end{flalign}

\begin{flalign}
    J_2^{R,4} =
    8^{2 \epsilon -1} \pi ^{5-2 \epsilon } \left(m_3^2\right)^{-2 \epsilon}  \left(\frac{m_3^2}{m_1^2}\right)^{\tfrac{3}{2}} 
    \frac{ \Gamma (2-4 \epsilon ) \Gamma \left(\epsilon +\frac{1}{2}\right) \Gamma \left(2 \epsilon -\frac{1}{2}\right)}{\Gamma \left(\frac{5}{2}-2 \epsilon \right) \Gamma \left(\frac{1}{2}-\epsilon \right)}  
    {}_2F_1 \left( \begin{array}{c}
		\frac{1}{2},\epsilon +\frac{1}{2}  \\
		\frac{5}{2}-2 \epsilon \\
	\end{array}	\bigg| \frac{m_3^2}{4m_1^2} \right)
    &&
\end{flalign}

\subsection{Residues of integral $J_4$}

The expressions listed below need be multiplied by a factor of  $-2^{-4 \epsilon-1} \pi ^{\frac{3}{2}-2 \epsilon} \left(m_1^2\right)^{-2 \epsilon}$ to get the residue.

\subsubsection*{Cone 1}

\begin{flalign}
	 J_4^{1,1} &= \frac{\Gamma \left(\frac{3}{2}-2 \epsilon \right) \Gamma (1-\epsilon )^2 \Gamma (\epsilon ) \Gamma (2 \epsilon -1)}{\Gamma (2-3\epsilon)} &&
\end{flalign}

\begin{flalign}
	 J_4^{1,2} &= - \frac{\sqrt{\pi}}{2} \Gamma (\epsilon -1) \Gamma (\epsilon ) \left(\frac{16 m_1^2}{m_2^2}\right)^{\epsilon} {}_3F_2 \left( \begin{array}{c}
		1,\frac{3}{2}-\epsilon ,\epsilon \\
		2-2 \epsilon ,2-\epsilon \\
	\end{array}	\bigg| \frac{4 m_1^2}{m_2^2} \right)
&& \end{flalign}

\begin{flalign}
	 J_4^{1,3} &= -\sqrt{\pi } \Gamma (\epsilon ) \Gamma (1-\epsilon ) \Gamma (2 \epsilon -1) \left(\frac{4 m_1^2}{m_2^2}\right)^{2 \epsilon -1} \bigg[ {}_2F_1 \left( \begin{array}{c}
		\frac{1}{2},2 \epsilon -1 \\
		1-\epsilon \\
	\end{array}	\bigg| \frac{4 m_1^2}{m_2^2} \right) -1 \bigg]
&& \end{flalign}

\begin{flalign}
	 & J_4^{1,4} = \frac{m_2^2}{m_3^2} \left(\frac{4 m_1^2}{m_2^2}\right)^{\epsilon} \Gamma (\epsilon )^2 \Gamma (1-\epsilon )^2 \nonumber \\
	 &  \times \sum_{m,n=0}^{\infty} \left(\frac{m_2^2}{m_3^2}\right)^m \left(\frac{4 m_1^2}{m_3^2}\right)^n \frac{\Gamma (m+n+1) \Gamma \left(n-\epsilon +\frac{3}{2}\right) \Gamma (m+n-\epsilon +2)}{\Gamma (m+1) \Gamma (n+1) \Gamma (m-\epsilon +2) \Gamma (n-2 \epsilon +2) \Gamma (n-\epsilon +2)}
&& \end{flalign}

\begin{flalign}
	 & J_4^{1,5} = -\frac{m_2^2}{m_3^2} \left(\frac{4 m_1^2}{m_3^2}\right)^{\epsilon} \left(\frac{4 m_1^2}{m_2^2}\right)^{\epsilon } \Gamma (\epsilon )^2 \Gamma (1-\epsilon )^2 \nonumber \\
	 &  \times \sum_{m,n=0}^{\infty} \left(\frac{m_2^2}{m_3^2}\right)^m \left(\frac{4 m_1^2}{m_3^2}\right)^n \frac{\Gamma \left(n+\frac{3}{2}\right) \Gamma (m+n+2) \Gamma (m+n+\epsilon +1)}{\Gamma (m+1) \Gamma (n+2) \Gamma (m-\epsilon +2) \Gamma (n-\epsilon +2) \Gamma (n+\epsilon +1)}
&& \end{flalign}

\begin{flalign}
	 & J_4^{1,6} = \frac{m_2^2}{m_3^2} \left(\frac{4 m_1^2}{m_3^2}\right)^{2\epsilon} \Gamma (\epsilon )^2 \Gamma (1-\epsilon )^2 \nonumber \\
	 &  \times \sum_{m,n=0}^{\infty} \left(\frac{m_2^2}{m_3^2}\right)^m \left(\frac{4 m_1^2}{m_3^2}\right)^n \frac{\Gamma \left(n+\frac{3}{2}\right) \Gamma (m+n+\epsilon +2) \Gamma (m+n+2 \epsilon +1)}{\Gamma (m+2) \Gamma (n+2) \Gamma (m+\epsilon +1) \Gamma (n-\epsilon +2) \Gamma (n+\epsilon +1)}
&& \end{flalign}

\begin{flalign}
	 & J_4^{1,7} = -\frac{m_2^2}{m_3^2} \left(\frac{4 m_1^2}{m_3^2}\right)^{\epsilon} \Gamma (\epsilon )^2 \Gamma (1-\epsilon )^2 \nonumber \\
	 &  \times \sum_{m,n=0}^{\infty} \left(\frac{m_2^2}{m_3^2}\right)^m \left(\frac{4 m_1^2}{m_3^2}\right)^n \frac{\Gamma (m+n+2) \Gamma \left(n-\epsilon +\frac{3}{2}\right) \Gamma (m+n+\epsilon +1)}{\Gamma (m+2) \Gamma (n+1) \Gamma (m+\epsilon +1) \Gamma (n-2 \epsilon +2) \Gamma (n-\epsilon +2)}
&& \end{flalign}

\subsubsection*{Cone 3}

\begin{flalign}
	 J_4^{3,1} &= \sqrt{\pi} \left(\frac{m_2^2}{4 m_1^2}\right)^{1-2 \epsilon } \Gamma (2\epsilon -1) \Gamma (1-\epsilon ) \Gamma (\epsilon )
&& \end{flalign}

\begin{flalign}
	 J_4^{3,2} &= -\sqrt{\pi} \left(\frac{m_3^2}{4 m_1^2}\right)^{-\epsilon} \Gamma (\epsilon )  \Gamma (\epsilon-1) \left(\frac{m_2^2}{4 m_1^2}\right)^{1-\epsilon} {}_2F_1 \left( \begin{array}{c}
		1,\epsilon \\
		2-\epsilon \\
	\end{array}	\bigg| \frac{m_2^2}{m_3^2} \right)
&& \end{flalign}

\begin{flalign}
	 J_4^{3,3} &= -\sqrt{\pi } \left(\frac{m_3^2}{4 m_1^2}\right)^{-2\epsilon } \left(\frac{m_2^2}{4 m_1^2}\right) \Gamma (1-\epsilon) \Gamma (\epsilon ) \Gamma (2\epsilon ) {}_2F_1 \left( \begin{array}{c}
		1,2\epsilon \\
		2 \\
	\end{array}	\bigg| \frac{m_2^2}{m_3^2} \right)
&& \end{flalign}

\begin{flalign}
	 J_4^{3,4} &= -\sqrt{\pi} \left(\frac{m_2^2}{4 m_1^2}\right)^{\frac{3}{2}-2 \epsilon} \frac{ \Gamma \left(\frac{3}{2}-2 \epsilon \right) \Gamma (1-\epsilon ) \Gamma \left(\epsilon +\frac{1}{2}\right) \Gamma \left(2 \epsilon -\frac{1}{2}\right)}{\Gamma \left(\frac{5}{2}-2 \epsilon \right)} {}_2F_1 \left( \begin{array}{c}
		\frac{1}{2},\epsilon +\frac{1}{2}  \\
		\frac{5}{2}-2 \epsilon \\
	\end{array}	\bigg| \frac{m_2^2}{4 m_1^2} \right)
&& \end{flalign}

\begin{flalign}
	 & J_4^{3,5} = \pi \left(\frac{m_3^2}{4 m_1^2}\right)^{\frac{1}{2}-\epsilon} \left(\frac{m_2^2}{4 m_1^2}\right)^{1-\epsilon}  \Gamma (\epsilon ) \Gamma (1-\epsilon ) \nonumber \\
	 & \times \sum_{m,n=0}^{\infty} \left(\frac{\text{me}^2}{4 m_1^2}\right)^{m} \left(\frac{m_2^2}{4 m_1^2}\right)^{n} \frac{\Gamma \left(m+n+\frac{1}{2}\right) \Gamma \left(m+n-\epsilon +\frac{3}{2}\right) \Gamma \left(m+n+\epsilon +\frac{1}{2}\right)}{\Gamma \left(m+\frac{1}{2}\right) \Gamma (n+1) \Gamma (m+n+1) \Gamma \left(m-\epsilon +\frac{3}{2}\right) \Gamma (n-\epsilon +2)}
&& \end{flalign}

\begin{flalign}
	 & J_4^{3,6} = \left(\frac{m_2^2}{4 m_1^2}\right) \left(\frac{m_3^2}{4 m_1^2}\right)^{-2 \epsilon +\frac{1}{2}} \Gamma (\epsilon ) \Gamma (1-\epsilon ) \Gamma \left(2 \epsilon -\frac{1}{2}\right) \Gamma \left(\frac{3}{2}-2 \epsilon \right) \nonumber \\
	 & \times \sum_{m,n=0}^{\infty} \left(\frac{m_3^2}{4 m_1^2}\right)^m	 \left(\frac{m_2^2}{4 m_1^2}\right)^n \frac{\Gamma \left(m+n+\frac{1}{2}\right) \Gamma \left(m+n-\epsilon +\frac{3}{2}\right) \Gamma \left(m+n+\epsilon +\frac{1}{2}\right)}{\Gamma (n+2) \Gamma (m+n+1) \Gamma \left(m-2 \epsilon +\frac{3}{2}\right) \Gamma \left(m-\epsilon +\frac{1}{2}\right) \Gamma (n+\epsilon +1)}
&& \end{flalign}

\begin{flalign}
	 & J_4^{3,7} = \pi \left(\frac{m_3^2}{4 m_1^2}\right)^{1-\epsilon} \left(\frac{m_2^2}{4 m_1^2}\right)^{1-\epsilon} \Gamma (\epsilon ) \Gamma (1-\epsilon ) \nonumber \\
	 & \times \sum_{m,n=0}^{\infty} \left(\frac{m_3^2}{4 m_1^2}\right)^m \left(\frac{m_2^2}{4 m_1^2}\right)^n \frac{\Gamma (m+n+1) \Gamma (m+n-\epsilon +2) \Gamma (m+n+\epsilon +1)}{\Gamma (m+1) \Gamma (n+1) \Gamma \left(m+n+\frac{3}{2}\right) \Gamma (m-\epsilon +2) \Gamma (n-\epsilon +2)}
&& \end{flalign}

\begin{flalign}
	 & J_4^{3,8} = -2 \pi  \left(\frac{m_2^2}{4 m_1^2}\right) \left(\frac{m_3^2}{4 m_1^2}\right)^{1-\epsilon} \Gamma (\epsilon ) \Gamma (1-\epsilon ) \nonumber \\
	 & \times \sum_{m,n=0}^{\infty} \left(\frac{m_3^2}{4 m_1^2}\right)^m \left(\frac{m_2^2}{4 m_1^2}\right)^n\frac{\Gamma (m+n+2) \Gamma (m+n+\epsilon +1) \Gamma (m+n+2 \epsilon +1)}{\Gamma (m+1) \Gamma (n+2) \Gamma (m-\epsilon +2) \Gamma (n+\epsilon +1) \Gamma \left(m+n+\epsilon +\frac{3}{2}\right)}
&& \end{flalign}

\begin{flalign}
	 & J_4^{3,9} = -\left(\frac{m_3^2}{4 m_1^2}\right) \left(\frac{m_2^2}{4 m_1^2}\right) \frac{\Gamma \left(\frac{3}{2}-2 \epsilon \right) \Gamma (1-\epsilon )^2 \Gamma (\epsilon )^2 \Gamma \left(2 \epsilon -\frac{1}{2}\right)}{\Gamma (1-3 \epsilon) \Gamma (3 \epsilon)}	\nonumber \\
	 &  \times \sum_{m,n=0}^{\infty} \left(\frac{m_3^2}{4 m_1^2}\right)^m \left(\frac{m_2^2}{4 m_1^2}\right)^n	  \frac{\Gamma (m+n+\epsilon +2) \Gamma (m+n+2 \epsilon +1) \Gamma (m+n+3 \epsilon +1)}{\Gamma (m+2) \Gamma (n+2) \Gamma (m+\epsilon +1) \Gamma (n+\epsilon +1) \Gamma \left(m+n+2 \epsilon +\frac{3}{2}\right)}
&& \end{flalign}

\begin{flalign}
	 & J_4^{3,10} = -2 \pi  \left(\frac{m_3^2}{4 m_1^2}\right) \left(\frac{m_2^2}{4 m_1^2}\right)^{1-\epsilon} \Gamma (\epsilon ) \Gamma (1-\epsilon )	\nonumber \\
	 & \times \sum_{m,n=0}^{\infty} \left(\frac{m_3^2}{4 m_1^2}\right)^m \left(\frac{m_2^2}{4 m_1^2}\right)^n \frac{\Gamma (m+n+2) \Gamma (m+n+\epsilon +1) \Gamma (m+n+2 \epsilon +1)}{\Gamma (m+2) \Gamma (n+1) \Gamma (m+\epsilon +1) \Gamma (n-\epsilon +2) \Gamma \left(m+n+\epsilon +\frac{3}{2}\right)}
&& \end{flalign}

\begin{flalign}
	 & J_4^{3,11} = -\left(\frac{m_3^2}{4 m_1^2}\right)^{-\epsilon -\frac{1}{2}} \left(\frac{m_2^2}{4 m_1^2}\right)^{2-\epsilon} \frac{\Gamma (\epsilon ) \Gamma (1-\epsilon )}{\Gamma \left(\frac{1}{2}-\epsilon \right) \Gamma \left(\epsilon +\frac{1}{2}\right)}	\nonumber \\
	 &  \times \sum_{m,n=0}^{\infty} \left(\frac{m_2^2}{m_3^2}\right)^m \left(\frac{m_2^2}{4 m_1^2}\right)^n \frac{\Gamma \left(m+\frac{3}{2}\right) \Gamma \left(n+\frac{1}{2}\right) \Gamma \left(m+\epsilon +\frac{1}{2}\right) \Gamma \left(n-\epsilon +\frac{3}{2}\right) \Gamma \left(n+\epsilon +\frac{1}{2}\right)}{\Gamma (n+1) \Gamma (m+n+2) \Gamma (m+n-\epsilon +3)}
&& \end{flalign}

\begin{flalign}
	 & J_4^{3,12} = -\left(\frac{m_3^2}{4 m_1^2}\right)^{-2\epsilon -\frac{1}{2}} \left(\frac{m_2^2}{4 m_1^2}\right)^{2} \frac{\Gamma (\epsilon ) \Gamma (1-\epsilon )}{\Gamma \left(\frac{1}{2}-\epsilon \right) \Gamma \left(\epsilon +\frac{1}{2}\right)}	\nonumber \\
	 & \times \sum_{m,n=0}^{\infty} \left(\frac{m_2^2}{m_3^2}\right)^m \left(\frac{m_2^2}{4 m_1^2}\right)^n \frac{\Gamma \left(n+\frac{1}{2}\right) \Gamma \left(m+\epsilon +\frac{3}{2}\right) \Gamma \left(m+2 \epsilon +\frac{1}{2}\right) \Gamma \left(n-\epsilon +\frac{3}{2}\right) \Gamma \left(n+\epsilon +\frac{1}{2}\right)}{\Gamma (n+1) \Gamma (m+n+3) \Gamma (m+n+\epsilon +2)}
&& \end{flalign}

\section{Residues of $H_{\{1,1,1\}} \left(m_1^2, m_2^2, m_2^2; m_3^2 \right)$ \label{Sec:H211Results}}

\renewcommand{\theequation}{E-\arabic{equation}}
  \setcounter{equation}{0}  

\subsection{Residues of integral $K_1$}

\begin{flalign}
    K_1 = \frac{\pi^{4-2\epsilon}}{\left(-m_3^2\right)^{2\epsilon-1}} \frac{\Gamma (1-\epsilon )^3 \Gamma (2 \epsilon -1)}{\Gamma(3-3\epsilon)}
    &&
\end{flalign}

\subsection{Residues of integral $K_2$}

\subsubsection*{LHS}

\begin{flalign}
	 K_2^{L,1} &= \frac{\pi^{4-2\epsilon}}{\left(-m_3^2\right)^{2\epsilon-1}} \left( \frac{m_1^2}{m_3^2} \right)  \frac{\Gamma (1-\epsilon)^2 \Gamma(-\epsilon) \Gamma(2\epsilon)}{\Gamma(2-3\epsilon)} {}_3F_2 \left( \begin{array}{c}
	   2\epsilon, 3\epsilon-1, 1 \\
	   2, \epsilon+1 \\
	\end{array}	\bigg| \frac{m_1^2}{m_3^2} \right)
&& \end{flalign}

\begin{flalign}
	 K_2^{L,2} &=
     \frac{\pi^{4-2\epsilon}}{\left(-m_3^2\right)^{2\epsilon-1}} \left(-\frac{m_1^2}{m_3^2}\right)^{1-\epsilon}
     \frac{\Gamma(1-\epsilon)^3 \Gamma(\epsilon)^2 \Gamma(2\epsilon-1)}{\Gamma(1-2\epsilon) \Gamma(2-\epsilon) \Gamma(2\epsilon)}
     {}_2F_1 \left( \begin{array}{c}
	   \epsilon, 2\epsilon-1 \\
	   2-\epsilon \\
	\end{array}	\bigg| \frac{m_1^2}{m_3^2} \right)
&& \end{flalign}

\subsubsection*{RHS}

\begin{flalign}
	 K_2^{R,1} &= -\frac{\pi^{4-2\epsilon}}{\left(-m_3^2\right)^{2\epsilon-1}}
     \frac{\Gamma (1-\epsilon )^3 \Gamma (2 \epsilon -1)}{\Gamma (3-3 \epsilon )}
&& \end{flalign}

\begin{flalign}
	 K_2^{R,2} &= -\frac{2\pi^{4-2\epsilon}}{\left(-m_3^2\right)^{2\epsilon-1}}
      \left(-\frac{m_3^2}{m_1^2}\right)^{2\epsilon-1}
      \Gamma(1-\epsilon) \Gamma(\epsilon) \Gamma(2\epsilon-2) 
    {}_2F_1 \left( \begin{array}{c}
	   \epsilon ,2 \epsilon -1 \\
	   2-\epsilon \\
	\end{array}	\bigg| \frac{m_3^2}{m_1^2} \right)
&& \end{flalign}

\subsection{Residues of integral $K_3$}

The residues for the integral $K_3$ are given below in terms of a multiplicative prefactor 
$k_I = (m_1^2)^{1-\epsilon} (m_2^2)^{-\epsilon} \pi^{\tfrac{9}{2}-2\epsilon}/2$.

\subsubsection*{Cone 1}

\begin{flalign}
	 K_3^{1,1} &= - k_I \frac{\Gamma(2-2\epsilon) \Gamma(1-\epsilon) \Gamma(\epsilon) \Gamma(2\epsilon-1)}{\Gamma \left(\frac{3}{2}-\epsilon\right) \Gamma(2-\epsilon)} \left(\frac{m_1^2}{4m_2^2}\right)^{-\epsilon} 
     {}_2F_1 \left( \begin{array}{c}
	   \epsilon, 2\epsilon-1 \\
	   2-\epsilon \\
	\end{array}	\bigg| \frac{m_3^2}{m_1^2} \right)
&& \end{flalign}

\begin{flalign}
    & K_3^{1,2} = k_I \Gamma (1-\epsilon) \Gamma (\epsilon) \left(\frac{m_1^2}{4 m_2^2}\right)^{\epsilon-1} \nonumber \\ 
    & \times \sum_{m,n=0}^{\infty} 
     \left(\frac{m_3^2}{4 m_2^2}\right)^m 
    \left(\frac{m_1^2}{4 m_2^2}\right)^{n}
     \frac{\Gamma (m+n+1) \Gamma (m+n+\epsilon ) \Gamma (m+n+2 \epsilon -1)}{\Gamma (m+1) \Gamma (n+1) \Gamma (m-\epsilon +2) \Gamma (n+\epsilon ) \Gamma \left(m+n+\epsilon +\frac{1}{2}\right)} 
&& \end{flalign}

\begin{flalign}
	& K_3^{1,3} = - k_I \Gamma(\epsilon) \Gamma(1-\epsilon) \nonumber \\
    & \times \sum_{m,n=0}^{\infty} \left(\frac{m_3^2}{4 m_2^2}\right)^m \left(\frac{m_1^2}{4 m_2^2}\right)^n
     \frac{\Gamma (m+n+1) \Gamma (m+n-\epsilon+2) \Gamma (m+n+\epsilon)}{\Gamma (m+1) \Gamma (n+1) \Gamma \left(m+n+\frac{3}{2}\right) \Gamma (m-\epsilon +2) \Gamma (n-\epsilon +2)}
     &&
\end{flalign}

\subsubsection*{Cone 2}

\begin{flalign}
	 K_3^{2,1} &= k_I \left(\frac{m_1^2}{4 m_2^2}\right)^{-\epsilon}
     \left(-\frac{m_3^2}{m_1^2}\right)^{1-2\epsilon}
     \frac{\Gamma (1-2\epsilon) \Gamma (1-\epsilon)^2 \Gamma (2\epsilon)  }{\Gamma (3-3\epsilon) \Gamma \left(\frac{3}{2}-\epsilon \right)}
    {}_2F_1 \left( \begin{array}{c}
	   2 \epsilon -1,3 \epsilon -2 \\
	   \epsilon \\
	\end{array}	\bigg| \frac{m_1^2}{m_3^2} \right)
&& \end{flalign}

\begin{flalign}
	K_3^{2,2} &= k_I \frac{4}{\sqrt{\pi}} \frac{\Gamma^2(1-\epsilon) \Gamma^2(\epsilon)}{\Gamma(3-2\epsilon)}
    \left(-\frac{m_3^2}{m_2^2}\right)^{-\epsilon}
    {}_2F_1 \left( \begin{array}{c}
	   2\epsilon-1, \epsilon \\
	   2-\epsilon \\
	\end{array}	\bigg| \frac{m_1^2}{m_3^2} \right)
&& \end{flalign}

\subsubsection*{Cone 3}

\begin{flalign}
	 K_3^{3,1} &= - k_I \frac{4}{\pi} \frac{m_2^2}{m_1^2} \left(-\frac{m_3^2}{m_1^2}\right)^{-\epsilon}     \frac{\Gamma(1-\epsilon)^3 \Gamma(\epsilon )^3}{\Gamma (2-2\epsilon) \Gamma (2\epsilon-1)} \nonumber \\ 
     & \times \sum_{m,n=0}^{\infty} \left(\frac{4 m_2^2}{m_3^2}\right)^m \left(\frac{m_1^2}{m_3^2}\right)^n \frac{\Gamma \left(m+\frac{1}{2}\right) \Gamma (m+n+\epsilon ) \Gamma (m+n+2 \epsilon -1)}
     {\Gamma (m+1) \Gamma (n+1) \Gamma (m-\epsilon +2) \Gamma (m+\epsilon ) \Gamma (n+\epsilon)}
&& \end{flalign}

\begin{flalign}
	 K_3^{3,2} &= - \frac{k_I}{\pi} \Gamma(1-\epsilon )^2 \Gamma(\epsilon )^2 \left(\frac{4 m_2^2}{m_3^2}\right) \nonumber \\
     & \times \sum_{m,n=0}^{\infty} \left(\frac{4 m_2^2}{m_3^2}\right)^m \left(\frac{m_1^2}{m_3^2}\right)^n \frac{\Gamma \left(m+\frac{1}{2}\right) \Gamma (m+n+1) \Gamma (m+n+\epsilon)}
     {\Gamma (m+1) \Gamma (n+1) \Gamma (m-\epsilon +2) \Gamma (m+\epsilon ) \Gamma (n-\epsilon +2)}
&& \end{flalign}

\begin{flalign}
	 K_3^{3,3} &= - k_I \left(\frac{m_1^2}{4 m_2^2}\right)^{-\epsilon-1} \left(-\frac{m_3^2}{m_1^2}\right)^{-2\epsilon} 
     \frac{\Gamma (1-2\epsilon) \Gamma(1-\epsilon)^2 \Gamma(\epsilon)^2 \Gamma(2\epsilon)}
     {\Gamma(1-3\epsilon) \Gamma\left(\frac{1}{2}-\epsilon \right) \Gamma(3\epsilon) \Gamma \left(\epsilon +\frac{1}{2}\right)} \nonumber \\
     & \times \sum_{m,n=0}^{\infty} \left(\frac{4 m_2^2}{m_3^2}\right)^m \left(\frac{m_1^2}{m_3^2}\right)^n
     \frac{\Gamma \left(m+\epsilon +\frac{1}{2}\right) \Gamma (m+n+2\epsilon) \Gamma(m+n+3\epsilon-1)}
      {\Gamma(m+2) \Gamma(n+1) \Gamma(m+\epsilon+1) \Gamma(m+2\epsilon) \Gamma(n+\epsilon)}
&& \end{flalign}

\begin{flalign}
	 K_3^{3,4} &= - k_I \frac{\Gamma(1-\epsilon)^2 \Gamma(\epsilon)^2}{\Gamma\left(\frac{1}{2}-\epsilon\right) \Gamma\left(\epsilon+\frac{1}{2}\right)} \left(-\frac{m_3^2}{4m_2^2}\right)^{-\epsilon-1} \nonumber \\
    & \times \sum_{m,n=0}^{\infty} \left(\frac{4 m_2^2}{m_3^2}\right)^m \left(\frac{m_1^2}{m_3^2}\right)^n
     \frac{\Gamma\left(m+\epsilon+\frac{1}{2}\right) \Gamma(m+n+\epsilon+1) \Gamma(m+n+2\epsilon)}
     {\Gamma(m+2) \Gamma(n+1) \Gamma(m+\epsilon+1) \Gamma(m+2\epsilon) \Gamma(n-\epsilon+2)}
&& \end{flalign}

\begin{flalign}
	 K_3^{3,5} &= -k_I\frac{\Gamma(1-\epsilon)^2 \Gamma(\epsilon)^2}{2\pi}
     \left(\frac{m_1^2}{4 m_2^2}\right)^{\epsilon-2} \left(\frac{m_1^2}{m_3^2}\right) \nonumber \\
     & \times \sum_{m,n=0}^{\infty} \left(\frac{4 m_2^2}{m_3^2}\right)^m \left(\frac{m_1^2}{m_3^2}\right)^n
     \frac{\Gamma(m+n+1) \Gamma\left(m-\epsilon+\frac{3}{2}\right) \Gamma(m+n+\epsilon)}
     {\Gamma(m+1) \Gamma(n+1) \Gamma (m-2\epsilon+3) \Gamma(m-\epsilon+2) \Gamma(n+\epsilon)}
&& \end{flalign}

\subsubsection*{Cone 4}

\begin{flalign}
	 K_3^{4,1} &= -\frac{k_I}{\pi} \Gamma(1-\epsilon)^2 \Gamma(\epsilon)^2
     \left(\frac{4 m_2^2}{m_1^2}\right) \nonumber \\
     & \times \sum_{m,n=0}^{\infty}
     \left(\frac{4 m_2^2}{m_1^2}\right)^m \left(\frac{m_3^2}{m_1^2}\right)^n
     \frac{\Gamma\left(m+\frac{1}{2}\right) \Gamma(m+n+1) \Gamma(m+n+\epsilon)}{\Gamma(m+1) \Gamma(n+1) \Gamma(m-\epsilon+2) \Gamma(m+\epsilon) \Gamma(n-\epsilon+2)}
&& \end{flalign}

\begin{flalign}
	 K_3^{4,2} &= \frac{k_I}{2\pi} \Gamma(\epsilon)^2 \Gamma(1-\epsilon)^2 \left(\frac{4 m_2^2}{m_1^2}\right)^{\epsilon+1} \nonumber \\
     & \times \sum_{m,n=0}^{\infty} \left(\frac{4 m_2^2}{m_1^2}\right)^m  \left(\frac{m_3^2}{m_1^2}\right)^n
     \frac{\Gamma \left(m+\epsilon +\frac{1}{2}\right) \Gamma (m+n+\epsilon +1) \Gamma (m+n+2 \epsilon)}{\Gamma (m+2) \Gamma (n+1) \Gamma (m+\epsilon +1) \Gamma (m+2 \epsilon ) \Gamma (n-\epsilon +2)}
&& \end{flalign}

\begin{flalign}
	 K_3^{4,3} &= \frac{k_I}{2\pi} \Gamma(1-\epsilon)^2 \Gamma(\epsilon)^2 \left(\frac{m_1^2}{4 m_2^2}\right)^{\epsilon-2} \nonumber \\
     & \times \sum_{m,n=0}^{\infty} \left(\frac{4 m_2^2}{m_1^2}\right)^m  \left(\frac{m_3^2}{m_1^2}\right)^n
     \frac{\Gamma(m+n+1) \Gamma\left(m-\epsilon+\frac{3}{2}\right) \Gamma(m+n-\epsilon+2)}{\Gamma(m+1) \Gamma(n+1) \Gamma(m-2\epsilon+3) \Gamma(m-\epsilon+2) \Gamma(n-\epsilon+2)}
&& \end{flalign}

\end{appendices}

\end{document}